\documentclass[fleqn,usenatbib]{mnras}

\usepackage{newtxtext,newtxmath}
\usepackage{mathtools}

\usepackage[T1]{fontenc}

\DeclareRobustCommand{\VAN}[3]{#2}
\let\VANthebibliography\thebibliography
\def\thebibliography{\DeclareRobustCommand{\VAN}[3]{##3}\VANthebibliography}

\usepackage{graphicx}	
\usepackage{amsmath}	
\usepackage{pdflscape}
\usepackage{ulem}

\newcommand{\beq}{\begin{equation}}
\newcommand{\eeq}{\end{equation}}
\newcommand{\bea}{\begin{eqnarray}}
\newcommand{\eea}{\end{eqnarray}}
\def\be{\begin{equation}}
\def\ee{\end{equation}}
\def\ba{\begin{eqnarray}}
\def\ea{\end{eqnarray}}

\def\nn{\nonumber}
\newcommand{\bi}{\begin{itemize}}
\newcommand{\ei}{\end{itemize}}
\newcommand{\efi}{\end{figure}}
\newcommand{\efib}{\end{figure}}
\newcommand{\efig}{\end{figure*}}
\newcommand{\no}{\nonumber}

\newcommand{\bfx}{\mbox{\boldmath$x$}}
\newcommand{\bfk}{\mbox{\boldmath$k$}}

\newcommand{\bfq}{\mbox{\boldmath$q$}}

\title[Emulator of Hybrid Model]{High precision accelerator for our hybrid model of the redshift space power spectrum}

\author[M. Icaza-Lizaola et al.]{M. Icaza-Lizaola$^{1}$\thanks{E-mail: icaza@kasi.re.kr},
Yong-Seon Song $^{1}$\thanks{E-mail: ysong@kasi.re.kr},
Minji Oh$^{2}$,
Yi Zheng$^{3,4}$
\\
\scriptsize $^{1}$Korea Astronomy and Space Science Institute,  776 Daedeok-daero,
Yuseong-gu, Daejeon 34055, Republic of Korea.\\
\scriptsize $^{2}$Chosun University, 375 Seosuk-dong, Dong-gu, Gwangjiu 501-759, Republic of Korea. \\
\scriptsize $^{3}$School of Physics and Astronomy, Sun Yat-sen University, 2 Daxue Road, Tangjia, Zhuhai, 519082, China. \\
\scriptsize $^{4}$CSST Science Center for the Guangdong-Hong kong-Macau Greater Bay Area, SYSU, Zhuhai, 519082, China. \\
}

\date{Accepted XXX. Received YYY; in original form ZZZ}

\pubyear{2022}

\begin{document}
\label{firstpage}
\pagerange{\pageref{firstpage}--\pageref{lastpage}}
\maketitle

\begin{abstract}

Upcoming Large Scale Structure surveys aim to achieve an unprecedented level of precision in measuring galaxy clustering. However, accurately modeling these statistics may require theoretical templates that go beyond second-order perturbation theory, especially for achieving precision at smaller scales. In our previous work, we introduced a hybrid model for the redshift space power spectrum of galaxies. This model combines second-order templates with N-body simulations to capture the influence of scale-independent parameters on the galaxy power spectrum. However, the impact of scale-dependent parameters was addressed by precomputing a set of input statistics derived from computationally expensive N-body simulations. As a result, exploring the scale-dependent parameter space was not feasible in this approach. To address this challenge, we present an accelerated methodology that utilizes Gaussian processes, a machine learning technique, to emulate these input statistics. Our emulators exhibit remarkable accuracy, achieving reliable results with just 13 N-body simulations for training. We reproduce all necessary input statistics for a set of test simulations with an error of approximately 0.1~per~cent in the parameter space within $5\sigma$ of the Planck predictions, specifically for scales around $k > 0.1$\,$h$\,Mpc$^{-1}$. Following the training of our emulators, we can predict all inputs for our hybrid model in approximately 0.2\,seconds at a specified redshift. Given that performing 13 N-body simulations is a manageable task, our present methodology enables us to construct efficient and highly accurate models of the galaxy power spectra within a manageable time frame.

\end{abstract}




\section{Introduction}
\label{sec:intro}

Over the last few decades cosmologists have allocated an enormous amount of resources for creating Large-Scale Structure (LSS) redshift surveys of galaxies \citep[e.g.][]{York_2000,10.1046/j.1365-8711.2001.04902.x,Eisenstein_2001,10.1117/12.2629843} where the redshifts of thousands to millions of galaxies are measured to build a three-dimensional map of matter tracers inside a large volume on the sky. One of the main scientific goals of these surveys is the measurement of per~cent accuracy clustering statistics that can be compared with theoretical templates of specific cosmological models. This is done with the hope of constraining the model free parameters and potentially discard those models that cannot reproduce observations. 

In particular, the Dark Energy Spectroscopic Instrument \cite[DESI,][]{2016arXiv161100036D}, has just finished its first year of observations and is conducting different surveys targeting galaxies at different redshifts $z$ to track the evolution of clustering statistics. These surveys include the Bright Galaxy Survey (BGS), which targets galaxies between $0 < z < 0.6$, the Luminous Red Galaxy (LRG) survey which samples galaxies between $0.3 < z < 1.0$, and the Emission Line Galaxy (ELG) survey which targets galaxies between $0.6 < z < 1.6$. 

Clustering statistics, like the matter power spectrum, are ideal for testing cosmological models as they have complex features that are sensitive to both the distribution of the matter and the nature of gravitational interactions within the model: The Baryon Acoustic Oscillations (BAO) \citep{Eisenstein_2005,2017MNRAS.470.2617A} make a significant imprint on the correlation function at scales close to 150\,Mpc and are present as a set of wiggles in the matter power spectrum. These BAO features can be used to constrain the expansion history of the universe. However, an appropriate choice of the equation of state can make different models have the same expansion history \citep{PhysRevD.72.043529}. To break this degeneracy, parameters that are correlated with the growth rate of cosmic structures, which in turn determines the average peculiar velocities of tracers, can be used. When these peculiar velocities have a component among the axis of observation they contribute to the Doppler shift of the object. This does not happen if the peculiar motion is perpendicular to the axis and therefore there is a distortion when using redshifts to measure the distance to tracers of matter, this is usually referred to as Redshift Space Distortions (RSD) \citep{Kaiser87,1996MNRAS.282..877B,1998ASSL..231..185H}. These distortions have a measurable effect on the clustering statistics as they become dependent on the angle of observation.

Theoretical templates of the power spectrum that capture the effects of RSD features have been extensively utilised in the analysis of LSS data, as they offer a theoretical prediction that can be compared with the survey observations. Given the impressive precision of modern surveys, these theoretical models need to be highly accurate, with errors below one~per~cent on the relevant scales. Arguably the most accurate way of generating these templates is using N-body simulations \citep{10.1111/j.1365-2966.2005.09655.x,10.1093/mnras/stt2071,10.1093/mnras/sty3427,10.1093/mnras/stu2301} to build {\it virtual universes}, where one finds the solutions to the equations of motions for a set of particles that represent the matter of our universe. While these simulations can accurately model the matter distribution of our universe, they are also some of the most expensive computational resources that are manufactured in cosmology.

Another prolific methodology for generating theoretical templates of clustering statistics is Perturbation Theory (PT) \citep[e.g.,][]{1984ApJ...279..499F,10.1093/mnras/227.1.1,Bernardeau02,RegPT,10.1111/j.1365-2966.2011.19379.x,PhysRevD.92.023534}. PT models calculates the evolution of small perturbations of density in the early universe that grow through gravitational interactions. In general, these equations are impossible to be solved analytically. PT models use the assumption that the physical magnitudes of interest can be expressed as a perturbation around their mean value (hence the name). If these perturbations are small, all terms that include high-order combinations of the perturbation terms can be discarded. Currently, it is possible to build second-order codes\footnote{Where quadratic combinations of the perturbations are kept, and higher-order corrections discarded.} that predict the matter power spectrum quickly and efficiently \citep[e.g.,][]{PhysRevD.82.063522}. However, next-generation surveys such as DESI will achieve remarkable precision at smaller scales. To maintain sub-per~cent accuracy at these scales, higher-order models that surpass the limitations of second-order perturbation theory might be required. Unfortunately this task is generally challenging, and most templates beyond second-order require a significant computational cost.

The cost required to compute either N-body simulations or PT templates affects our ability to constrain the cosmological parameters of our RSD models. This is mainly because the standard approach to do so is to explore the parameter space using, for example, a Monte Carlo Markov Chain, and find the regions of the space that are more likely to reproduce the data \citep[e.g.][]{2017MNRAS.470.2617A,2020MNRAS.492.4189I,2020MNRAS.498.2492G,2020MNRAS.499..210N}. This type of analysis usually requires more than $10^{6}$ evaluations of the model, therefore efficient templates that can be evaluated with little computational resources are a necessity.

In this work, we present a novel methodology that computes accurate theoretical templates of the  power spectra that go beyond second-order statistics in approximately 0.2\,seconds. We commence the discussion of our methodology by categorizing the cosmological parameters of a given universe model into two distinct groups. Throughout this work we refer to each group as {\it scale independent} 
and {\it scale dependent} parameters. 

The set of {\it scale dependent} parameters contains values that determine the shape of the power spectrum within the linear and quasi-linear regimes. Commonly encountered examples of such parameters in numerous models include the density parameters for dark matter ($\omega_\mathrm{cdm}$) and baryons ($\omega_\mathrm{b}$), and the spectral tilt ($n_\mathrm{s}$).

The set of {\it scale independent} parameters corresponds to parameters that primarily affect the amplitude, stretch or displace the power spectrum, but do not alter its shape, a common example being the amplitude of the scalar mode parameter ($A_\mathrm{s}$).When working on Mpc units, the Hubble constant $h$ parameter only affects the amplitude of the power spectrum, and therefore we include it into our group of \textit{scale independent} parameters. 

As density fluctuations are smoothed out at entering horizon in most dark energy model, the effect of most exotic dark energy models on the shape of spectrum is coherent in scale. Therefore parameters that determine the effects of dark energy in these models fall into this category, an example being the $w_{0}$ and $w_\mathrm{a}$ parameters of the linear parametrisation of \citet{2001IJMPD..10..213C} and \citet{PhysRevLett.90.091301}. Although it is worth keeping in mind that models in which dark energy clusters \citep[e.g.][]{2001PhRvD..63j3510A} or some models that do not follow general relativity \citep[e.g.][]{2010RvMP...82..451S} would be an exception.

In \cite{2016JCAP...08..050Z,2018JCAP...07..018S}, we present a methodology that efficiently predicts how a power spectrum varies when modifying {\it scale independent} parameters and computes accurate theoretical templates of the matter power spectrum that extend beyond second-order PT in this modified cosmology. 

Let us denote a target modified cosmology as $\theta_{\text{new}}$. Let us also suppose that in a given fiducial cosmology $\theta_{\text{fid}}$, with different values for our {\it scale independent} parameters, we know the value of the power spectrum with great accuracy, e.g.\ from an N-body simulation. If we also compute a second-order PT model in $\theta_{\text{fid}}$, which can be cheap to compute, we can estimate the higher-order corrections at $\theta_{\text{fid}}$ by comparing the residuals between the PT model and the N-body simulation power spectrum. Our method employs a set of linear scaling relations to determine the values of these corrections in $\theta_{\text{new}}$ from their values in $\theta_{\text{fid}}$.The scaling relations are functions of the growth functions of the density and velocity fields, denoted as $G_{\delta}$ and $G_{\Theta}$, which are treated as free parameters in our methodology. These scaling relations are particularly useful as they allow us to use the power spectra of an N-body simulation constructed for a specific cosmology to predict how the power spectra will change for different values of any \textit{scale-independent} parameter.

We refer to this methodology as a {\it hybrid model} to emphasize the fact that we are using results from both PT and N-body simulations. We have also introduced an extension of the TNS model of \cite{2010PhRvD..82f3522T} to describe the non-linear mapping from real to redshift space. This model is shown to have per~cent accuracy at reproducing the power spectrum even at $k < 0.18\,h\,\mathrm{Mpc}^{-1}$. Additionally, in \cite{ZhengY2019,2021PhRvD.104d3528S} we expand our model to account for the galaxy biases which lets us model the galaxy power spectrum from the matter power spectrum. Given that the model considers higher-order statistics it is adequate for the analysis of surveys like DESI.

Our hybrid methodology allows us to explore the parameter space of {\it scale independent} parameters efficiently, however, the effect of {\it scale dependent} parameters on the power spectrum is set by our N-body simulations. Given the substantial computational costs associated with running these simulations, we need a new effective methodology to investigate the effects of {\it scale dependent} parameters.

Recently, schemes that use Machine Learning (ML) methods which learn to reproduce theoretical templates have gained popularity \citep[e.g.][]{2007ApJ...654....2F,2009ApJ...705..156H,Agarwal_2014,Spurio_Mancini_2022,DeRose_2022}. These methods are conjointly referred to as emulators or surrogate models and can be very useful as some ML models can generate almost instantaneous outputs once trained. 

Some widely used ML methodologies for building clustering statistics emulators are Gaussian Processes (GP) \citep[e.g.][]{2023MNRAS.519.2962E,Lawrence_2017,Moran_2022}, which are Bayesian interpolators that assume that values of clustering statistics at different points in parameter space have a multivariate normal distribution. This assumption can be used to predict the statistics at an unsampled point by making a linear interpolation with nearby points of the training data set. There are many advantages of GP over other methodologies, one of them is that it is a none-parametric approach where one does not require to define a functional shape for the clustering statistics. It also works comparatively well with high dimensional parameter spaces and a relatively small number of training data. However, it is also known that building GP emulators becomes complicated when large training data sets are used. The algorithm predicts a normal distribution, where its mean is the best guess of the new value and its width measures the uncertainty that the methodology has in its prediction. This can be used to determine which regions of parameter space should be explored next when optimizing a survey design.

In this work we use a GP methodology to accelerate the evaluation of the input statistics that we require to build an hybrid model template. Given that our hybrid methodology already deals with variations of the {\it scale independent} parameters, they do not need to be explored with our emulators. This reduces the dimensionality of the parameter space that we emulate significantly when compared to other works, which results in emulators that are faster to train.

This paper is organized as follows. In Section~\ref{sec:HybridModel}, we introduce our hybrid model methodology and enhancements to our emulators that enable efficient modeling of the matter power spectrum as a function of both {\it scale dependent} and {\it scale independent} parameters. Throughout this section, we present the theoretical background used to model the power spectra in real space, introduce our methodology for constructing emulators, and describe the process for computing our scaling relations. Then, in Section~\ref{sec:Performance}, we present a series of tests to evaluate the performance and accuracy of our emulators. We introduce the methodology we followed to select the points in parameter space where our emulators are trained, and we present several tests that help us understand how factors such as the number of free parameters and test points influence the accuracy of our emulators. 

The work presented up until the end of Section~\ref{sec:Performance} allows us to predict the underlying redshift space dark matter power spectrum and does not require us to select a specific mapping between dark matter and galaxies. In Section~\ref{chap:Tracers}, we choose a specific mapping and introduce and exemplify the methodology to compute the galaxy power spectrum within that mapping.

\section{Emulating hybrid model}
\label{sec:HybridModel}

Throughout this chapter we introduce  our hybrid models methodology that we enhanced with our GP emulators to be able to explore the parameter space of {\it scale dependent} parameters and compute matter power spectra templates that extend beyond second-order perturbation theory to incorporate higher-order statistics. As stated above, this work is a continuation of the hybrid model proposed by \cite{2016JCAP...08..050Z,2018JCAP...07..018S} that uses a set of scaling relations to model the effects of {\it scale independent} parameters in the power spectrum.

We present our model in multiple stages. We commence in Section~\ref{RSD_power_spectrum} by introducing a theoretical template of the power spectrum in redshift space. This power spectrum can be expressed as a combination of several terms. In our approach, each of these terms is computed using a set of GP emulators trained on N-body simulations.

In Section~\ref{subsec:Model_description}, we present the methodology used to construct emulators. These emulators are trained on data from a set of N-body simulations generated at various points sparsely distributed across our parameter space. As previously mentioned, they are employed to estimate the influence of {\it scale dependent} parameters on our power spectra.

In Section~\ref{beyond_emulator_parameter_space}, we introduce our scaling relations methodology to model variations in {\it scale independent} parameters. As we will explain in more detail below, these scaling relations take as input a set of statistics computed using second-order PT predictions. The codes used to compute these predictions are already relatively efficient and take approximately 15 seconds to complete. However, we have also developed emulators of these statistics to reduce their computational time to only a small fraction of a second.

\subsection{RSD power spectrum}
\label{RSD_power_spectrum}
The density power spectrum in redshift space can be expressed in terms of real space items as \citep{PhysRevD.82.063522}:

\begin{equation}
\label{none_perturbed_PS}
P^s(k,\mu)=\int d^3\vec{x} e^{i\vec{k}\cdot\vec{x}}<e^{j_1 A_1} A_2,A_3 >\, ,
\end{equation}
where $\vec{x}=\vec{r}-\vec{r'}$ and $\vec{r}$ are real-space position vectors, and we have used the following definitions:
\begin{align}
\label{second_order_def}
j_1  &= -ik\mu, \\
A_1 &= u_z(\vec{r}) - u_z(\vec{r}'), \\
A_2 &= \delta(\vec{r}) + \nabla_z u_z(\vec{r}), \\
A_3 &= \delta(\vec{r}') + \nabla_z u_z(\vec{r}').
\end{align}

Here $\vec{u}=-\vec{v}/aH$, where $\vec{v}$ is the velocity vector. In general equation \ref{none_perturbed_PS} is not fully perturbative, specially at smaller scales. Following~\cite{PhysRevD.82.063522,2016JCAP...08..050Z}, we divide equation \ref{none_perturbed_PS} into the product of a  perturbative and a non-perturbative contribution, which is a valid approximation in the weekly non-linear regime requiered by DESI. The perturbative contribution is Taylor expanded so that equation \ref{none_perturbed_PS} becomes: 
\begin{equation}
\label{Matter_PS_equation}
\begin{multlined}
P^s(k,\mu)=D^{\rm FoG}(k\mu\sigma_p)[P_{\delta \delta}(k)+2\mu^2 P_{\delta \Theta}(k)+\mu^4P_{\Theta \Theta}(k) \\
+A(k,\mu)+B(k,\mu)+F(k,\mu)+T(k,\mu)]\,,
\end{multlined}
\end{equation}
where $P_{\delta\delta}$, $P_{\Theta\Theta}$ and $P_{\delta\Theta}$ are the auto power spectrum of the density field, of the velocity divergence field ($\Theta=\nabla \vec{u}$) and the density and velocity divergence cross power spectra respectively. The term $D^{\rm FoG}(k\mu\sigma_p)$ corresponds to the non-perturbative correction of our model and $\sigma_p$ is a free parameter that models the velocity dispersion of matter and effectively absorbs model inaccuracies of the perturbative terms. Throughout this work we use the functional form $D^{\rm FoG}(x)=e^{-x^2/H^2}$\citep{2017JCAP...05..030Z}. The $ABFT$ terms correspond to all higher order terms of our model and are defined as:
\begin{align}
\label{terms}
A(k,\mu)  &=j_1 \int d^3\vec{x} e^{i\vec{k}\cdot\vec{x}}<A_1A_2A_3 >_c, \\
B(k,\mu)  &= j_1^2 \int d^3\vec{x} e^{i\vec{k}\cdot\vec{x}}<A_1 A_2>_c<A_1 A_3 >_c, \\
\label{terms_T}
T(k,\mu)  &= \frac{1}{2} j_1^2\int d^3\vec{x} e^{i\vec{k}\cdot\vec{x}}<A_1^2 A_2A_3 >_c,\\
F(k,\mu)  &= -j_1^2\int d^3\vec{x} e^{i\vec{k}\cdot\vec{x}}<u_z(\vec{r}) u_z(\vec{r'}))><A_2A_3 >_c,
\end{align}

Our goal in this work is to compute all the individual terms of equation \ref{Matter_PS_equation}, which includes $P_{\delta\delta}$, $P_{\delta\Theta}$, $P_{\Theta\Theta}$, and the higher-order corrections, with a level of precision that exceeds second-order PT, making the final power spectrum valuable for the scales relevant to DESI. 

 It's important to highlight that all these components can be derived from the results of an N-body simulation. Let us note that the $ABTF$ terms, are n-point correlation functions at the first glance, and can be calculated by 2-point estimators from simulations~\citep{2016JCAP...08..050Z}. It is useful to replace the T term with the $M$ term defined as:
\begin{align}
M(k,\mu)  &= \frac{1}{2} j_1^2\int d^3\vec{x} e^{i\vec{k}\cdot\vec{x}}<A_1^2 A_2A_3 >\,,
\end{align}

whose ensemble average $<>$ includes both correlated and un-correlated terms. The T term only account for the correlated terms, by definition. Therefore we can rewrite the $M$ term to be (Appendix B of~\cite{ZhengY2019})
\begin{align}
\begin{multlined}
\label{M_definition}
    M(k,\mu) = j_1^2\sigma_z^2[P_{\delta \delta}(k)+2\mu^2 P_{\delta \Theta}(k)+\mu^4P_{\Theta \Theta}(k)]\\
    + B(k,\mu) + F(k,\mu) + T(k,\mu)\,.
\end{multlined}
\end{align}
Here $\sigma_{z}$ is the line of sight velocity dispersion which can be measured from an N-body simulation or estimated directly from the linear power spectrum. In order to keep all higher order terms of our expansion separated we rename the first term of equation \ref{M_definition} as $W(k,\mu)$:
\begin{align}
\begin{multlined}
\label{W_definition}
W(k,\mu)=-j_1^2\sigma_z^2[P_{\delta \delta}(k)+2\mu^2 P_{\delta \Theta}(k)+\mu^4P_{\Theta \Theta}(k)]. 
\end{multlined}
\end{align}
We note that $j_1$ is a function of $\mu$, and therefore $W(k,\mu)$ should also depend on $\mu$. Then equation~\ref{Matter_PS_equation} can be rewritten in terms of $M$ and $W$ as:
\begin{equation}
\label{Matter_PS_equation2}
\begin{multlined}
P^s(k,\mu)=\{[P_{\delta \delta}(k)+2\mu^2 P_{\delta \Theta}(k)+\mu^4P_{\Theta \Theta}(k)] \\
+W(k,\mu)+A(k,\mu)+M(k,\mu)\}D^{\rm FoG}(k\mu\sigma_p)\,,
\end{multlined}
\end{equation}
Note that we have separated $W(k,\mu)$ from the term in the squared bracket, this has been done to emphasize that this term constitutes a higher-order correction. However, we highlight that both terms are determined by the same statistics. Consequently, both of these terms can be estimated using the same set of emulators. 

We will use this equation to represent the general RSD matter power spectrum formulation adopted by our Hybrid model. Throughout this study, we construct a series of N-body simulations and calculate the value of each individual term in each simulation. Subsequently, we utilise these values to train a set of emulators capable of accurately and efficiently predicting each of these individual components.

\subsection{Emulating power spectrum of particles}
\label{subsec:Model_description}

We have established a set of terms that can be included into equation~\ref{Matter_PS_equation2} to construct a model of the galaxy power spectrum in redshift space, which accurately captures their behavior in the weakly non-linear regime. Additionally, we have emphasized that these statistics are computable from N-body simulations.

These N-body simulations are the bottleneck of the hybrid model methodology as computing one simulation takes around 13 CPU hours with 128 processors.  The aim of this work is to emulate our input statistics so that they can be computed with a negligible amount of computational effort and time.

Given that each N-body simulation takes a significant amount of resources  we can generate tens of templates but we would struggle to generate more than 100 templates. Therefore, there is an incentive to keep the number of simulations to a minimum, this requires building an efficient simulation design. 

In Section~\ref{simulations} we introduce our methodology for building N-body simulations. Then in Section~\ref{subsec:Model_characteristics} we introduce the characteristics of the parameter space that we explore. Finally, in Section~\ref{subsec:Intro_GP} we describe the methodology that we follow to build GP emulators.

\subsubsection{Simulations}
\label{simulations}

N-body simulations are regarded as the most precise methodology for replicating the intricate non-linear behavior of matter clustering, especially at the smaller scales. Here, we use the publicly available code {\tt Gadget2} \citep{Springel:2005} to construct a series of cosmological N-body simulations, which serve as the training set of our emulators. At a given redshift, this code utilises the TreePM algorithm to evolve a given particle distribution and compute the forces acting on them within a comoving periodic box. The construction of a single simulation realization requires approximately 13 CPU hours with 128 processors.

The initial conditions of our simulations are generated using Second-order Lagrangian Perturbation Theory based on \texttt{2LPTic} \citep{Crocce:2006}. This method relies on an input matter power spectrum, which we obtain using the publicly available code \texttt{CAMB} \citep{Lewis:1999bs} (Code for Anisotropies in the Microwave Background). CAMB utilises the linear perturbation theory to accurately predict the matter power spectra.

The particle distribution of our initial conditions cannot be determined solely by the CAMB spectra. The code also generates two random numbers: the first determines the amplitude of the fluctuations of the given power spectrum, while the second determines the random phase of the density field in Fourier space. In this work, we discard the first random number by fixing the amplitude to that of CAMB, this is done so that the particle distribution preserves the input clustering power at the scales of interest. In total our methodology for selecting the initial conditions takes about 2 CPU minutes with 32 processors for one initial condition.

Each of the simulations presented in this work are generated by pairing together two different {\tt Gadget2} runs, the simulations have a fixed-amplitude but opposite-phases. The methodology followed  to paring the simulations at a target redshift can be found in \cite{10.1093/mnrasl/slw098}.

The fiducial values of our cosmological parameters are taken from \cite{2020A&A...641A...6P}, in this work we use the Plank predictions made excluding lensing (fourth column of table 2 of \cite{2020A&A...641A...6P}).

Table~\ref{tab:simulation} presents the settings of our simulations. The first section of the table displays the central values of the cosmological parameters, which vary across different simulations. It is important to note that each N-body simulation utilises distinct values for these parameters, as detailed in Section~\ref{subsec:Model_characteristics} below.

Once the simulation is completed, we compute their matter power spectra using the cloud-in-cell mass assignment scheme with 1024 grids for fast Fourier transform. Additionally, we correct for shot noise and mass assignment. Each power spectrum calculation takes approximately 3 CPU minutes on a single processor.

\begin{table}
\scriptsize
\begin{tabular}{@{}lll}
\hline
Parameter & Physical meaning & Value \\
\hline
\multicolumn{3}{|c|}{Central value of emulated cosmological parameter} \\
\hline
$\Omega_\mathrm{m}$  & matter density at $z=0$ in units of the critical density & $0.315$ \\
$\Omega_\mathrm{b}$ & baryon density at $z=0$ in units of the critical density & $0.049$\\
$n_{s}$ & primordial power spectral index & $0.9649$ \\
\hline
\multicolumn{3}{|c|}{Fixed Cosmological parameter} \\
\hline
$\Omega_{\Lambda}$ & $1-\Omega_\mathrm{m}$ & $0.685$ \\
$\Omega_{\nu}$ & massive neutrino density at $z=0$ in units of the critical density & $0.0$\\
$N_{\nu}$ & effective number of mass-less neutrinos & $3.046$\\
$h$ & $H_{0}/(100$\,km\,s$^{-1}$\,Mpc$^{-1})$ & $0.6727$ \\

$A_\mathrm{s}$ & amplitude of scalar primordial fluctuation & $2.101\times 10^{-9}$ \\
$k_{\rm pivot}$ & pivot scale  & 0.05\,Mpc$^{-1}$ \\
\hline
\multicolumn{3}{|c|}{Simulation specification} \\
\hline
$L_{\rm box}$ & simulation box size & 1890\,Mpc\,h$^{-1}$\\
$N_{\rm p}$ & simulation particle number & $1024^{3}$\\
$m_{\rm p}$ & simulation particle mass & $5.5\times 10^{11}M_{\odot}h^{-1}$\\
$N_{\rm snap}$ & number of output snapshots & $13$ \\
$N_{\rm mesh}$ & number of particle mesh in long-range force computation & $2048$ \\
$\epsilon$ & softening length for gravity & $92.28$\,kpc \\
$z_{\rm ini}$ & redshift when simulation starts & $49.0$ \\
$z_{\rm final}$ & redshift when simulation finishes & $0.0$ \\
\hline
\end{tabular}

\caption{This table presents the parameter values for the fiducial cosmology and the specifications used in our N-body simulations. The first section displays the central values of the parameters that we are emulating over. The second section lists the cosmological parameters that remain fixed across all simulations. Finally, the last section provides the configuration values used in the simulations.}
\label{tab:simulation}
\end{table}

\subsubsection{Parameter space}
\label{subsec:Model_characteristics}

We have stated that we are interested in building  emulators to accelerate the creation of these inputs. We leave the description of our emulator methodology for subsection \ref{subsec:Intro_GP} below. Here, we introduce the characteristics of the emulators that we build, emphasizing the parameter space that we explore and our redshifts of interest.

To build our emulators we require two sets of templates at different points of the parameter space, the first set contains points that are used for training our Gaussian emulators and should be optimized in such a way that the parameter space of interest is thoroughly sampled, we refer to this parameter space points as our {\it training set}. The second set of points should be disjointed from the training set and contain points that are used to test our emulators on new data, these points are referred to as the {\it test set}. Throughout this work, we refer to the points in parameter space that we select to be part of our training set as our {\it simulation design}. Our methodology for selecting simulation designs is presented in Section~\ref{subsec:grid-selection}.

Within the hybrid model paradigm, the impact of \textit{scale-independent} parameters on the power spectrum can be calculated as posterior corrections introduced into our models through a set of scaling relations. The detailed methodology for this is provided in Section~\ref{beyond_emulator_parameter_space}. For now, we emphasize that our N-body simulations exclusively explore the space of \textit{scale-dependent} parameters of interest. For this study, these parameters include the density of dark matter and baryons at $z=0$ ($w_\mathrm{cdm}=\Omega_\mathrm{cdm} h^{2}$ and $w_\mathrm{b}=\Omega_\mathrm{b} h^{2}$) and the power spectral index ($n_\mathrm{s}$). 

In here we explore the region within $5 \sigma$ of the \cite{2020A&A...641A...6P} predictions, which corresponds to the following values:

\begin{itemize}
\label{Plank_space}
    \item $w_\mathrm{cdm}=0.1202 \pm 5(0.0014) $
    \item $w_\mathrm{b}=0.02236 \pm 5(0.00015)$
    \item $n_\mathrm{s}= 0.9649 \pm 5(0.0044)$.
\end{itemize}

As stated above, the rest of the cosmological parameters are fixed to the values given in the second section of Table~\ref{tab:simulation} when running our N-body simulation.

We note that these constrains are already quite small even at $5 \sigma$. As we explore in Section~\ref{sec:Performance} below, this, along with the fact that we only need to explore three parameters, translates in emulators that can achieve sub-per~cent accuracy with a relatively small set of training points. This simplifies our work significantly as the number of N-body simulations that we need to build becomes quite manageable. Having a small training set also means that training our emulators is not computationally expensive. Most emulators presented in this work can be trained in around 3 minutes on a personal computer.

We note that for each redshift that we explore we require to build a new emulator. In this work, our emulators are built in five different redshifts $z=0.3, 0.5, 0.7, 0.9, 1.1$. The emulator predictions of any intermediate redshift can be computed with an interpolation between the predictions from these five redshifts. We select these particular redshifts as they include the regions of interest for DESI BGS, LRG, and ELG surveys respectively \citep{2016arXiv161100036D}.

\subsubsection{Gaussian Process}
 \label{subsec:Intro_GP}

We now introduce the methodology we follow to construct Gaussian Process emulators. Throughout this work we follow the methodology of \cite{2007PhRvD..76h3503H,2009ApJ...705..156H}, which we summarise below for completeness.  For now, we assume that a simulation design has previously been determined. The detailed explanation of the methodology used for selecting these designs will be provided in Section~\ref{subsec:grid-selection}.

The first step is to compute the statistic we are emulating $p_i(k,z)$ for the $N_\theta$ points ($\theta_0, \ldots,\theta_i, \ldots, \theta_{N_\theta}$) within the simulation design. This data is utilised to train the emulators. 

Most of the statistics we emulate in this work are power spectra (except for the $A$ and $M$ higher order correction terms). When emulating power spectra it is a common practice \citep[e.g.,][]{2009ApJ...705..156H,2017ApJ...847...50L,2023MNRAS.520.3443M} to transform it in such a way that the BAO features are emphasized prior to training the emulator, we use the following equation:

\begin{equation}
P_i(k,z)=\ln{\frac{k^{3/2}p_i(k,z)}{4\pi ^3}}.
\end{equation}

We also normalize our data so that all training models have a similar scale. Let $P(k,z)$ be the subset of all $N_\theta$ values of $P_i(k,z)$ for a given $k$ and $z$. Then our normalization is done with the following equation:

\begin{equation}
\bar{P}_i(k,z)=\frac{P_i(k,z)-\mu(P(k,z))}{\sigma(P(k,z))}
\end{equation}

Where $\mu$ and $\sigma$ are the mean and standard deviation operators respectively.

We note that each power spectrum contains hundreds of points in $k$-space. If we were to directly emulate it, we would need to make one prediction for each $k$ value. An alternative and more efficient approach is to reduce the dimensionality of the problem by decomposing the matter power spectrum into a set of principal components and their corresponding weights. Since the majority of the information in the power spectrum is captured by the most relevant principal components, we can retain those and discard the rest, thereby reducing the number of predictions required.

For each redshift $z$ we define the $N_k \times N_\theta$ matrix $\bar{P}_z=[\bar{P}(\theta_1,k,z),..,\bar{P}(N_\theta,k,z)]$. In order to find the principal components we use the Singular Value decomposition (SVD) methodology \citep{doi:10.1073/pnas.97.18.10101}:

\begin{equation}
   \bar{P}_z=UdV,
\end{equation}

where U is a $N_k \times N_\theta$ matrix, $d$ is a diagonal matrix such that each singular values  its a diagonal element, and they are ordered from larger to smaller. V is a $N_\theta \times N_\theta$ matrix corresponding to the weights associated with each singular value. 

We define the matrix of eigenvalues $\Phi=U \cdot D \sqrt{N_\theta}$ and $W=V^T\sqrt{N_\theta}$. As stated only the largest eigenvalues are relevant for our model, let $N_{PC}$ be the number of relevant principal components, we can truncate these matrixes so that they only contain the first $N_{PC}$  rows: $\Phi=[\phi_1(k,z),...,\phi_{N_{PC}}(k,z)]$ and $W=[w_1(\theta),..,w_{N_{PC}}(\theta)]$. Then, we define the principal component decomposition of our power spectra as:

\begin{equation}
\label{PS_decomposition}
    \bar{P}(\theta,k,z)=\sum_{\alpha=1}^{N_{PC}}\phi_\alpha(k,z)w_\alpha(\theta)+\epsilon.
\end{equation}

The error $\epsilon$ is a consequence of not using all principal components but only the first $N_{PC}$. Note that the principal components $\phi_i(k,z)$ are completely determined during  our SVD and are independent of the cosmology. The dependence on the cosmological parameters $\theta$ is only present on the weight function $w_\alpha(\theta)$. Therefore, the computation of these weights is the only task required when considering a new fiducial cosmology. Consequently, our problem is reduced to modeling the scalar functions $w_\alpha(\theta)$, where $\alpha=1,\ldots,N_{PC}$. 

We model each $w_\alpha(\theta)$ function using a Gaussian Process (GP) emulator \citep{10.7551/mitpress/3206.001.0001}. This method uses the behavior of training points within the simulation design to interpolate predictions for new points in parameter space. The methodology assumes that the target function varies smoothly.

As shown in Section~\ref{subsec:Nbody_performance} below, employing the initial five principal components is sufficient for constructing precise emulators in this work. Consequently, in this study, we set $N_{PC}=5$ for all emulators.

The goal of our emulator is to estimate the values of $w_\alpha^*=w_\alpha(\theta^*)$ for a new point in parameter space $\theta^*$. We define the weight vector $\Vec{w}=[w_1(N_{1}),..,w_{N_{PC}}(N_{\theta})]$ of length $N_{PC} \times N_\theta$, that results from flattening the matrix W, and the vector $\Vec{P}$ of length $N_\theta \times N_k$ that results from flattening out the matrix $\bar{P_z}^T$. We also define the matrix $\Phi_M$ as the matrix that satisfies\footnote{$\Phi_M=[I_{N_\theta}\otimes \phi_1,....,I_{N_\theta}\otimes \phi_{N_{PC}}]$} $\Vec{P} \sim \Phi_M \times \Vec{w}$. $\Phi_M \times \Vec{w}$ is not equal to $\Vec{P}$ because of  the truncation error that we impose by not using all principal components. We define the $N_{PC} \times N_\theta$ vector $\Vec{w}'=\Phi_M \Phi_M \Phi^T_M \Vec{P}$, that corrects for this error, note that $\Vec{P} = \Phi_M \times \Vec{w}'$.

For each principal component, we introduce a kernel function $\psi_\alpha(\theta^1,\theta^2)$ that measures similarities between two points in parameter space,

\begin{equation}
\psi_\alpha(\theta_1,\theta_2)=\frac{1}{\lambda_{w_\alpha}} \prod_{i=1}^{N_\theta} r_{w_\alpha,i}^{4(\theta^1_i-\theta^2_i)^2}
\end{equation}

$\psi_\alpha$ depends on a set of hyperparameters, the first is $\lambda_{w_\alpha}$ that parameterizes the marginal precision associated with emulating $w_\alpha$. The other free parameters are the scalars that form the vector $r_{w_\alpha}=[r_{w_\alpha,0},..,r_{w_\alpha,N_\theta}]$ each parameter controls the correlation length of the $i^{th}$ dimension in parameter space.

The GP methodology suggests that $[w_\alpha',w_\alpha^*]$ are distributed as an $N_\theta+1$ multivariate normal distribution:

\begin{equation}
\label{matrix_dist}
\begin{bmatrix}
w_\alpha'\\ w_\alpha^*
\end{bmatrix}
=N\left ( 
\begin{bmatrix}
0\\ 0
\end{bmatrix} ,
\begin{bmatrix}
 \Psi_\alpha(\theta,\theta)+E & \Psi_\alpha(\theta,\theta^*) \\ 
 \Psi_\alpha(\theta^*,\theta) &  \Psi_\alpha(\theta^*,\theta^*)
\end{bmatrix}
\right )
\end{equation}

Where $\Psi_\alpha(\theta,\theta)$ is the covariance matrix of  $i^{th}$ row and $j^{th}$ columns defined as $\Psi^{ij}_\alpha(\theta,\theta)=\psi_\alpha(\theta_i,\theta_j)$. $E$ is an estimate of the error in our methodology given by $E=(\lambda_\epsilon \Phi_M^T\Phi_M)^{-1}$ and $\lambda_\epsilon$ is a free parameter that characterizes the precision of the error in the methodology as a whole. 

We refer to the construction of a GP emulator as the process of determining the free parameters $\lambda_{\epsilon}$, $w_{\alpha}$, and $r_\alpha$. In their work, \cite{2007PhRvD..76h3503H} provides an expression (equation~21) for the posterior distribution of the free parameters based on the training set vector of weights $\Vec{w}$. To obtain our free parameters, we identify the values that maximize this expression. Throughout the remainder of this study, we use the term {\it building a GP emulator} to describe this parameter determination procedure.

Equation \ref{matrix_dist} can be used to predict the mean and variance of the normal distribution associated with $w_\alpha^*$, the resulting expressions are as follows:

\begin{multline}
 \label{eq:Gp_prediction}
    \mu (w^*_\alpha)=\Psi_\alpha(\theta^*,\theta)(\Psi\alpha(\theta,\theta)+E)^{-1}w'_\alpha \\
    \sigma(w^*_\alpha)=\Psi_\alpha(\theta^*,\theta^*)-  \Psi_\alpha(\theta^*,\theta)(\Psi_\alpha(\theta,\theta)+E)^{-1} \Psi_\alpha(\theta,\theta^*). 
\end{multline}

Here, $\mu(w^*_\alpha)$ is interpreted as our best guess of the value of $w_\alpha(\theta^*)$ and $\sigma(w^*_\alpha)$ is considered an estimate of the uncertainty in our prediction.  We note that estimates of $\mu(w^*_\alpha)$ can be plugged into equation~\ref{PS_decomposition} to make a prediction of the matter power spectrum at a new point $\theta^*$ in normalized parameter space.

For a given parameter space point $\theta^*$, the estimated uncertainty $\sigma(w^*_\alpha)$ is proportional to the distance to points in the simulation design used to train the GP model. Regions close to any training point will have comparatively small errors while regions with larger uncertainties should correspond to areas in parameter space that are not thoroughly explored. 

\subsection{Cosmological model beyond emulator parameter space}
\label{beyond_emulator_parameter_space}

We have stated that within the hybrid model paradigm the effect of {\it scale independent} parameters on the power spectrum is modelled by a set of scaling relations and therefore we do not require to emulate their behaviour. In order to do so we first build theoretical templates of the power spectrum for a given value of our {\it scale independent} parameters using second order perturbation theory, we suggest that the higher order corrections between cosmologies with different values of {\it scale independent} parameters are related to each other through our scaling relations. We compute the higher order corrections in a fiducial cosmology using the N-body  emulators presented in Section~\ref{subsec:Model_description}, then use these scaling relations to transform the predictions into the new cosmology. 

The scaling relations are only dependent on the values of the density and velocity growth functions $G_{\delta}$ and $G_{\Theta}$ at the redshift of interest, which are included as two free parameters of our methodology, and account for RSD effects. These growth functions are related to the linear growth factor $D_+$ and linear growth rate $f\equiv d\ln D_+/d\ln a$ through $G_\delta=D_+$ and $G_\theta=f\,D_+$. We emphasize that the emulators we construct can be used without modification to explore any {\it scale independent} parameters, such as the dark energy parameters $w_0$ and $w_a$ introduced above. Provided these parameters can be varied within our PT templates. 

While the statistics computed from Nobody simulations or PT templates are typically presented in distance units of Mpc\,$h^{-1}$ in our methodology, we convert these outputs into Mpc units before calculating our scaling relations. This conversion serves to simplify the impact of $h$ (as well as parameters dependent on it) on the power spectrum. In Mpc units, $h$ only influences the amplitude of the power spectrum, and is a  \textit{scale-independent} parameter.

Throughout this section we introduce our scaling relations methodology. As we have stated, the first step of our method is building a second order perturbation theory predictions of the power spectra in the new cosmology, this is achieved by building emulators to predict the outputs of several statistics of the RegPT model. In Section \ref{subsec:RegPT} we introduce the RegPT model and the emulators that we build. Then in Section \ref{Projected_power_spectra} we introduce the methodology and scaling relations that we use for computing the higher order corrections of these models. 

\subsubsection{Gaussian process for theoretical power spectra}
\label{subsec:RegPT}

Our second-order templates are constructed using the RegPT model from \cite{2012PhRvD..86j3528T}. We employed their two-loop order models to calculate $P_{\delta\delta}$, $P_{\delta\Theta}$, and $P_{\Theta\Theta}$, which are the necessary statistics of our model, as explained in Section \ref{Projected_power_spectra} below. RegPT utilises a pre-computed value of the linear power spectrum $P_{\rm lin}$ at a given cosmology to model the second-order power spectrum. In this work, we compute $P_{\rm lin}$ using the publicly available Code for Anisotropies in the Microwave Background CAMB\citep{Lewis:1999bs,Lewis:2002ah}. The statistics are computed using equation $B6$ of \cite{2012PhRvD..86j3528T}: 
\begin{equation}
    \label{RegPT_2loop}
    \begin{multlined}
        P_{XY}^{\rm 2-loop}(k;D_+)=\\
        D_+^2e^{-2\alpha_k}\left[\left\{1+\alpha_k+\alpha^2_k/2
        +D_+^2\bar{\Gamma}^{\rm (1)}_{\rm 1-loop}(k)(1+\alpha_k)\right.\right.\\
        \left.+D_+^4\bar{\Gamma}^{\rm (1)}_{\rm 2-loop}(k)\right\}^2P_{\rm lin}(k)+D_+^2\left\{(1+\alpha_k)^2P^{\rm (2)tree-tree}_{\rm corr}(k) \right.\\
       \left.\left. +D_+^2(1+\alpha_k)P^{\rm (2)tree-1loop}_{\rm corr}(k)+D_+^4P^{\rm (2)1loop-1loop}_{\rm corr}(k)\right\}\right.\\
       \left.+D_+^4P^{\rm (3)tree-tree}_{\rm corr}(k)\right]\,.
    \end{multlined}
\end{equation}

Here $\alpha_k$ is given by $\alpha_k=k^2\sigma_{\rm d}^2D_+^2/2$ with $\sigma_{\rm d}$ being the dispersion of displacement filed.

From the expression above, we note that in order to make a prediction RegPT utilises $P_{\rm lin}$ along with 6 statistics that it computes, namely $\bar{\Gamma}^{\rm (1)}_{\rm 1-loop}$, $\bar{\Gamma}^{\rm (1)}_{\rm 2-loop}$, $P^{\rm (2)tree-tree}_{\rm corr}$, $P^{\rm (2)tree-1loop}_{\rm corr}$, $P^{\rm (2)1loop-1loop}_{\rm corr}$, $P^{\rm (3)tree-tree}_{\rm corr}$. The detailed expressions of the statistics are shown in the appendix 4c of \cite{2012PhRvD..86j3528T}. As elaborated in detail in Appendix~\ref{appendix:hybrid}, our methodology employs these seven terms along with  $P_{\rm lin}$, that is an input to calculate a set of statistics, enabling us to express the scaling relationships of our second order theoretical templates of $P_{XY}$. 

Each RegPT prediction requires significantly fewer resources than the N-body simulations, with each template taking around 5 seconds to run in our personal computers 
(15\,seconds considering that we need three different power spectra). However, given that an average MCMC exploration of the parameter space for RSD analysis can take more than $10^{6}$ likelihood estimations, it is still convenient to reduce the evaluation time of each RegPT template.

With this in mind,  we build one individual emulator for each of the terms that we compute using the RegPT and CAMB public codes. Given that we have 7 terms and that we require to compute $P_{\delta\delta}$, $P_{\delta\Theta}$ and $P_{\Theta\Theta}$ we require to build a total of 21 emulators per redshift of interest.  The emulators are built using the methodology described in Section~\ref{subsec:Intro_GP}.  Our emulator methodology reduces the evaluation time from around 15 seconds to a fraction of a second.

\subsubsection{Power spectrum scale relations}
\label{Projected_power_spectra}

We now present the description of the scaling relations that we use for transforming the higher order corrections of our hybrid model. In Section~\ref{RSD_power_spectrum} we introduced a set of statistics that we emulate from our N-body simulations. Each of them has their own scaling relation that allows us to compute their value in a different cosmology, the final redshift space matter power spectra can be computed by plugging these terms into equation~\ref{Matter_PS_equation2}.  First we introduce the corrections of the terms $P_{XY}(k)$,
where $X$ and $Y$ are either $\delta$ or $\Theta$. As stated above, the first step in the methodology is to use a second-order perturbation theory template. This template is developed using the RegPT emulators presented in Section~\ref{subsec:RegPT}. However, as mentioned earlier, we are interested in models that go beyond second-order theory.  For a given fiducial cosmology, we use our N-body simulation emulators to compute the higher-order corrections of our power spectrum. This is done by relating these higher-order terms with the residual of the power spectrum of the PT model and the N-body simulation. which we define as:

\begin{equation}
\label{N_body-regpt}
\bar{P}_{XY}^{\rm res}(k,z)=\bar{P}_{XY}^{\rm PT}(k,z)-\bar{P}_{XY}(k,z)\,,
\end{equation}
$\bar{P}_{XY}^{\rm PT}(k,z)$ is the power spectrum that we get from the perturbation theory emulators and $\bar{P}_{XY}$ is the one that we get from the N-body simulation emulators. Throughout this work, bared quantities correspond to a fiducial cosmology.

The scaling relation for the higher order corrections of the power spectrum in a new cosmology are a function of $G_X$ and $G_Y$ and are given by:

\begin{equation}
\label{residual_scaling_relation}
P_{XY}^{\rm res}(k,z)=\frac{G_XG_YG_\delta^4}{\bar{G}_{X}\bar{G}_{Y}\bar{G}_{\delta}^4}\bar{P}_{XY}^{\rm res}(k,z)\,.
\end{equation}

We call that we use to compute $P^{PT}_{XY}$ within the hybrid model is given in equation~\ref{RegPT_2loop}, which is derived from equation~\ref{eq:pk_RegPT} in Appendix~\ref{appendix:hybrid}. As explained in more detail there, the value of $P^{PT}_{XY}$ in a given cosmology is fully determined as a function of  the $\Gamma_\Theta^{(n)}$ and $\Gamma_\delta^{(n)}$ point propagators, and in our hybrid model the relation between point propagators in different cosmologies is given by equations~\ref{point_prop_new_cosmo_1} and \ref{point_prop_new_cosmo_2}. Note that these expressions, along with equation~\ref{N_body-regpt} can be combined trough equation~\ref{residual_scaling_relation} into a prediction $P_{XY}$.

So far, we have explained the methodology we follow to compute the $P_{XY}$ predictions within the hybrid model. The discussion on the methodology to build the predictions for the higher order $A$ $M$ corrections in equation~\ref{Matter_PS_equation2} is detailed in Appendix~\ref{appendix:higher_order}. Where, we introduce equations~\ref{eq:estimatedAn}, which are the scaling relations necessary to transform the higher-order correction terms into a new cosmology.

\section{precision test}
\label{sec:Performance}

In this chapter, we introduce the various Gaussian emulators that we construct and assess their accuracy in predicting the statistics of a predefined set of test points. We begin in Section~\ref{subsec:grid-selection} by introducing our simulation designs within our parameter space. These simulations serve as the training set we use to build our emulators.

We have introduced our methodology for predicting redshift space power spectra of dark matter within a given fiducial cosmology. This process involves individually forecasting all the terms specified in equation~\ref{Matter_PS_equation2} using our emulators. The necessary terms that we emulate consist of three power spectra: $P_{\delta\delta}$, $P_{\delta\Theta}$, and $P_{\Theta\Theta}$, along with the $A$ and $M$ higher-order correction terms. In Sections~\ref{subsec:Nbody_performance} and \ref{subsec:highorder_performance}, we present and assess the accuracy of our emulators for these statistical predictions.

We have also introduced our hybrid methodology that allows us to compute power spectra in a new cosmology. This involves using a set of scaling relations to transform emulator outputs from one fiducial cosmology into a new cosmology with different values of our {\it scale independent} parameters. To achieve this, we require a set of emulators for the set of seven RegPT statistics that are used in the computation of  $P^{PT}_{XY}$ in Section~\ref{subsec:RegPT}. The results involving our hybrid model specifically  can be found in Section \ref{hybrid_model_tests}. We begin in Section~\ref{subsec:RegPT_performance}, where we assess the accuracy of the RegPT emulators. Subsequently, in Section~\ref{subsec:accuracy_test}, we present several tests that examine how the number of free parameters and test points influence the accuracy of our emulators.

\subsection{Test on emulated spectra}

\subsubsection{Simulation grid design}
\label{subsec:grid-selection}

In this section we summarise the methodology followed to select our simulation designs. As mentioned above there are different computational requirements on the number of RegPT theoretical templates and  N-body simulations that we can build. Consequently, this necessitates distinct approaches for selecting parameter space points to train each emulator.

We begin by summarizing the methodology we follow for selecting our N-body simulation design. We have stated that due to the high computational cost of running a single N-body simulation there is a strong incentive to generate the minimal amount required to build accurate emulators. Given that Gaussian Process are local interpolators, the accuracy of the model at predicting a new point is correlated with its distance to points in the training set. Therefore, in order to minimize the number of simulations needed we require a sparse simulation designs, that enables exploration of the parameter space regions using a limited number of points.

With this in mind, we follow the methodology suggested in \cite{2009ApJ...705..156H} to construct our simulation designs in the three-dimensional parameter space of Section \ref{subsec:Model_characteristics}. Firstly, we build a Symmetric Latin Hypercube (SLH) \citep{doi:10.1080/01621459.1993.10476423}. Subsequently, we use an Annealing algorithm \citep{MORRIS1995381} to explore the space of potential SLHs that can be created by moving some points within the grid. The algorithm explores the parameter space of possible resulting SLHs, to identify the SLH that achieves the highest level of sparsity in the grid. The details of the methodology are summarised for completeness in Appendix~\ref{apendix:LatinHypercube}.

To determine the appropriate grid size for our emulators, we conducted  as set of tests using the CAMB code introduced above \citep{Lewis:1999bs,Lewis:2002ah}. In this test study, we utilised a specific version of CAMB that incorporates the halofit model \cite{2003MNRAS.341.1311S}, which introduces non-linear corrections to the predicted matter power spectra. 

We first build a set of optimized simulation designs of different grid sizes using our Annealing methodology. We then build CAMB power spectra at the resulting training points of each design, this spectra can be used to train an emulator. We then test the accuracy of these emulators by predicting the power spectra at 100 random  points of parameter space, and comparing the predictions with the actual CAMB outputs. We note that when building simulation designs somewhere around 11-13 grid points the emulators we build become quite accurate. With this in mind we select a simulation design of 13 points and build one N-body simulation on each of them. 

We are also interested in running extra N-body simulation in new points of the parameter space that are to be used to test the accuracy of our methodology. Since we are utilizing a local interpolator, we aim to select test points located in regions of the parameter space that are relatively distant from the training points. In this region, our emulator should have lower confidence in the accuracy of its predictions. By focusing on these areas, which are likely to be the emulator's weakest performing regions, we can demonstrate that if our emulator proves accurate there, it will likely be accurate in the rest of the parameter space as well.

Throughout this work, we have chosen two distinct test points, each serving a specific purpose in evaluating the model's accuracy within different regions of interest inside the parameter space.

The first point, referred to as $\theta_{1}$ throughout this work, is intended to assess the model's accuracy within the 1 $\sigma$ region of the Planck predictions. This region holds particular significance as it corresponds to the region with cosmological parameters with the highest likelihood. Consequently, during MCMC analysis, a significant portion of the exploration time is expected to be spent in this area. By conducting tests on $\theta_1$, we aim to evaluate how well our methodology can model the region we are most interested in modeling accurately. 

We select $\theta_1$ to be close to the border of the 1 $\sigma$ region of Planck and relatively far for all points in our training set. $\theta_1$ is determined as follows:  we set the values of $\Omega_{cdm}$ and $\Omega_b$ to match the estimates from \cite{2020A&A...641A...6P}, and we fix $n_s$ to $0.9649 + 0.0044$, which corresponds to the 1$\sigma$ limit from Planck. Subsequently, we determine the value of $h$ such that $w_{cdm}$ equals $0.1202 + 0.0014$, which also corresponds to the 1$\sigma$ limit established by Planck. This leads us to set $h$ equal to 0.6766, and we employ this value to compute $w_b$.

Our second test point is selected around 5 $\sigma$ of the planck estimates, this point is very near to the edge of our emulator priors and corresponds to the region where we are less interested in the prediction. We select this point to be as far as possible to all training set points. By selecting this test point, we are interested in quantifying the accuracy of our methodology in the most remote region of our parameter space, where we expect our emulator to perform the worst.

The values of the cosmological parameters of our final training set design are shown in the first section of table \ref{tab:data}. The second section shows the exact values of $\theta_1$ and $\theta_5$.

\begin{figure*}
\includegraphics[width=120mm,height=120mm,trim={0 3cm 0 0},clip]{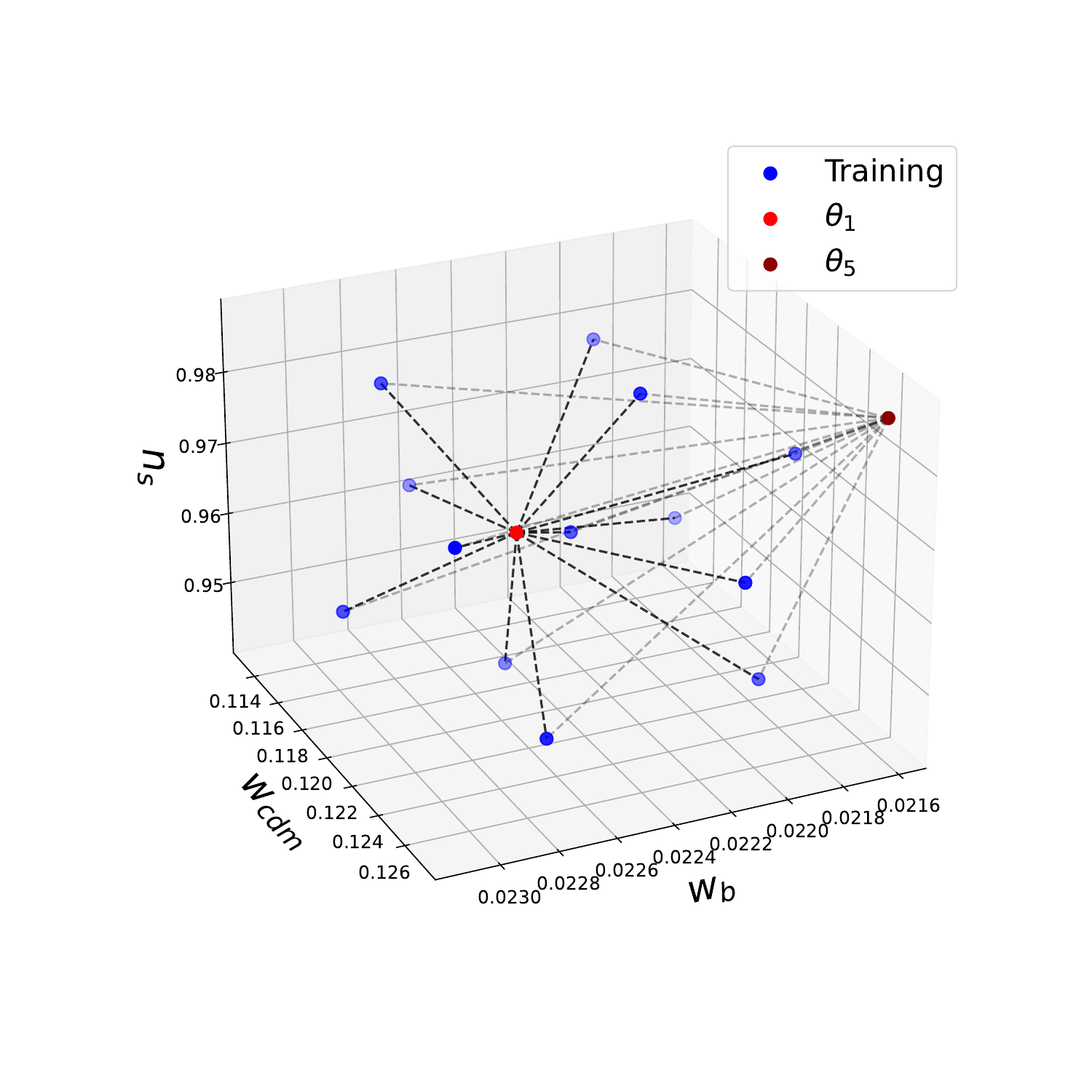}
\caption[ ]{This grid displays the positions of our training data points (blue dots) within our parameter space. The light and dark red dots represent the positions of our $\theta_1$ and $\theta_5$ points, respectively. The plot aims to illustrate the distribution of training points in relation to one another. Since GP is a local interpolator method, the proximity to training points determines the accuracy when emulating a test point.}
\label{Grid_points_visualisation}
\end{figure*}

Fig.~\ref{Grid_points_visualisation} presents a visual representation of our training simulation. Each of the 13 blue dots in the plot corresponds to a distinct training point. The plot aims to illustrate the sparse distribution of these dots throughout the parameter space, and to highlight that no region is more densely populated than the rest. The dark red dot indicates the position of $\theta_5$, and the dashed gray lines show its distance from the training set points. It is evident from the plot that $\theta_5$ is isolated in one corner of our parameter space, far removed from most training points. Therefore, we anticipate that our emulators will struggle to model this test point, and we can use it to assess the emulator's precision in regions where it performs poorly.

The light red dot marks the position of $\theta_1$, and the figure shows that this dot is positioned near the center of the grid in an area with a high likelihood. This test point will enable us to evaluate the precision of our emulators in the region of interest, where we aim to make predictions as accurate as possible.

We previously mentioned that approximately 13 points were sufficient for constructing accurate emulators of power spectra derived from CAMB. However, CAMB and our N-body simulations are different models with different accuracy and we should test if our simulation design can make accurate N-body emulators. In a hypothetical scenario where we determine the necessity for a larger training dataset, we can utilise the uncertainty estimate outlined in equation~\ref{eq:Gp_prediction} to pinpoint areas requiring further exploration and subsequently select new training points in those regions. However, as we are about to show, such an expansion is not required, as our emulators built in our 13 points design are quite accurate.

\begin{figure*}

\includegraphics[width=80mm,height=80mm]{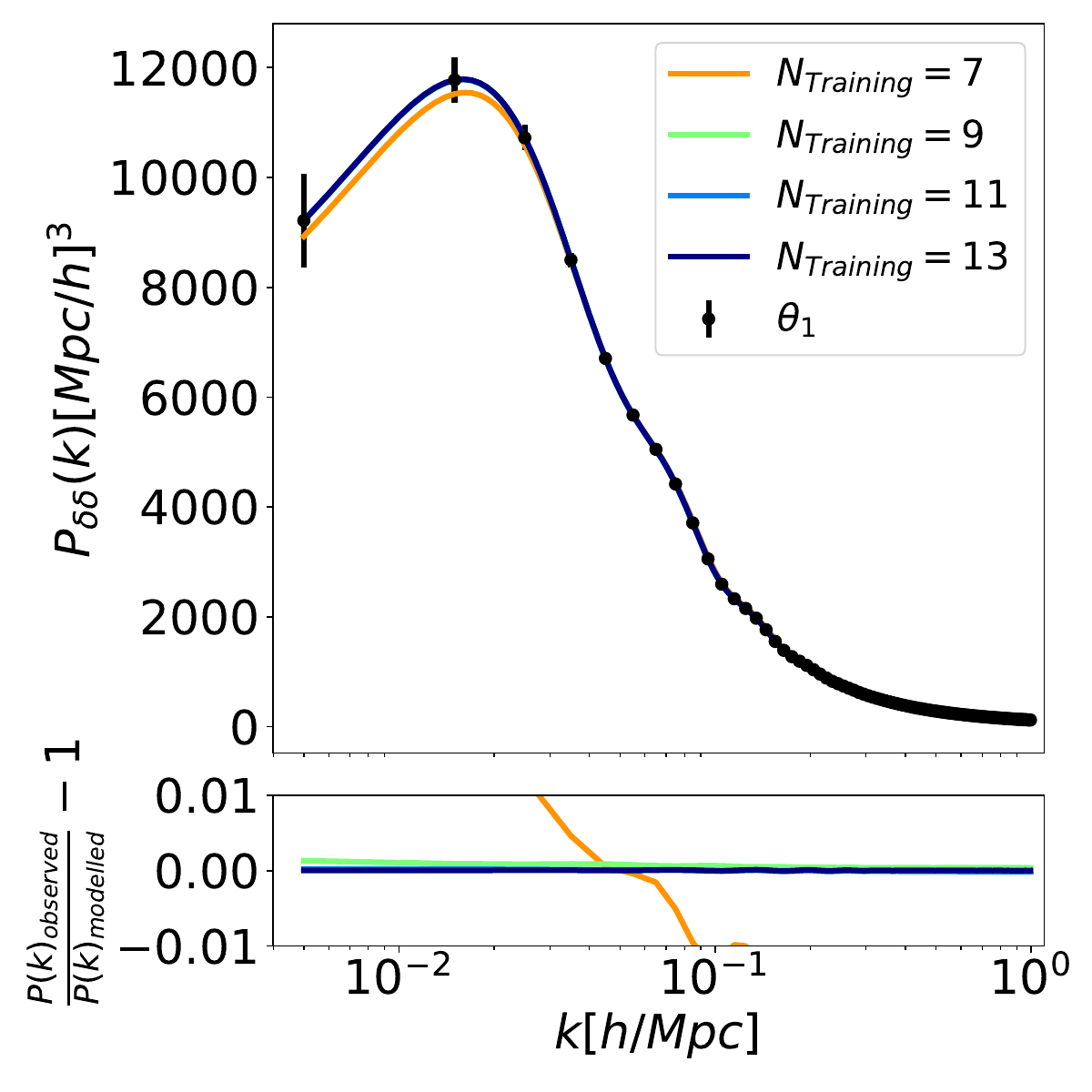}
\includegraphics[width=80mm,height=80mm]{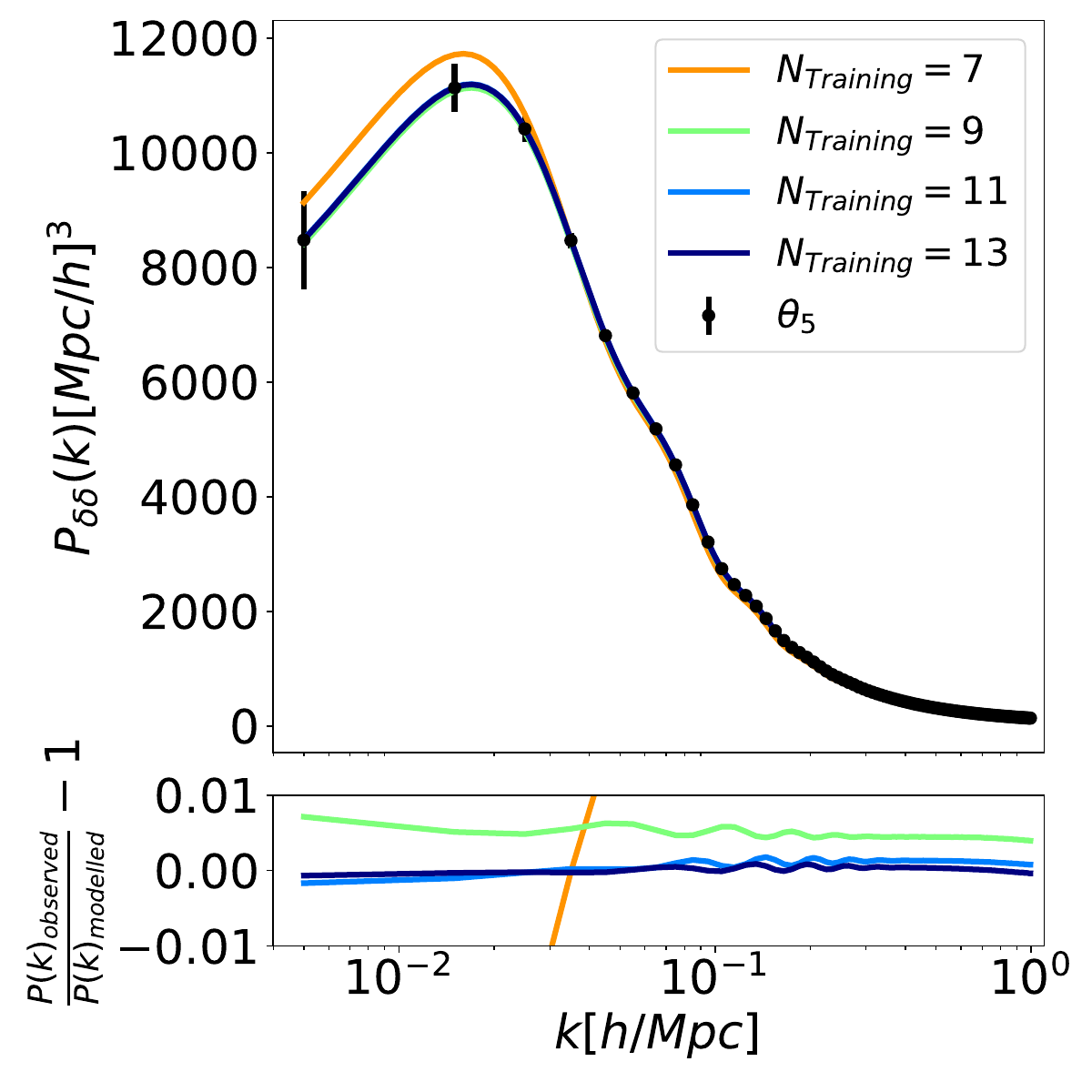}
\caption{ The top panels show the value of $P_{\delta \delta}$ at $z=0.7$, directly measured from the N-body simulation (black dots), for our $\theta_1$ (left) and $\theta_5$ (right) test points, with errors computed using equation \ref{error_equation}. The colored lines correspond to predictions made by emulators built with simulation designs of different sizes as mentioned in the figure label. The bottom panels display the per~cent error of each model when compared to the N-body simulations. The figure shows that models with more than 7 test points have an accuracy below $1\%$ for both points. }
\label{Nsim_needed}
\end{figure*}

To illustrate the relation between the accuracy of our N-body emulators and the number of test points, we commence by constructing an emulator for $P_{\delta\delta}$ using our N-body outputs at $z=0.7$. Subsequently, we systematically remove two points from our design and build a new emulator with the remaining test points. With each successive removal of two points, the emulator's performance should deteriorate, eventually the accuracy of the resulting emulators exceeds $1\%$, and becomes insufficient  for our purposes, this happens when the emulator is trained on only 7 points. We determine which points to remove at each step by identifying the point and its reflection (we use a symmetric hypercube) that, when removed, maximizes the average distance between all remaining points in the training set. In this sense, we aim to eliminate the points that would result in a sparser  simulation design and therefore in the best possible emulator.

Fig.~\ref{Nsim_needed} highlights the accuracy of the resulting emulators in reproducing the value of $P_{\delta\delta}$. The errors of the N-body simulations on the top panels (black dots) are computed using the following expression

\begin{equation}
\label{error_equation}
\sigma_{P(k)}=\frac{P(k)}{\sqrt{N_{modes}}}
\end{equation}
where $N_{modes}$ is the number of independent k-modes  in a particular bin.

By looking at the bottom panels of both plots, we observe that models with 9 training points or more are able to accurately reproduce the value of $P_{\delta\delta}$ for both our $\theta_1$ and $\theta_5$  test points with below per~cent accuracy. However, when using only 9 points, the predictions for $\theta_5$ are barely below the $1\%$ threshold at large scales. We note that the model with 11 points reaches accuracies of around $0.1\%$ but still exhibits systematic shifts at that error size for high and low values of $k$. This becomes evident when we consider the blue line in the right bottom panel, which corresponds to the errors estimated using 11 training points and is not centered around zero.

With our 13 points, there seems to be no deviation from zero, at least to the accuracy of around $0.1\%$. We conclude that both our emulators with 11 and 13 points are sufficiently accurate to reproduce $P_{\delta\delta}$ for both of our test points. Given that we have only two test points and we have not tested our full parameter space, we take the conservative decision of keeping 13 points to build our emulators.

We note that 13 simulations is a relatively small and quite manageable number of simulations. Given that this is the bottleneck of our Hybrid Model, our emulator methodology reduces the main computational cost of making beyond second order theoretical templates to that of running a manageable subset of N-body simulations. This is further discussed in Section~\ref{subsec:Nbody_performance}, where we analyse the accuracy of our N-body simulation emulators.

\begin{table}
\centering
\begin{tabular}{|c|c|c|c|}
\hline
\multicolumn{4}{|c|}{Training} \\
\hline
Number & $w_{cdm}$ & $w_b$ & $n_s$ \\
\hline
1 & 0.1202 & 0.0224 & 0.9649 \\
2 & 0.1225 & 0.0222 & 0.9869 \\
3 & 0.1179 & 0.0225 & 0.9429 \\
4 & 0.1167 & 0.0229 & 0.9832 \\
5 & 0.1237 & 0.0219 & 0.9466 \\
6 & 0.1190 & 0.0231 & 0.9576 \\
7 & 0.1214 & 0.0216 & 0.9722 \\
8 & 0.1155 & 0.0217 & 0.9539 \\
9 & 0.1249 & 0.0229 & 0.9759 \\
10 & 0.1132 & 0.0226 & 0.9612 \\
11 & 0.1272 & 0.0221 & 0.9686 \\
12 & 0.1260 & 0.0227 & 0.9502 \\
13 & 0.1144 & 0.0219 & 0.9796 \\
\hline
\multicolumn{4}{|c|}{Test} \\
\hline
$\theta_1$ & 0.1216 & 0.0226 & 0.969\\
$\theta_5$ & 0.1272&0.0216 & 0.986\\

\hline
\end{tabular}
\caption[ ]{Cosmological parameters of the simulation design used for training our N-body simulation emulator and for our test model.}
\label{tab:data}
\end{table}

In contrast to N-body simulations, RegPT templates are more cost-effective to generate, allowing us to create a large number of them in a relatively short time. We utilise two sets of RegPT templates in our study. The first set is the training set of our Gaussian emulator and consists of 50 points in the parameter space. As mentioned earlier, even with just 11 to 13 points, we can obtain accurate models of the matter power spectrum. Therefore, having 50 points is certainly more than sufficient. The second set serves as a test set, comprising 100 points that we employ in Section~\ref{subsec:RegPT_performance} to assess the accuracy of our methodology. Our 50 training points are selected using the Annealing methodology presented in Appendix~\ref{apendix:LatinHypercube}, while our test set consists of 100 random points within our parameter space.

 \subsubsection{The emulated power spectra}
\label{subsec:Nbody_performance}
Our N-body simulation emulators are built using the simulation design of Section \ref{subsec:grid-selection}. As stated we build 13 simulations that we use to train a GP emulator of the power spectrum. We also run two independent simulations which we now use to test the accuracy of our emulator.

\begin{figure*}
\includegraphics[width=80mm,height=70mm]{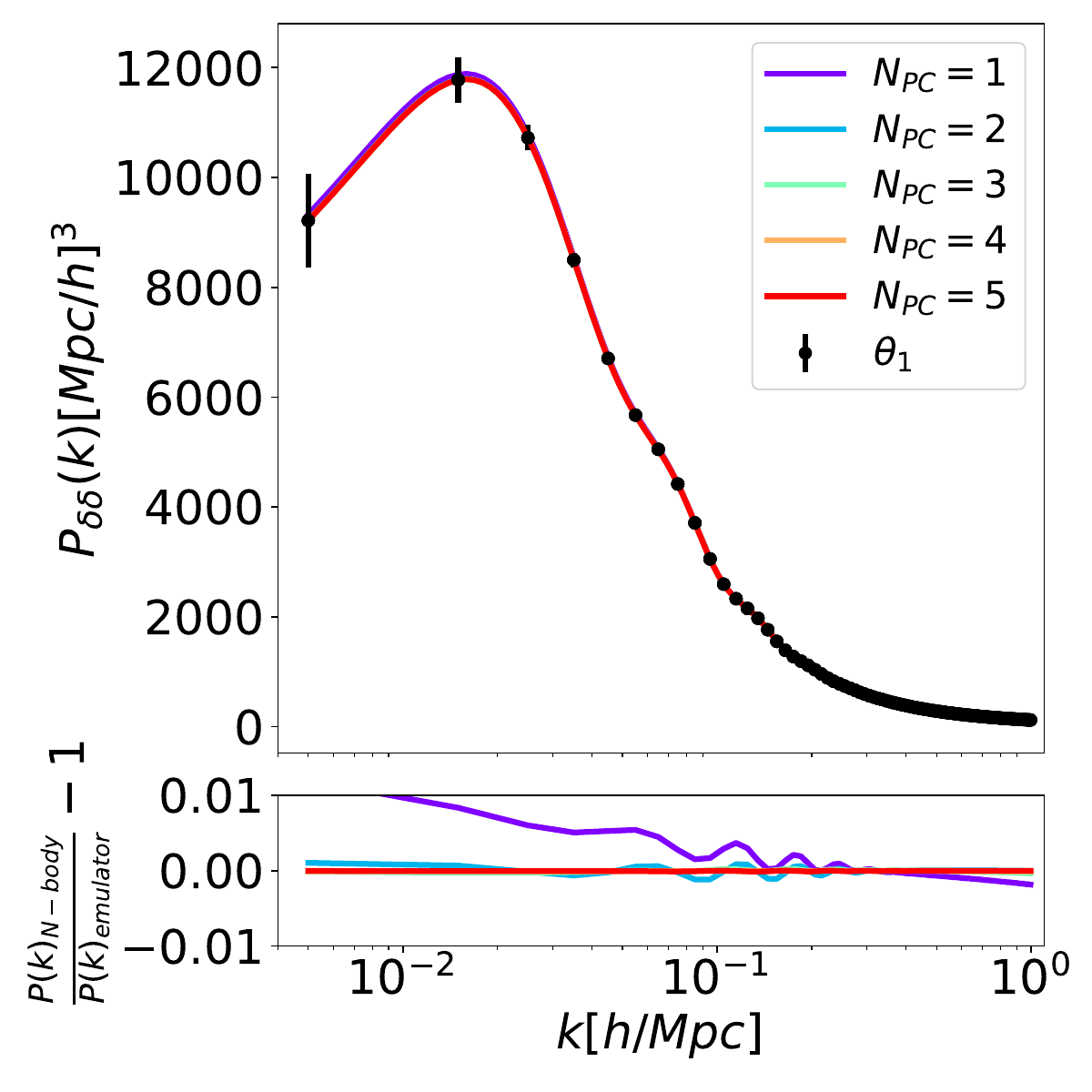}
\includegraphics[width=80mm,height=70mm]{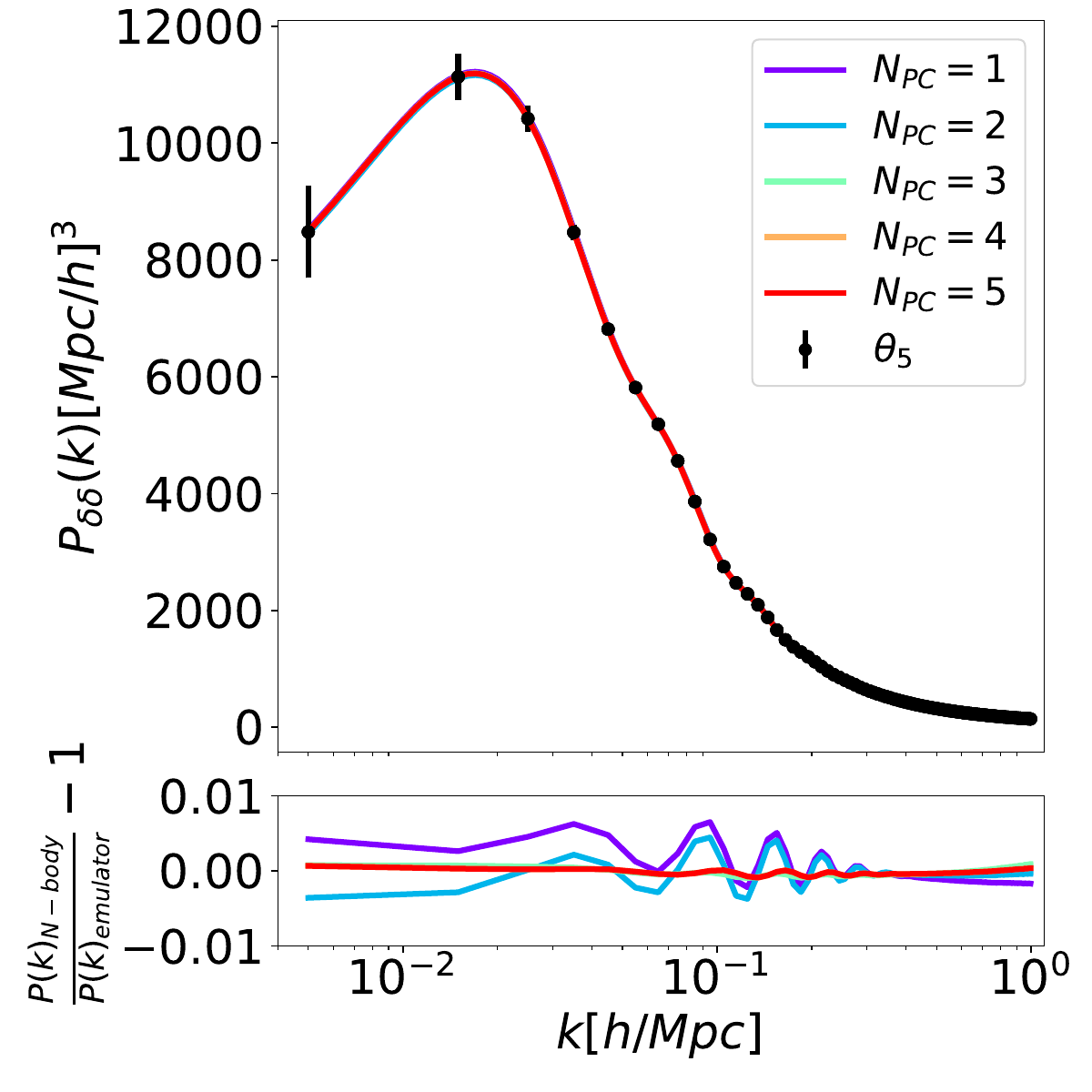}
\caption[ ]{The top panel of the plot shows the matter power spectra obtained from our test N-body simulation (black dots), and the predictions from our emulator using different numbers of principal components (colored lines). The left plot shows the simulation and predictions for the $\theta_{1}$ test point and the right column show those for the $\theta_{5}$ test point. The percentile errors of each model are shown in comparison to the N-body simulation. When utilizing  three or more principal components, the prediction accuracy is below 0.1~per~cent for both test points. The errors on the N-body statistics are computed using equation \ref{error_equation}.}
\label{Nbody_models_plot}
\end{figure*}

When introducing our Gaussian process methodology in Section~\ref{subsec:Intro_GP}, we emphasized that the power spectrum primarily relies on the most significant principal components. We now test that we retain enough  components so that we can build accurate models with below per~cent accuracy when reproducing $P_{\delta\delta}$ for both of our test points. Fig.~\ref{Nbody_models_plot} shows the  values of $P_{\delta\delta}$ measured measured from the test points N-body simulation (black dots) and  predicted with our emulators constructed using various values of $N_{PC}$ (colored line), which we defined as the number of principal components employed within a given model.

The bottom panel of the plot shows the percentile error of the power spectrum of the emulator models when compared to the measurement from our N-body simulation at $z=0.7$. The plot demonstrates that when using three or more principal components, the error is below 0.1~per~cent for both test points. This level of accuracy is an order of magnitude below the per~cent level accuracy that we aimed to achieve at all scales of interest. Throughout this work, we made the conservative decision of selecting $N_{PC}=5$. This choice was mainly driven by the fact that the computational cost increases when adding two more principal components is not substantial. Additionally, considering that we have tested our methodology on only two test point, this decision provides us with a larger error margin.

\begin{figure*}
\includegraphics[width=80mm,height=70mm]{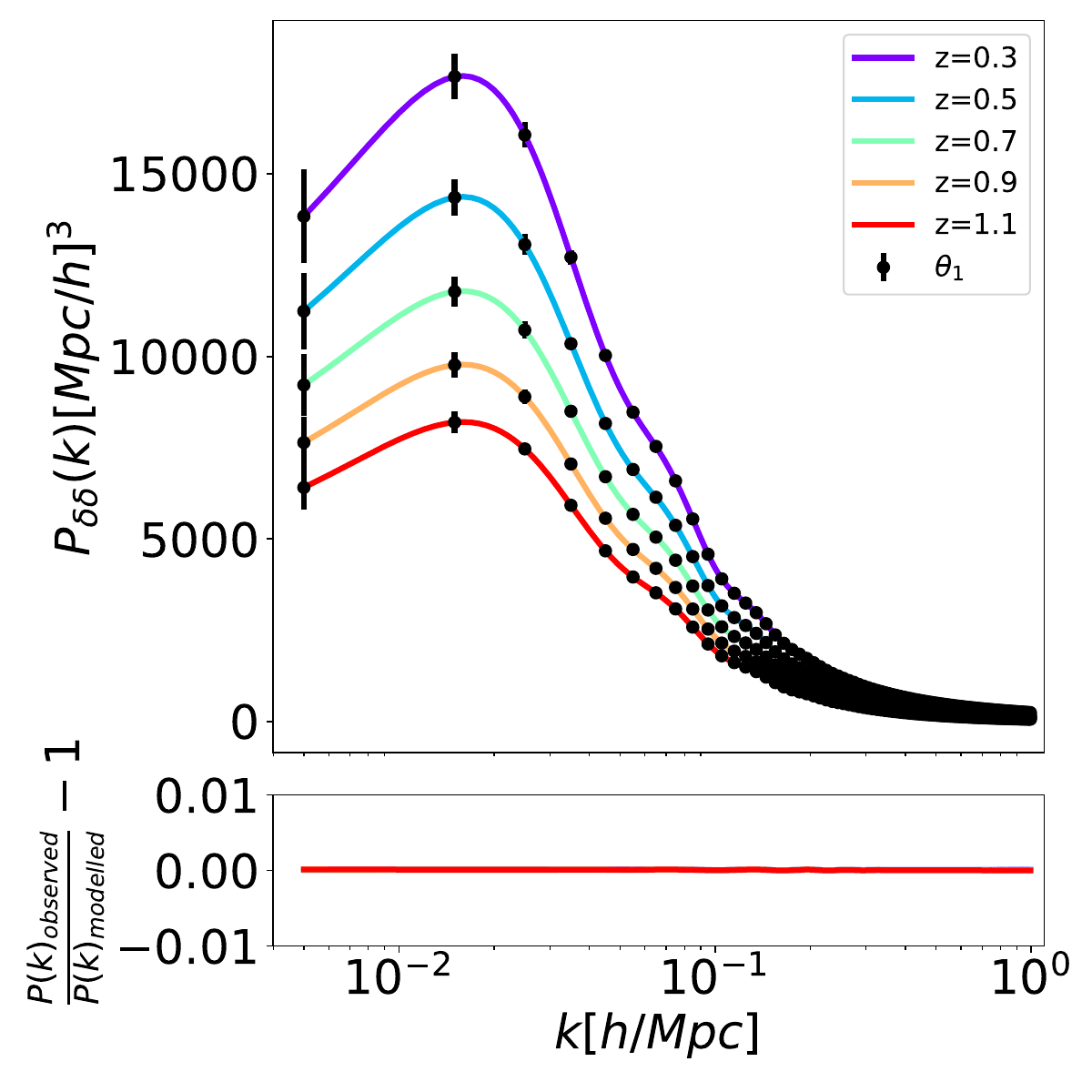}
\includegraphics[width=80mm,height=70mm]{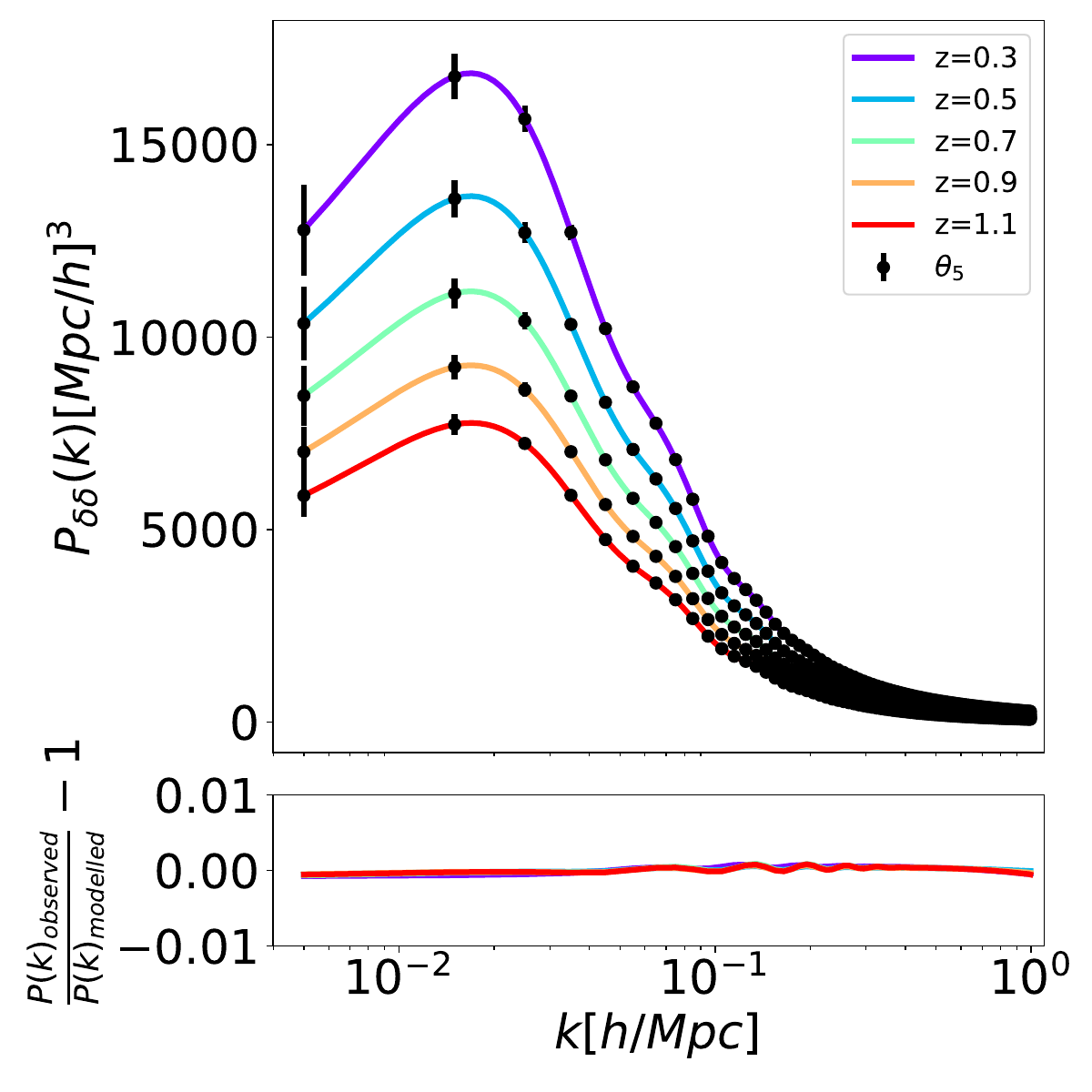}
\includegraphics[width=80mm,height=70mm]{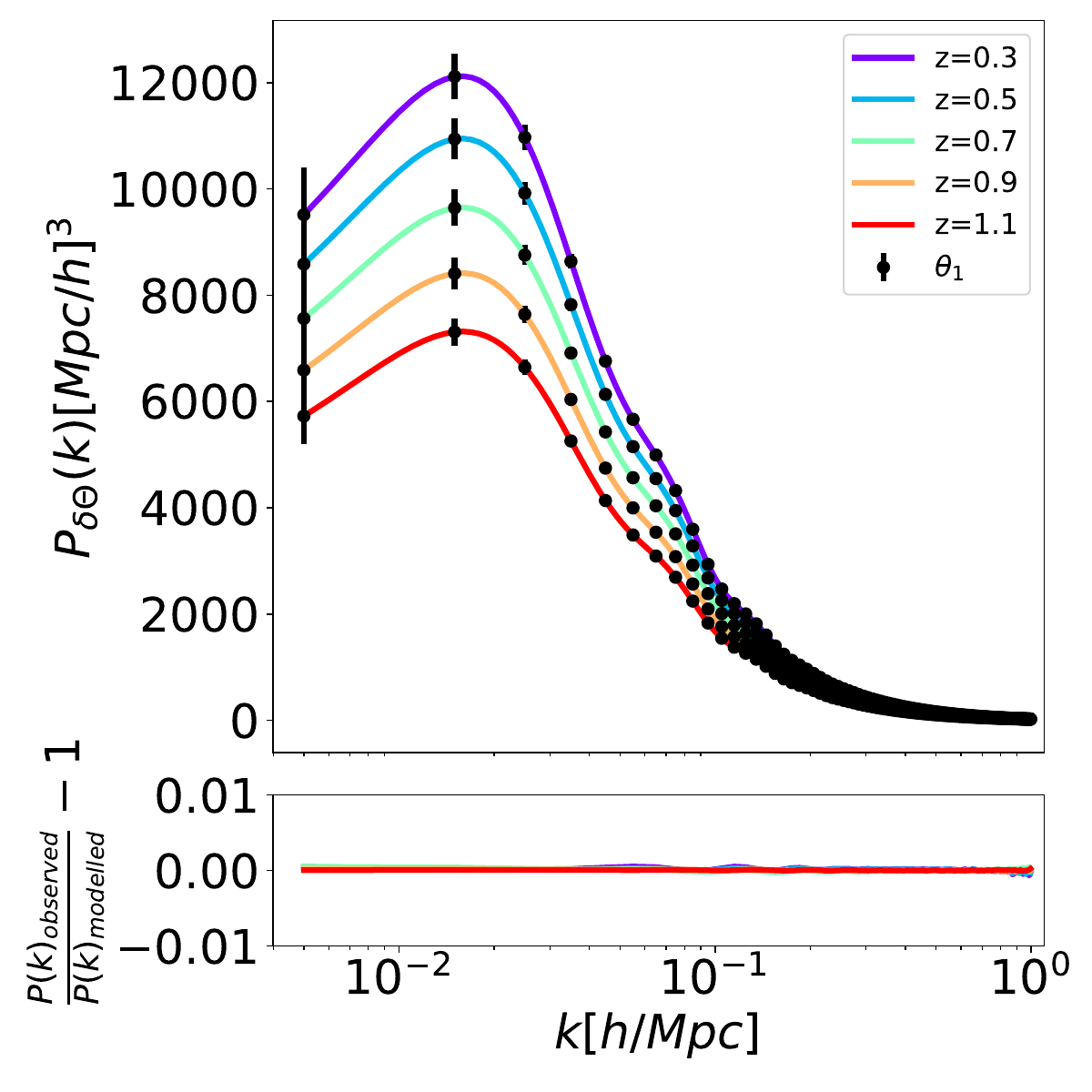}
\includegraphics[width=80mm,height=70mm]{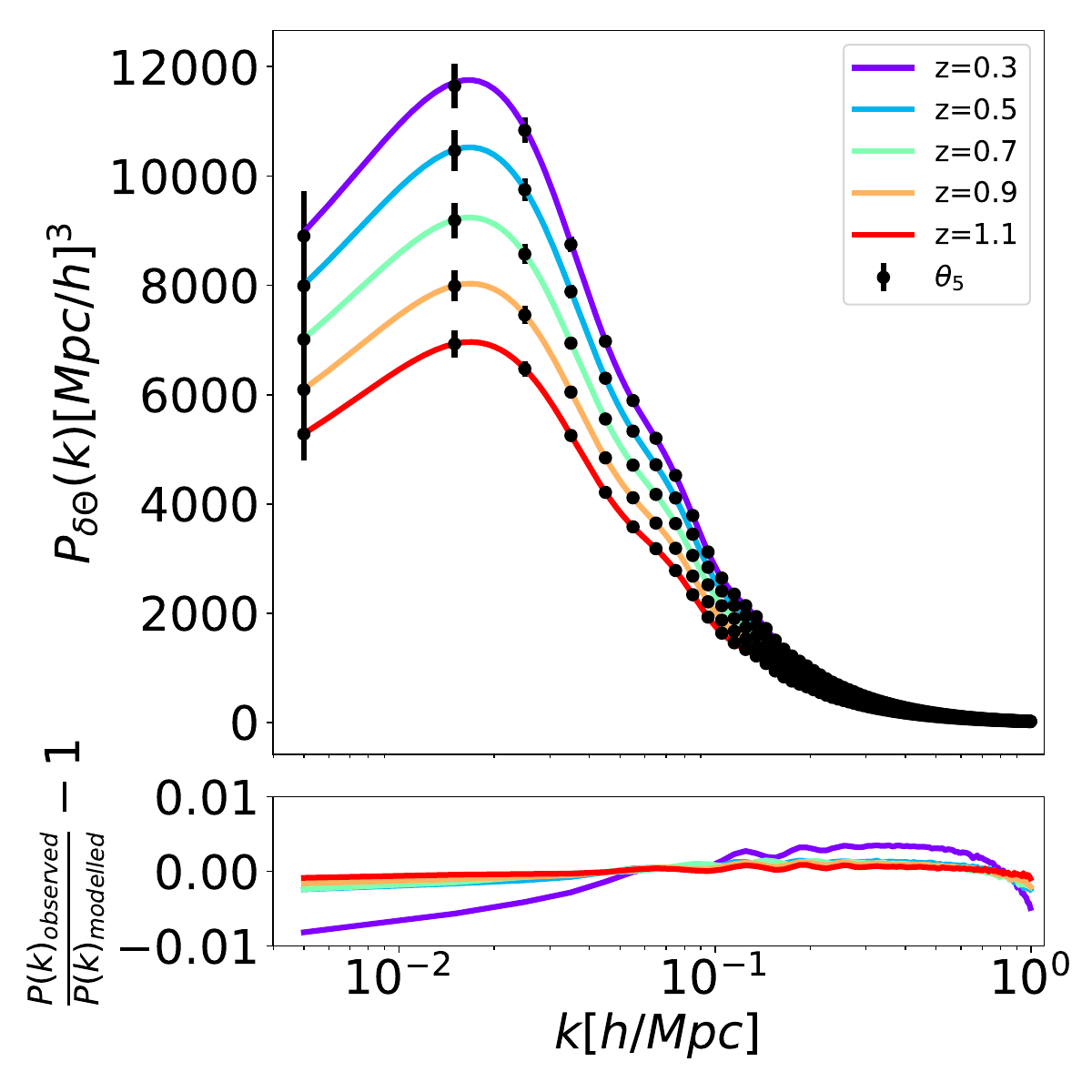}
\includegraphics[width=80mm,height=70mm]{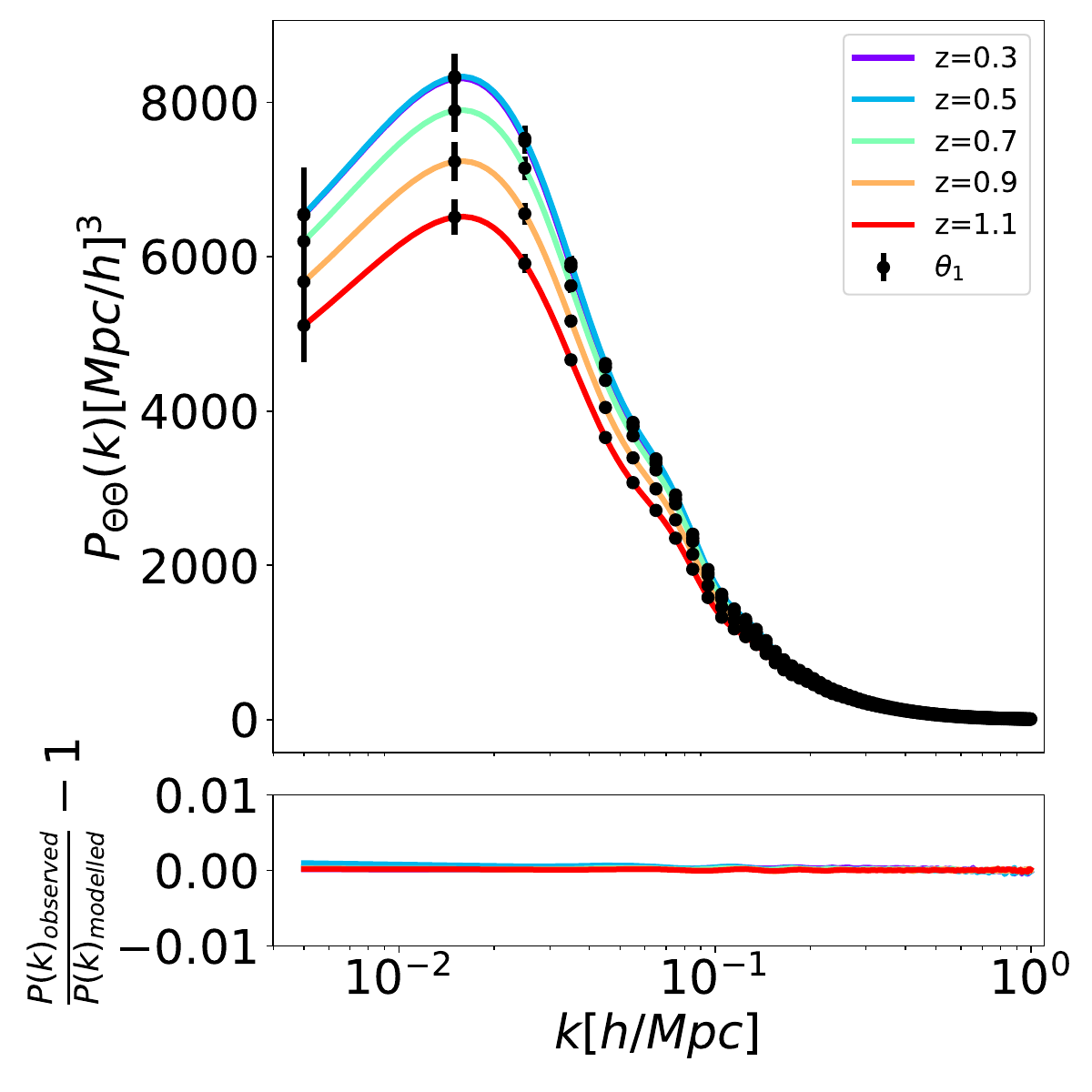}
\includegraphics[width=80mm,height=70mm]{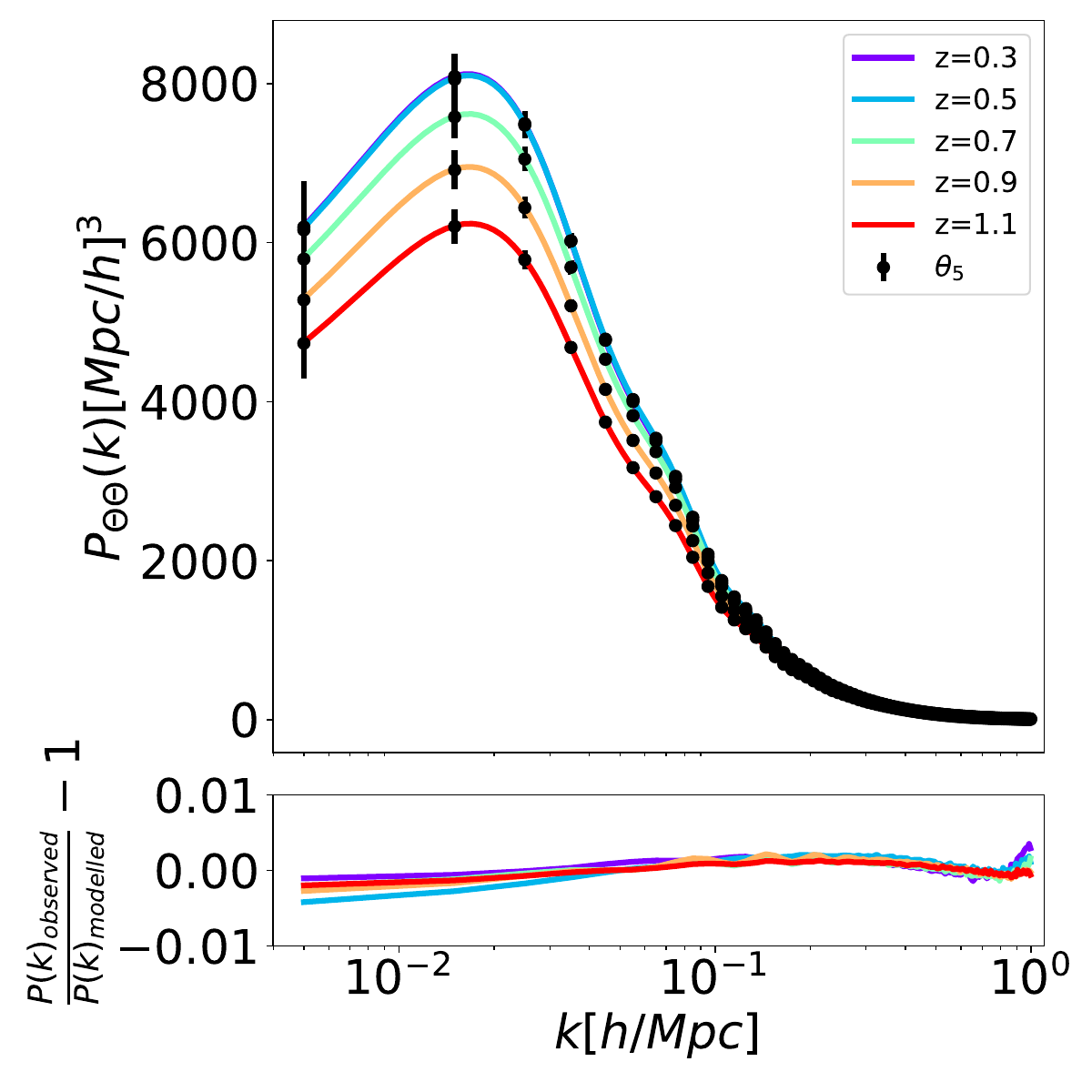}
\caption{The black dots show the density-density (top),  density-velocity (middle) and velocity -velocity (bottom) power spectra of our test N-body simulations, with the $\theta_{1}$ simulation on the left column and the $\theta_{5}$ on the right. The errors associated to each dot are computes using equation~\ref{error_equation}. Alongside we include the predictions of our different emulators built at the corresponding redshifts (colored line). The bottom panel shows the percentile error of the predictions of all emulators for the corresponding test point. We note that the accuracy is below 1~per~cent for all six predictions. }
\label{Nbody_models_allz}
\end{figure*}

Fig.\ref{Nbody_models_allz} displays the N-body simulation results at our five redshifts of interest for three statistics: $P_{\delta\delta}$, $P_{\delta\Theta}$, and $P_{\Theta\Theta}$ (represented by black dots) for both of our test points. Additionally, we show the predictions generated by each of the 15 distinct emulators we developed, covering these five redshifts and three statistics (colored lines).

The lower panels of the figures illustrates the percentage accuracy achieved by our model. It is worth noting that all results for $\theta_1$ (found in the left column) exhibit an accuracy well below $0.1\%$, which is one order of magnitude better that our target accuracy. As mentioned earlier, $\theta_1$ resides within the region of high likelihood of the Planck predictions. Therefore, these plots suggest that our model achieves remarkable accuracy in the regions of parameter space that are most relevant to us.

The left panels in each row displays the results for our $\theta_5$ test point. The bottom panels  show that the errors in our models are slightly larger than their $\theta_1$ counterparts, which is especially noticeable in the prediction for $P_{\delta\Theta}$ at $z=0.3$. Nonetheless, it's important to note that all models exhibit errors smaller than 1~per~cent. $\theta_5$ was chosen from a region within the parameter space where we anticipated our emulator might perform poorly. Despite this, our emulators consistently deliver predictions with accuracy within a per~cent even in this challenging regions.

Our model seems to be quite accurate when predicting our test point power spectra. It is worth noticing that this is achieved with 13 N-body simulations, which we consider a manageable number of simulations that can be built with relatively low computational resources. In Section~\ref{subsec:accuracy_test} we explore the particularities of our methodology  highlighting the factors that enable us to achieve such accuracy with a modest number of simulations.

\subsubsection{Emulated higher order terms}
\label{subsec:highorder_performance}
We have shown that our emulators can predict $P_{\delta\delta}$, $P_{\delta\Theta}$, and $P_{\Theta\Theta}$ with below per~cent accuracy. However, in order to compute predictions of the matter power spectra in our fiducial cosmology trough equation \ref{Matter_PS_equation2} we also require to emulate the higher order terms $A(k,\mu)$, and $M(k,\mu)$.

The exact methodology used to compute the terms is presented in Appendix~\ref{appendix:higher_order} as we explain in more detail there, the $A$ and $M$ higher order corrections are expanded into three terms that can be combined trough equations~\ref{A_g_app} into $A$ and $M$ predictions. We build six higher order corrections emulators, one per each of these terms. With these emulators we complete our methodology for emulating the matter power spectra. During this section, we conduct tests to evaluate the performance of our emulator in reproducing these statistics. We compute these 6 statistics from each of our N-body simulations, which we use to train our emulators.

It is worth noting that these statistics are functions of a two-dimensional parameter space, namely $k$ and $\mu$. This is different from the power spectra discussed in Section~\ref{subsec:Nbody_performance}, which only relied on $k$. To address this difference, we adopted the approach of constructing 10 distinct emulators for each statistic, each emulator corresponding to different values of $\mu$. This approach is valid since, in general, the computation of the matter power spectrum as a function of $\mu$ is not necessary for any given value of $\mu$. And instead, in typical RSD clustering analyses, a set of $\mu$ values is utilised to calculate, for instance, a multipole expansion of the power spectra. Given that we have 10 values of $\mu$ and 6 statistics we need to build 60 emulators per redshift, and 300 in total for all of our five redshifts of interest.

\begin{figure*}
\includegraphics[width=80mm,height=70mm]{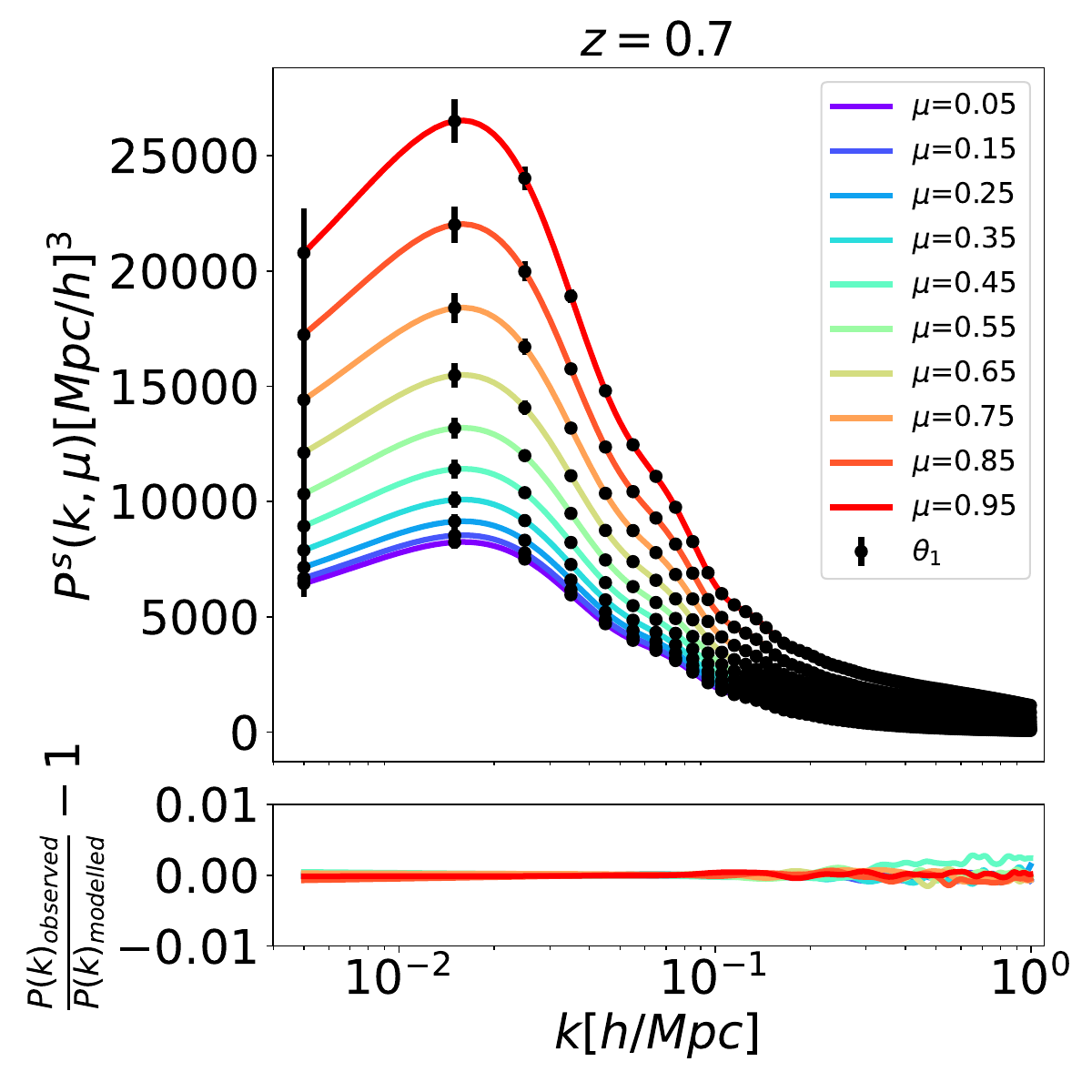}
\includegraphics[width=80mm,height=70mm]{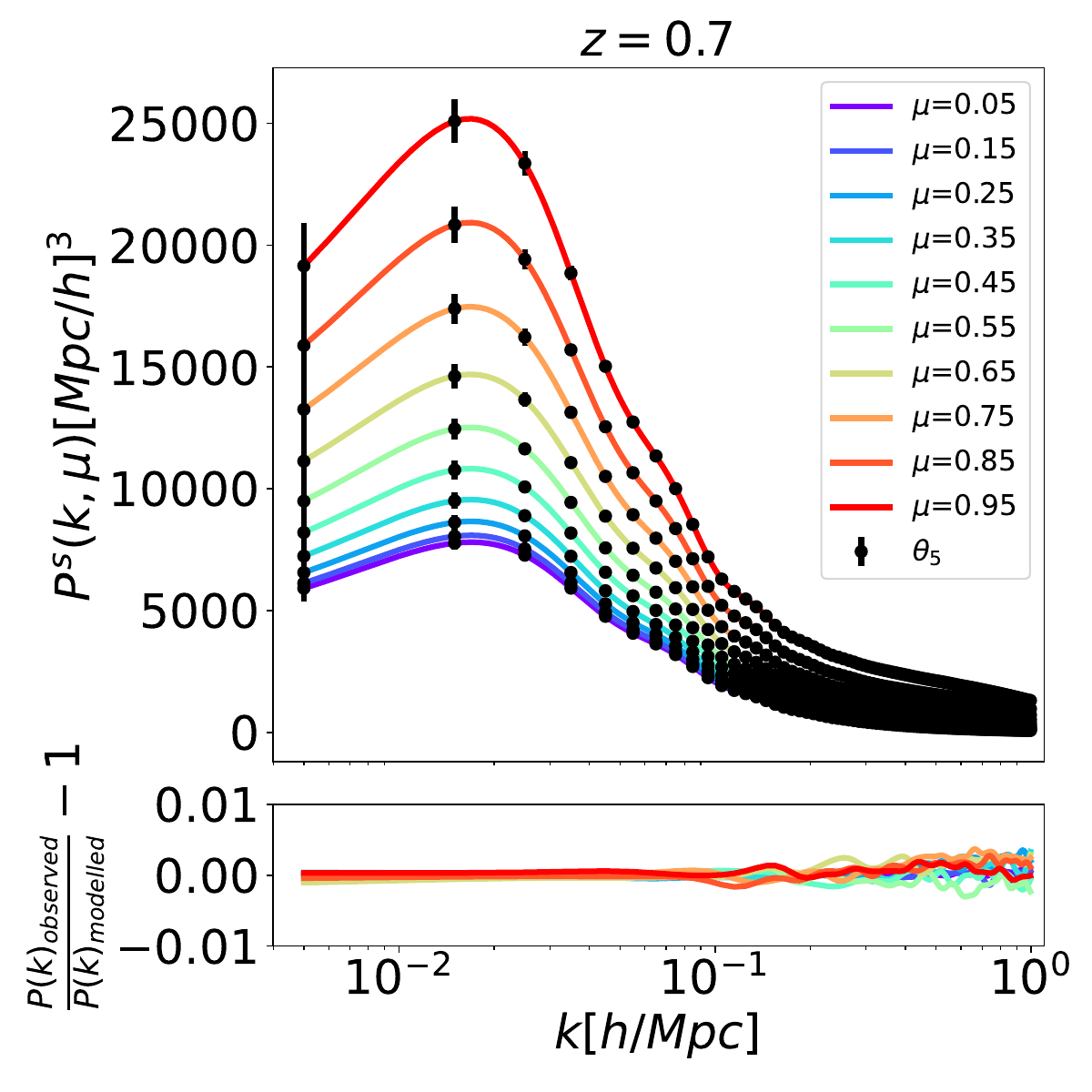}

\caption{The plot shows the matter power spectra of our N-body simulations, computed using equation~\ref{Matter_PS_equation2} (depicted as black dots). The error estimates are computed using equation~\ref{error_equation}. The colored lines represent predictions obtained by estimating the values of $A$ and $M$ with our emulators, when combined through equation~\ref{Matter_PS_equation2} with the values of $P_{\delta\delta}$, $P_{\delta\Theta}$, and $P_{\Theta\Theta}$ from our N-body simulations. Each distinct line corresponds to a different value of $\mu$ as highlighted in the figure's label. The models and predictions for our $\theta_1$ and $\theta_5$ test points are displayed in the left and right plots, respectively. The bottom panels show the percentage error of each prediction. We generated this plot with $\sigma_z=1$ and $\sigma_p=1$.
}
\label{ABFT_results}
\end{figure*}

The higher order correction terms have a small effect on the matter power spectra as they are second order effects. Therefore it might not be very informative to highlight the error estimates of each term directly but instead to show the effect that they have in the final prediction of the matter power spectra. 

This is illustrated in Fig.~\ref{ABFT_results}, where we calculate the full matter power spectra of our N-body simulation (depicted as black dots) by computing all individual terms of equation~\ref{Matter_PS_equation2} and then combining them together into a prediction of the matter power spectra. The colored lines are also generated using equation~\ref{Matter_PS_equation2}, but in this case, we include the higher-order correction term predictions from our emulators, while retaining the values of $P_{\delta\delta}$, $P_{\delta\Theta}$, and $P_{\Theta\Theta}$ from our simulations. This is done to demonstrate the impact of the errors of our high-order corrections emulators on the overall power spectra. The plot highlights that for both of our test points, the errors of our higher-order correction emulator are negligible, and they only have a below per~cent effect at $k > 0.1$\,$h$\,Mpc$^{-1}$.

\subsection{Precision test on hybrid RSD model}
\label{hybrid_model_tests}

 \subsubsection{RegPT}
 \label{subsec:RegPT_performance}

As mentioned in Section~\ref{subsec:RegPT} our PT theoretical template takes a set of seven statistics (equation~\ref{RegPT_2loop}) as inputs and uses them to compute the second order prediction of the matter power spectrum. These statistics are computed using the RegPT code and are a required input to our hybrid methodology.

We compute three RegPT power spectra $P_{\delta \delta}$, $P_{\delta \Theta}$, $P_{\Theta \Theta}$, at five different redshifts, which makes a total of 105 statistics to emulate. We build all models using our Gaussian emulators from Section~\ref{subsec:Intro_GP}. 

In order to test the accuracy of the methodology we use our models to predict the statistic of the 100 parameter space points in our test set. For each point $\alpha$ in our test set we compute the per~cent error ($PE$) of the GP predictions as 

\begin{equation}
\label{eq:percent_error}
    PE(\alpha,k,z)= \bigg | \frac{P_S^{GP}(\alpha,k,z)- P_S^\mathrm{RegPT}(\alpha,k,z)}{P^\mathrm{RegPT}(\alpha,k,z)} \bigg |
\end{equation}

\begin{figure*}
\includegraphics[width=70mm,height=55mm]{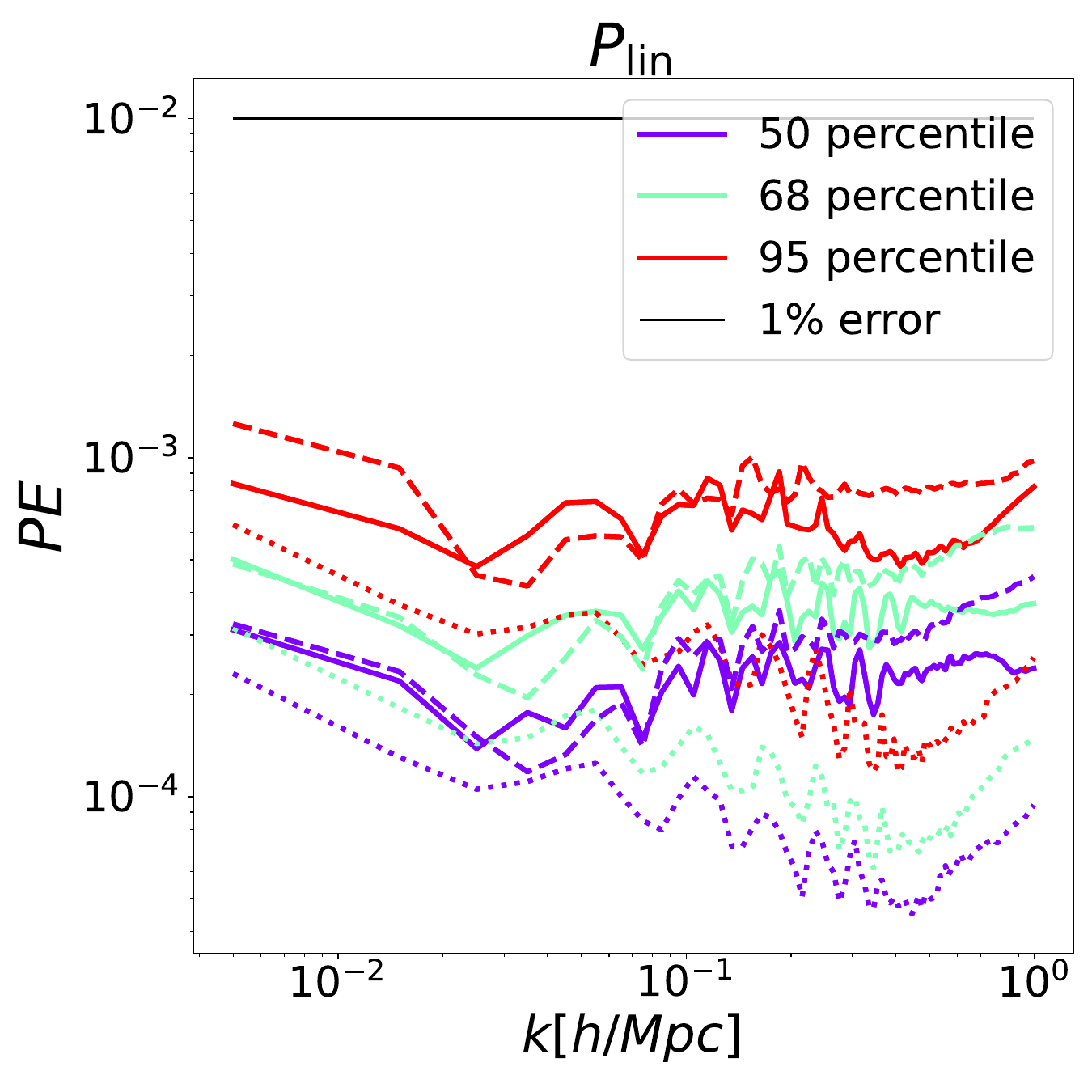}
\includegraphics[width=70mm,height=55mm]{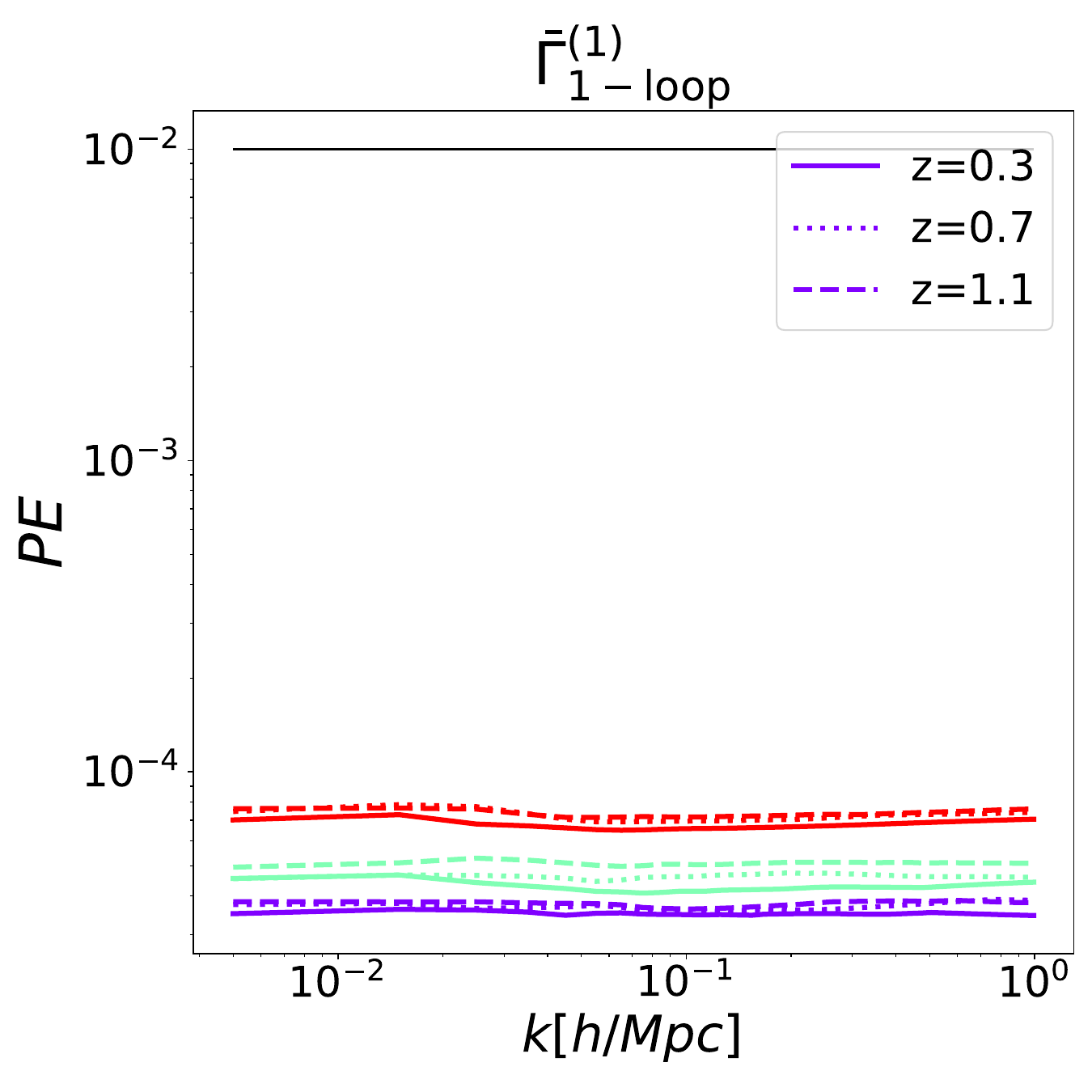}
\includegraphics[width=70mm,height=55mm]{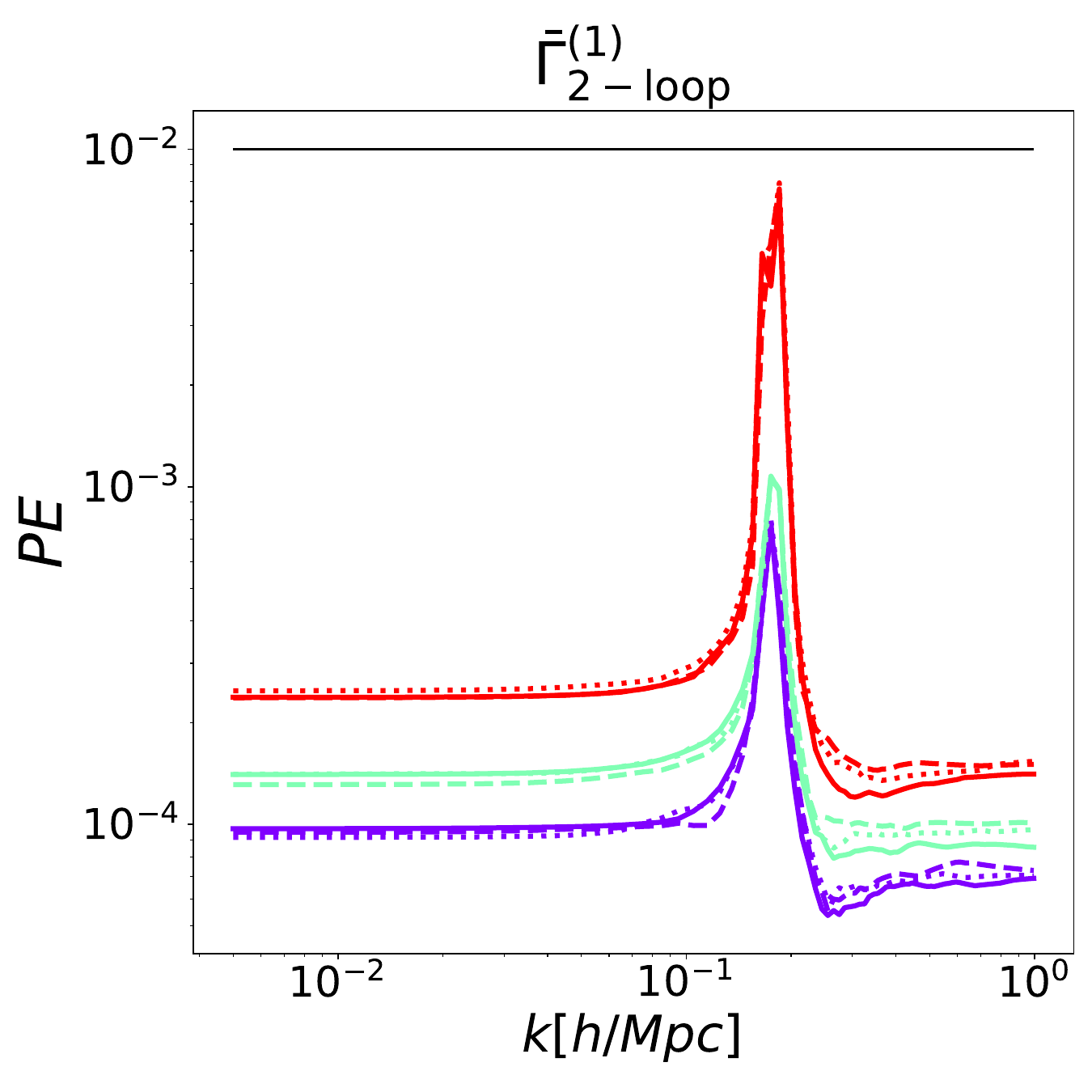}
\includegraphics[width=70mm,height=55mm]{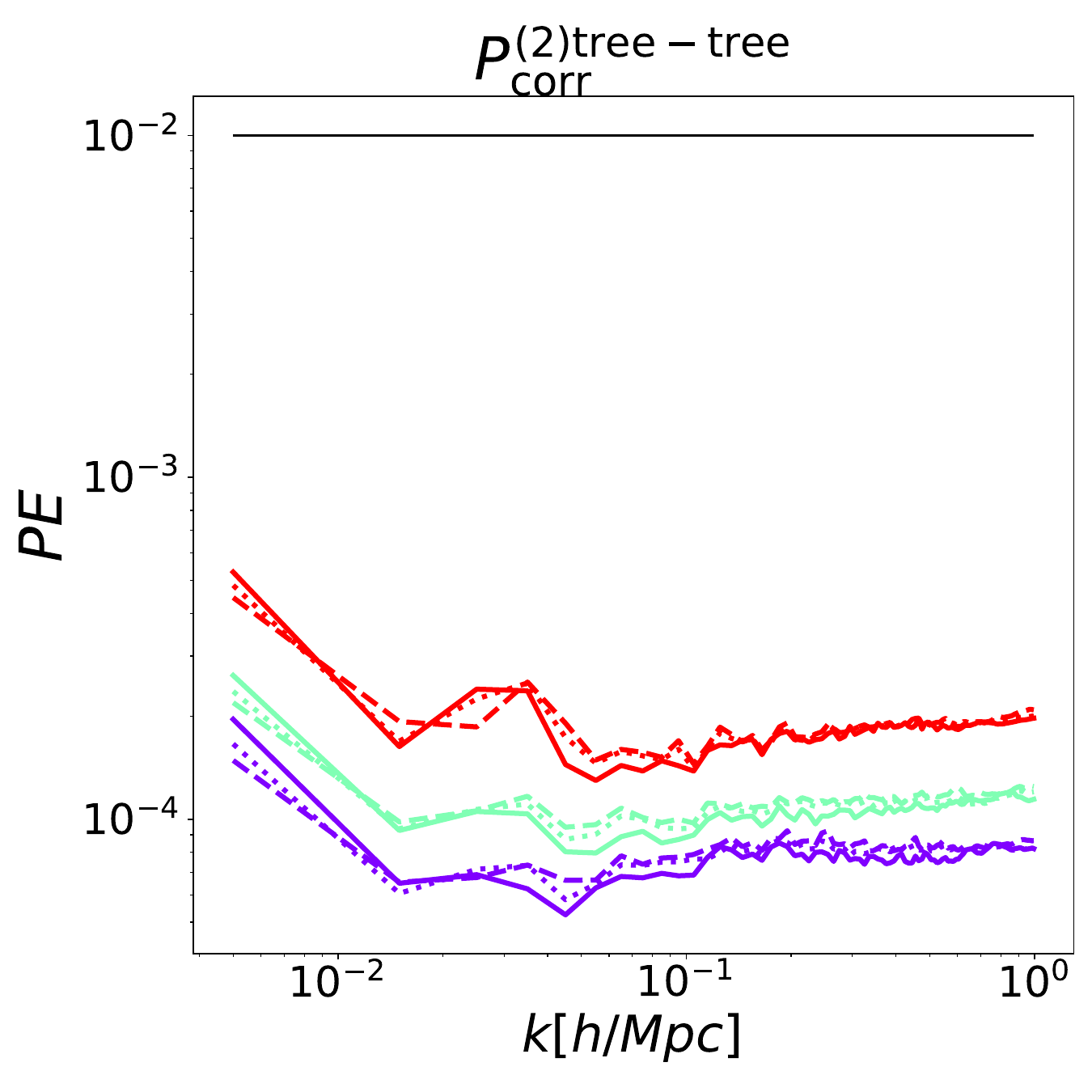}
\includegraphics[width=70mm,height=55mm]{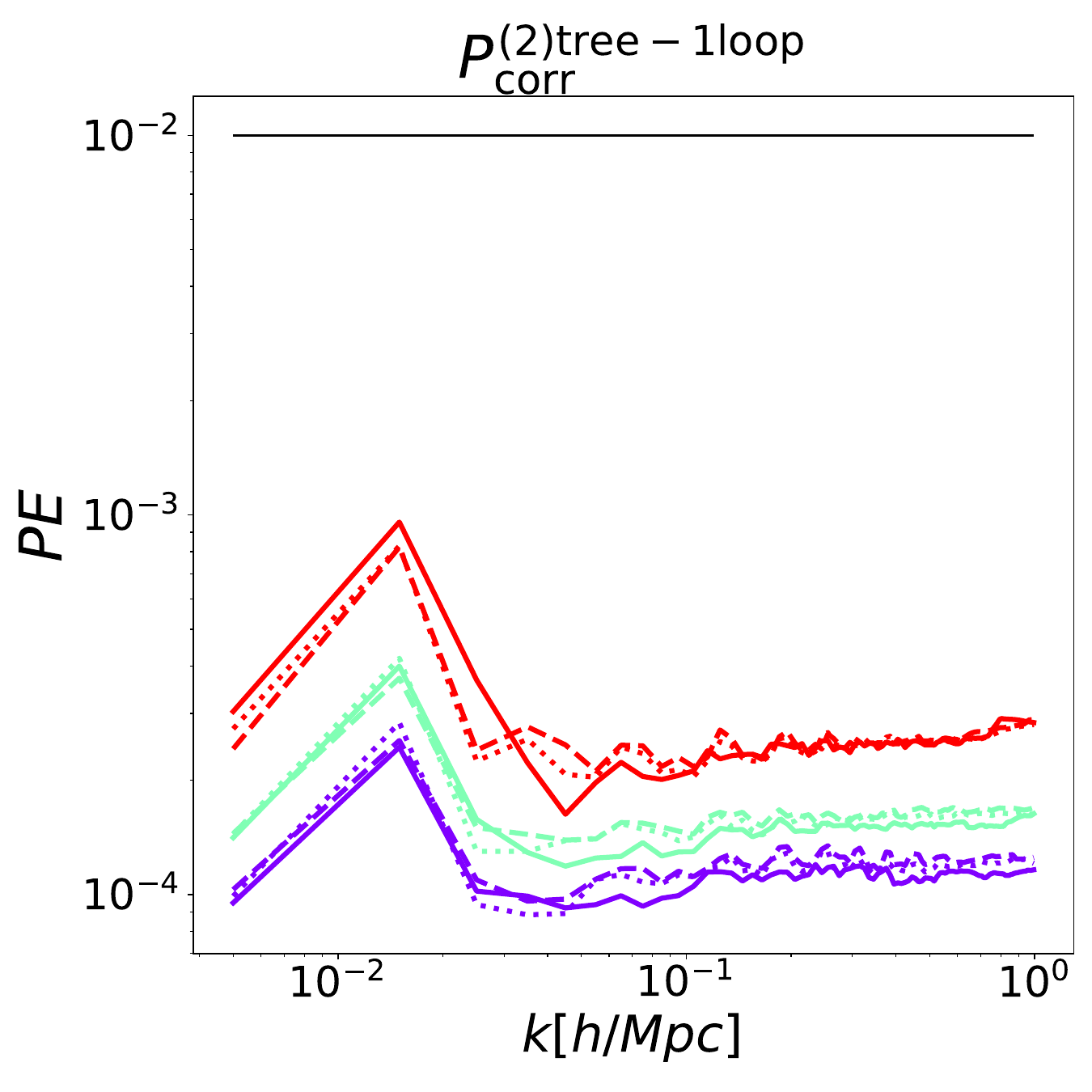}
\includegraphics[width=70mm,height=55mm]{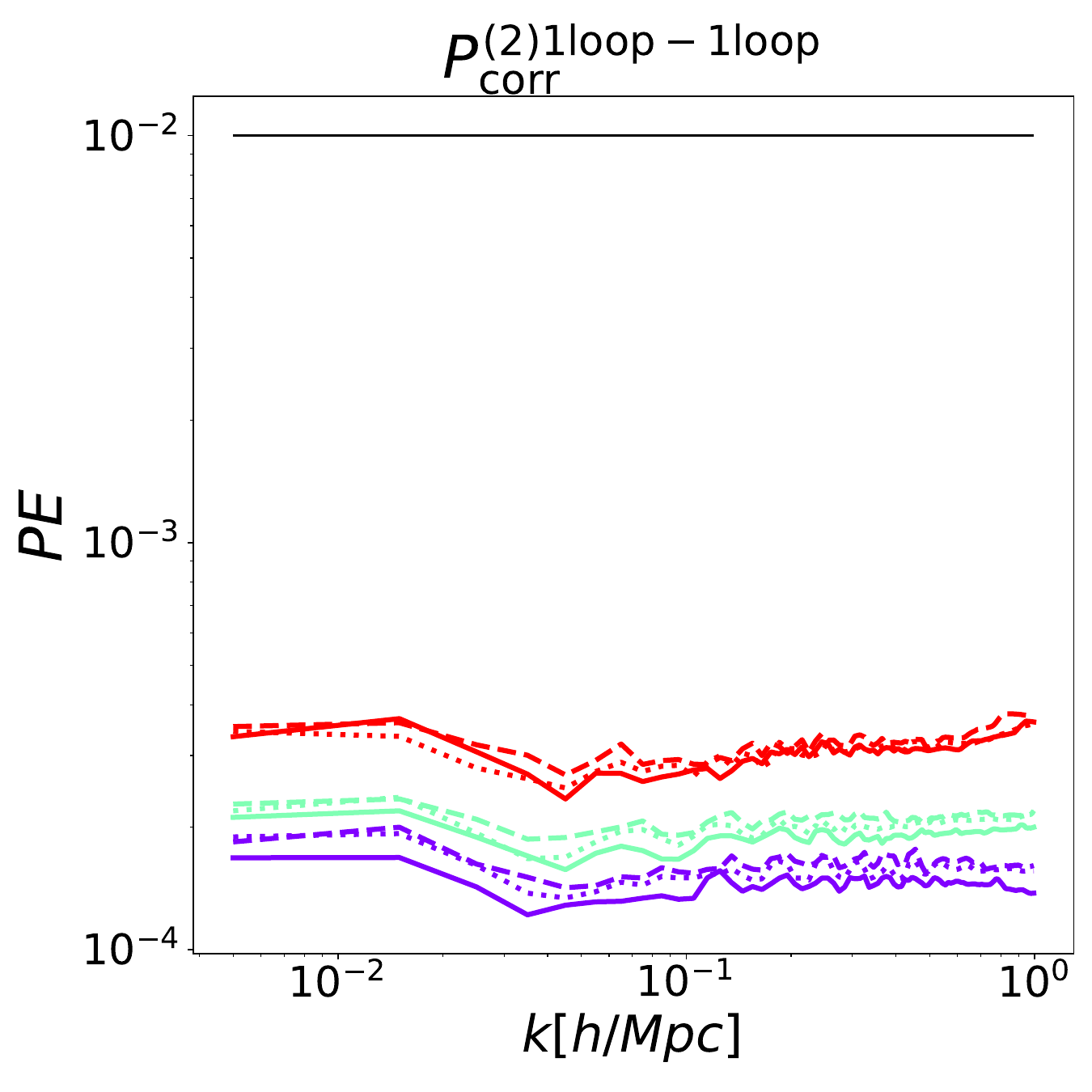}
\includegraphics[width=70mm,height=55mm]{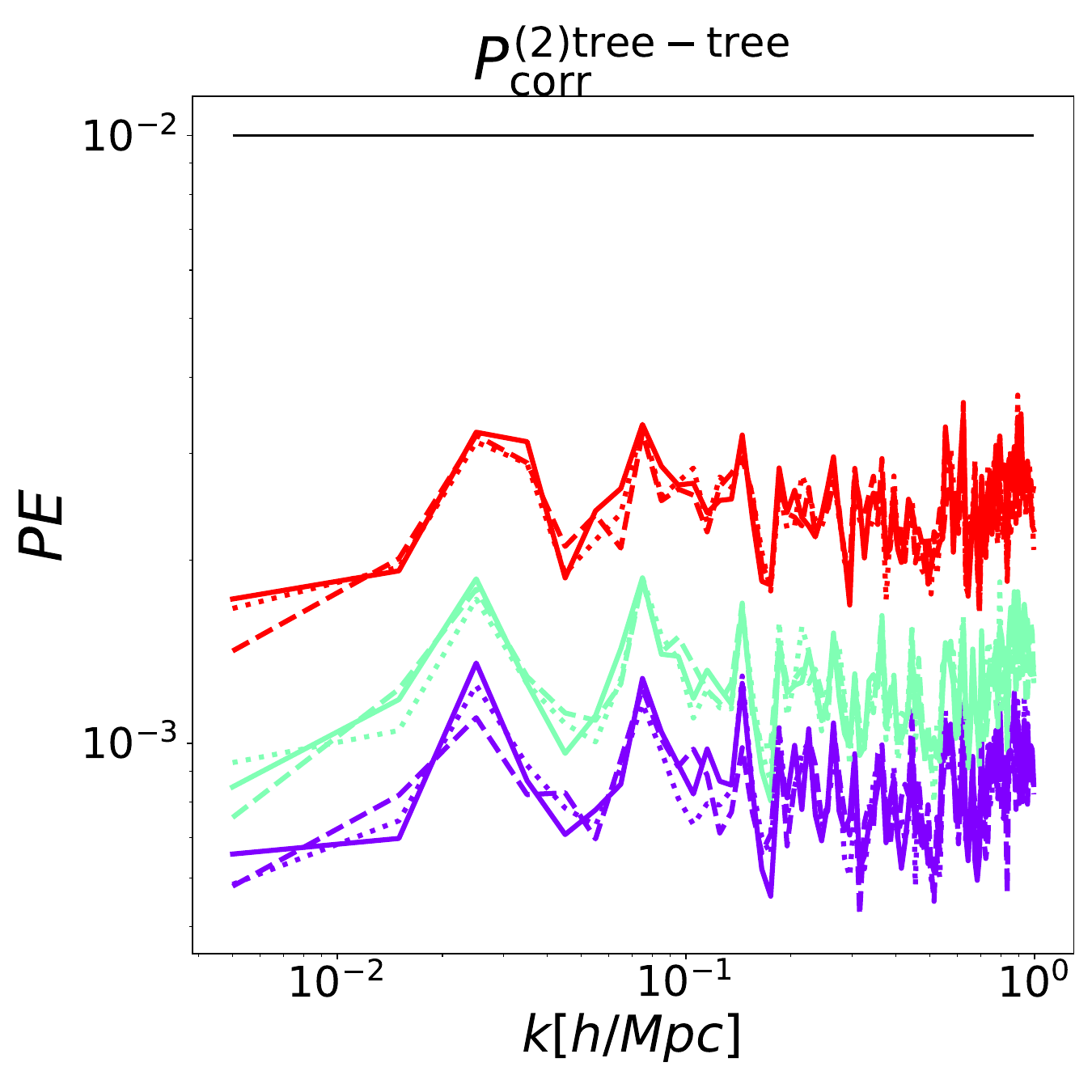}

\caption[ ]{The figure presents the percentile plots showcasing the percentile errors in our emulators predictions for all 7 RegPT statistics, as indicated in each figure title. The lines displayed are solid, dashed, and dotted, representing predictions at redshifts z=0.3, $z=0.7$, and $z=1.1$, respectively. Each line represents the boundaries of our percentile errors which signify the levels below which 
$50\%$ (blue), $68\%$ (red), and $95\%$ (green) of our percentile errors lie.}
\label{RegPT_models_plot}
\end{figure*}

Where $P_S^{GP}(k,z)$ is the power spectrum statistic predicted by our GP emulator and $P_S^\mathrm{RegPT}(k,z)$ is the value computed using the RegPT code on $\alpha$. Fig.~\ref{RegPT_models_plot} shows the percentile lines, below each line we find 50, 68 and 95~per~cent of the $PE$ values for all points in our test set. The plot shows all the seven statistics for $P_{\delta \delta}$  at three redshifts of interest. We do not include all five redshifts as the missing two do not provide additional information, and the plots are easier to interpret with less data. The black line corresponds to a $1\%$ error, as we can see our emulator can predict all RegPT statistics with much less than a one per~cent error for all $k$ values of interest and at all redshifts, in fact, most statistics are modelled with an accuracy of around 0.01~per~cent. We have generated an equivalent analysis for $P_{\delta \Theta}$, $P_{\Theta \Theta}$ and we have found similar accuracy for all of our 105 emulators. 

\subsubsection{Test for anisotropy spectrum}
\label{subsec:accuracy_test}

In this section, we investigate the impact of two factors on the accuracy of our predictions of the density matter power spectrum: the number of free parameters utilised and the size of our priors. First, we analyse the effect of the accuracy due to the fact that within the hybrid model framework, we do not require our emulators to account for variations in any \textit{scale independent} parameter. And we can therefore reduce the number of emulator free parameters to three.

We construct several simulation designs with five dimensions and of different varying grid sizes, where we add $A_s$ and $h$ to our list of free parameters. Similarly to the other three dimensions, we varied these parameters within $5\sigma$ of the predictions by \cite{2020A&A...641A...6P}. Thus, we set $h = 0.6727 \pm 5(0.0060)$ and $\ln(10^{10} A_s) = 3.045 \pm 5(0.016)$.

We then utilise the CAMB nonlinear code introduced in Section~\ref{subsec:grid-selection} to estimate the matter power spectra at each point in these designs, which we use to train one GP emulator per grid size. We also utilise our CAMB model to compute the power spectra at 100 randomly selected points within our five-dimensional priors. These spectra are used to evaluate the accuracy of the resulting emulators, this is done by computing the per~cent error of equation~\ref{eq:percent_error}.

\begin{figure*}
\includegraphics[width=80mm,height=80mm]{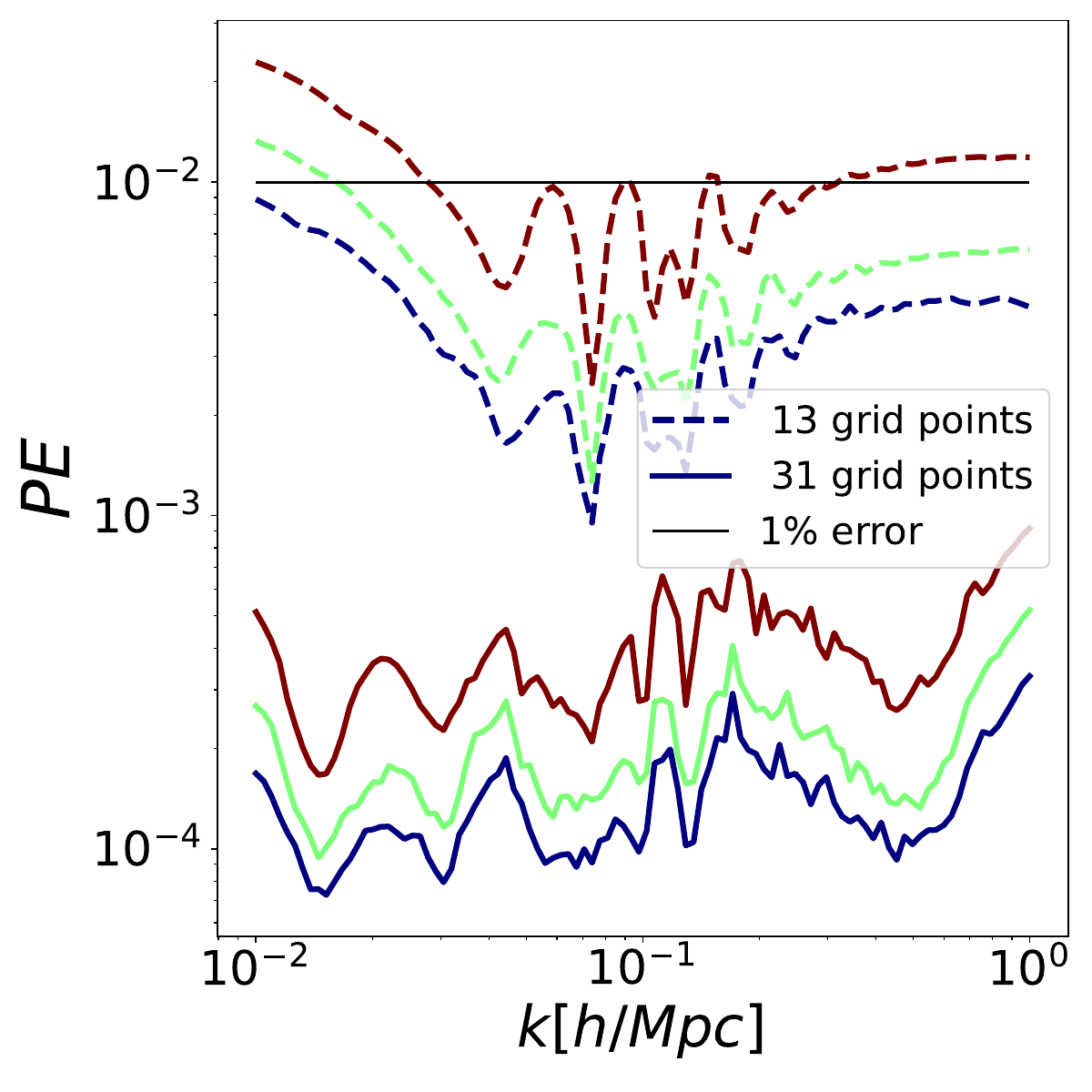}
\includegraphics[width=80mm,height=80mm]{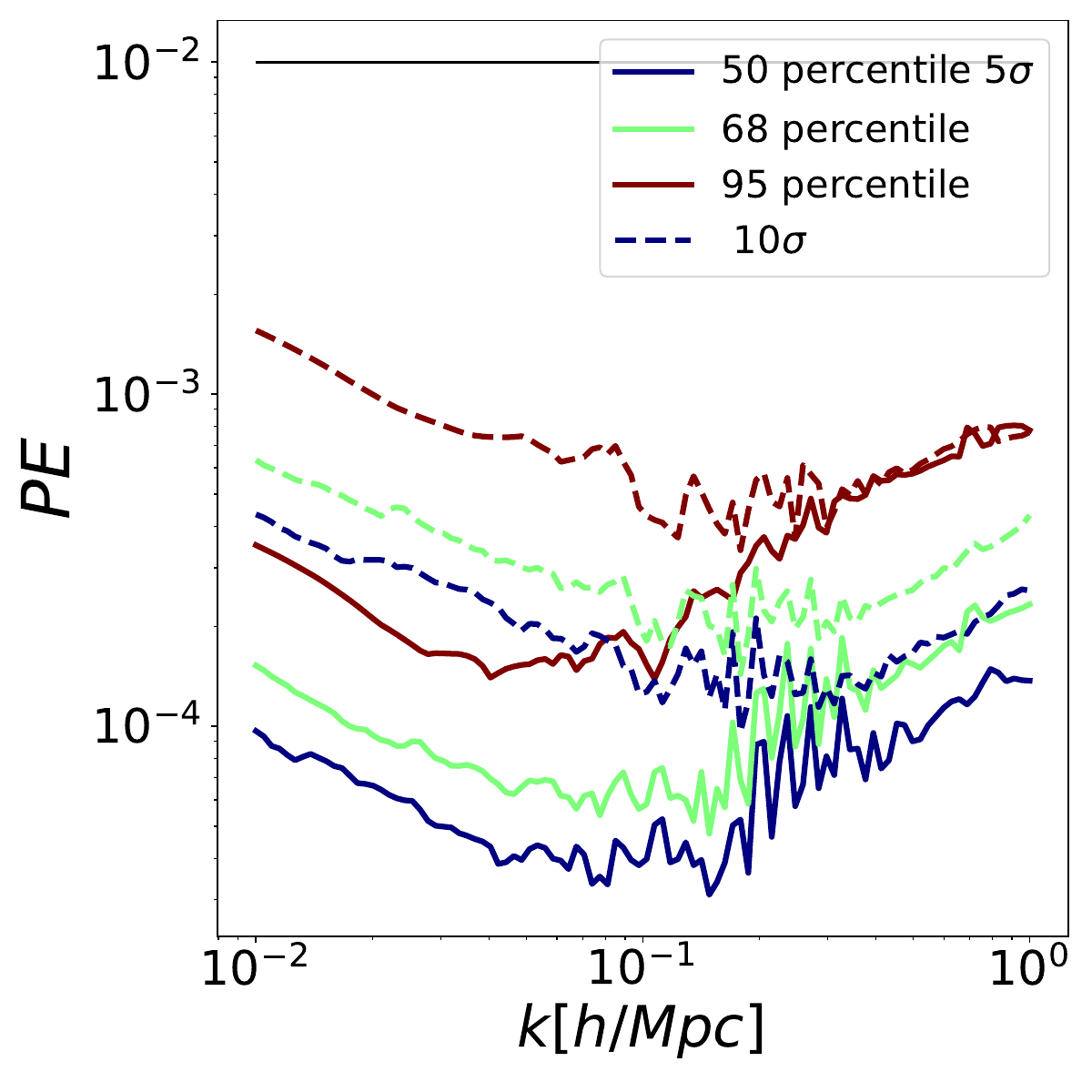}
\caption{ Percentile errors for the $50\%$, $68\%$, and $95\%$ percentiles for different models, similar to Fig.~\ref{RegPT_models_plot}. \textbf{Left:} Both models are trained on 5 parameters, but with a different number of training points. The dashed lines correspond to a model trained on  13 points, while the solid lines correspond to a model trained on a grid of 31 points. \textbf{Right:} Both models are trained on our standard grid of 13 points in 3 dimensions from Table \ref{tab:data} but with different priors. The dashed lines correspond to priors with a width of $10\sigma$ of the Plank 2018 predictions, while the solid lines correspond to $5\sigma$ priors.}
\label{5param_30vs13_PE}
\end{figure*}

The left side of Fig.~\ref{5param_30vs13_PE} shows the percentile errors for the 100 random test points using two different emulators, the dashed lines correspond to an emulator trained on 13 grid points, while the solid lines correspond to a model trained on a grid of 31 points. As mention when discussing Fig.~\ref{RegPT_models_plot}, this percentile lines corresponds to the thresholds below which the indicated percentage of the PE are. We note that the GP model trained with our canonical 13 points grid does not achieve the per~cent accuracy that we aim for (represented by the black line) in neither large nor small scales, and some models have a PE of around $4\%$ at most wavelengths. On the other hand, the model with 31 grid points achieves an accuracy below $0.1\%$ which is similar to what we get from our models in 3 dimensional space. From here we can infer that using the hybrid methodology to abandon variations on $A_s$ and $h$ allows us to reduce the number of N-body simulations by around two thirds.

The right side of Fig.~\ref{5param_30vs13_PE} shows the accuracy of two models. The first model is trained on our standard three-dimensional grid of 13 points from Table~\ref{tab:data}, which is sampled with priors of $5\sigma$ based on the Planck 2018 predictions. The second model is trained on a grid where the parameters have a width of $10\sigma$. It is important to note that both grids are identical in normalized parameter space, ensuring that the standardised distance between points remains the same. It is worth noting that although the $5\sigma$ model exhibits higher accuracy, both models fall significantly below our target per~cent accuracy represented by the black line. This suggests that there is room for flexibility in selecting our prior size, allowing for even larger priors that have already been strongly disfavored by Planck to be utilised in our models. The tight constraints on our cosmological parameters from \cite{2020A&A...641A...6P} result in relatively small variations between power spectra within our parameter space, thus facilitating the task of emulator modeling.

We note that our emulators only required 13 points to be built to their current accuracy, given that this is the bottleneck of the hybrid model methodology, our emulator methodology gives us a way of building a theoretical template of RSD power spectrum that go beyond second order perturbation theory and requires a manageable amount of N-body simulations to be made.

\section{Extension to diverse dark matter tracers}
\label{chap:Tracers}

The methodology presented so far models the clustering of the underlying matter distribution in redshift space. However, LSS surveys measure the positions of luminous objects like galaxies, which act as biased tracers of this matter distribution. 

Ideally, we would like to accurately replicate the observed galaxies in simulations, and use this simulations to build RSD emulators of the galaxy power spectrum trained in these simulations. However, in practice, there is no established method for precisely painting galaxies inside simulations that meets the elevated precision requirements of current LSS surveys. Therefore, within our emulator methodology we opted for using a theoretical galaxy bias that determines the tracers power spectra as a function of the  behavior of dark matter particles.

We begin this chapter in Section~\ref{Modleing_galaxy_pk} by introducing a specific theoretical model that maps the power spectrum of a tracer to that of the underlying dark matter distribution. This methodology only requires the emulator predictions that we have presented so far as input. It is worth noting that all the work presented before this section does not depend on the mapping between tracers and dark matter, and, in principle, the emulators presented so far should be useful in any alternative clustering model one desires to use.

In Section~\ref{Example_of_model}, we briefly exemplify the ability of our methodology to replicate the halo power spectra of a mock catalog of massive halos.

\subsection{Modeling the power spectrum of tracers of matter}
\label{Modleing_galaxy_pk}

The standard approach to computing theoretical models of the redshift space clustering statistics for any matter tracer is to model the required fields, mainly the density field ($\delta_g$), and the velocity field ($\vec{u}_g$), as functions of their matter space counterparts, here the subscript $g$ indicates that these statistics correspond to a biased dark matter tracer, this could be a halo or a type of galaxy.

Throughout this work, we assume that our tracers move with the same velocity of  the underlying matter distribution and therefore $\vec u_g=\vec{u}$.  Our expression for $\delta_g$ is given by
\bea
\label{eq:bias_x}
\delta_g(\bfx)=b_1\delta(\bfx)+\frac{1}{2}b_2[\delta(\bfx)^2-\sigma_2]+\frac{1}{2}b_{s2}[s(\bfx)^2-\langle s^2\rangle]\no\\
+{\rm higher\,\,order\,\, terms}.
\eea

Where $s(\bfx)$ is a function of the gravitational potential, $b_1$ and $b_2$ are two free parameters of the methodology used to model the mapping between tracers and the underlying dark matter, they are referred to as the linear bias parameter and the second-order local bias parameter, respectively. $b_{s2}$ is a second order non-local bias parameter.

 To adapt equation~\ref{Matter_PS_equation2} to the context of galaxy clustering, we note its inherent generality, which allows us to introduce the galaxy density ($\delta_g$) and velocity divergence ($\Theta_g$) fields as substitutes of the matter fields, and we rewrite the equation as:

\begin{equation}
\label{Galaxy_PS_equation}
\begin{multlined}
P_g^s(k,\mu)=D^{\rm FoG}(k\mu\sigma_{p,g})\{[P_{\delta_g \delta_g}(k)+2\mu^2 P_{\delta_g \Theta_g}(k)+\\
\mu^4P_{\Theta_g \Theta_g}(k)] +W_g(k,\mu)
+A_g(k,\mu)+M_g(k,\mu))\}\,.\end{multlined}
\end{equation}

 The full theoretical development of the Fourier space clustering analysis can be found in \citep{2009JCAP...08..020M,2015MNRAS.451..539G}. For this work, the important result is the individual expressions for the terms in equation~\ref{Galaxy_PS_equation}. We  now introduce the detailed formulation to calculate the density power spectrum of tracers $P_{\delta_g \delta_g}(k)$, which is one of the five terms required in equation~\ref{Galaxy_PS_equation}.

The detailed formulation to calculate the halo density power spectrum is \citep{2021PhRvD.104d3528S,ZhengY2019},

\bea
\label{eq:app_pdd_bias}
P_{\delta_g\delta_g}(k)=\left(b_1P_{\delta\delta}+b_2P_{b2,\delta}+b_{s2}P_{bs2,\delta} +b_{3\rm{nl}}\sigma_3^2P^{\rm{L}}_{\rm m} \right)^2/P_{\delta\delta} \, ,
\eea

where $P_{\delta\delta}$ is the dark matter density power spectrum that computed with our emulators, and $(P_{b2,\delta},P_{bs2,\delta},\sigma_3)$ are the estimated spectra using the linear matter power spectrum, where we assume that the density bias is local in Lagrangian space. $b_1$ and $b_2$ are free parameters of our methodology and the higher order bias parameters  $b_{s2}$ and $b_{3\rm{nl}}$ are estimated using the following consistency relation \citep{2012PhRvD..86h3540B,2014PhRvD..90l3522S},

\bea
b_{s2}=-\frac{4}{7}(b_1-1) \,,
&\quad&
b_{3\rm{nl}}=\frac{32}{315}(b_1-1)\,.\no
\eea

The rest of the power spectra terms in equation~\ref{eq:app_pdd_bias} are calculated as:

\bea
\label{eq:pb2d}
P_{b2,\delta}=\int \frac{d^3q}{(2\pi)^3}\,P_{\rm lin}(q)P_{\rm lin}(|\bfk-\bfq|)F_2(\bfq,\bfk-\bfq) \nonumber 
\eea
\bea
\label{eq:pbs2d}
P_{bs2,\delta}=\int \frac{d^3q}{(2\pi)^3}\,P_{\rm lin}(q)P_{\rm lin}(|\bfk-\bfq|)F_2(\bfq,\bfk-\bfq)S_2(\bfq,\bfk-\bfq) \nonumber 
\eea
\bea
\label{eq:sigma32}
\sigma_3^2=\int \frac{d^3q}{(2\pi)^3}\,P_{\rm lin}(q)\left[\frac{5}{6}+\frac{15}{8}S_2(\bfq,\bfk-\bfq)S_2(-\bfq,\bfk)\right. \nonumber \\ 
\left. -\frac{5}{4}S_2(\bfq,\bfk-\bfq)\right]. \nonumber 
\eea

The expressions above depend on three kernel functions which are defined as:

\bea
F_2(\bfk_1,\bfk_2)&=&\frac{5}{7}+\frac{1}{2}\frac{\bfk_1\cdot\bfk_2}{k_1k_2}\left(\frac{k_1}{k_2}+\frac{k_2}{k_1}\right) 
+\frac{2}{7}\left[\frac{\bfk_1\cdot\bfk_2}{k_1k_2}\right]^2\,, \nonumber  \\
G_2(\bfk_1,\bfk_2)&=&\frac{3}{7}+\frac{1}{2}\frac{\bfk_1\cdot\bfk_2}{k_1k_2}\left(\frac{k_1}{k_2}+\frac{k_2}{k_1}\right)+\frac{4}{7}\left[\frac{\bfk_1\cdot\bfk_2}{k_1k_2}\right]^2\,, \nonumber  \\
S_2(\bfk_1,\bfk_2)&=&\left[\frac{\bfk_1\cdot\bfk_2}{k_1k_2}\right]^2-\frac{1}{3}\,. \nonumber 
\eea

 It is important to highlight that the only term we need to construct with our emulators in order to compute $P_{\delta_g \delta_g}(k)$ using equation \ref{eq:app_pdd_bias} is $P_{\delta\delta}$, while the remaining terms are computed analytically within our methodology.

As mentioned earlier, we assume that the velocity of tracers of matter and the underlying matter structures is identical, and, therefore, $P_{\Theta_g \Theta_g} = P_{\Theta \Theta}$. For simplicity, we won't include the expressions for $P_{\delta_g\Theta_g}$, which can be found in the appendix of \cite{ZhengY2019}. However, similar to $P_{\delta_g \delta_g}(k)$, the only term that we need to compute with our emulators is $P_{\delta\theta}$, and the rest of the terms can be calculated analytically.

In summary, the emulators we have developed for predicting the values of $P_{\delta \delta}$, $P_{\delta\Theta}$, and $P_{\Theta\Theta}$ are the only inputs required to compute our predictions for the $P_{\delta_g \delta_g}$, $P_{\delta_g \Theta_g}$,  $P_{\Theta_g \Theta_g}$ and $W_g(k,\mu)$ terms in equation~\ref{Galaxy_PS_equation}.

The expressions necessary to compute the higher-order terms are discussed in Appendix~\ref{appendix:higher_order}. As concluded therein, the terms $A_g$ and $M_g$ can be expressed as combinations of six terms: $A_{b^2}$, $A_b$, $A_L$, $M_{b^2}$, $M_b$, and $M_L$, as shown in equations~\ref{A_g_app}. Within our methodology, we construct emulators for each of these six terms.

By plugging equation~\ref{A_g_app} into equation~\ref{Galaxy_PS_equation} we get to the following expression:

\bea
\label{Galaxy_PS_equation_final}
P_g^s(k,\mu)=D^{\rm FoG}(k\mu\sigma_{p,g})\{[P_{\delta_g \delta_g}(k)+\no\\2\mu^2 P_{\delta_g \Theta_g}(k)+\mu^4P_{\Theta_g \Theta_g}(k)]+ W_g(k,\mu)+
[b_1^2 A_{b^2}+bA_{b}+A_L]\no\\+[b_1^2 M_{b^2}+bM_{b}+M_L]\}\,.
\eea

In order to model the redshift space power spectra of galaxies, we compute the following terms: $P_{\delta \delta}$, $P_{\delta \Theta}$, $P_{\Theta \Theta}$, $A_{b^2}$, $A_{b}$, $A_{L}$, $M_{b^2}$, $M_{b}$, and $M_{L}$, where we recall that $W_g(k,\mu)$ can be computed as a function of the rest of the terms. These terms can then be inserted into equation~\ref{Galaxy_PS_equation_final}. Each of these terms can be measured from the matter distributions of an N-body simulation. In this work, we have developed emulators trained on a set of simulations that learn how to accurately and efficiently compute each of these terms without the need to run a new simulation.

\subsection{Precision test of our model}
\label{Example_of_model}

In this brief section, we provide an illustration of our methodology's capability to replicate the power spectra of tracers of the underlying matter distribution, such as galaxies or halos. This demonstration shows the kind of power spectrum templates we would develop when analyzing data from actual LSS surveys, like DESI.

We first generate a mock catalogue of halos to represent the observations from a real LSS survey. To accomplish this, we populate our fiducial N-body simulation (first line of Table~\ref{tab:data}) with halos using the  {\tt ROCKSTAR} group finder \cite{2013ApJ...762..109B}. We exclude all of the subhalos identified by the algorithm and arbitrarily select to retain only those halos with a mass ranging from $10^{13}$ to $10^{13.5}$ solar masses. This selection emulates a mass cut that might be applicable in a hypothetical LSS survey, where only galaxies with a certain luminosity range are considered.

\begin{figure}
\includegraphics[width=80mm,height=80mm]{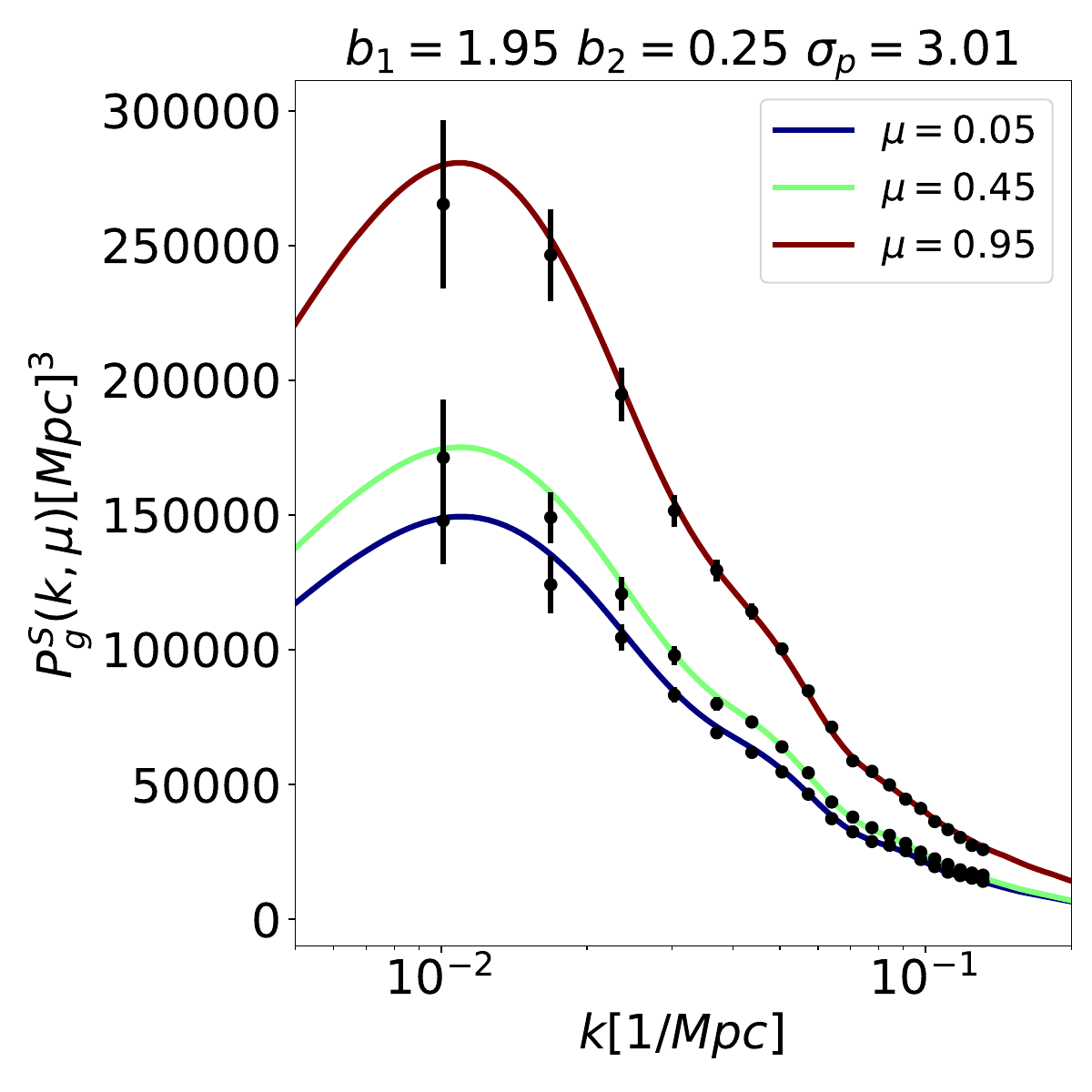}
\caption[ ]{The power spectra of the halos in our mock catalogue (black dots) are shown for various observation angles $\mu$ when compared to a theoretical (colored lines) template constructed using our models computed with equation~\ref{Galaxy_PS_equation_final}. The values of the nuisance parameters employed to derive these specific models are presented in the title of the plot.}
\label{pkmu_halo}
\end{figure}

The measured redshift space power spectrum for our selected halos is shown as the black dots of Fig.~\ref{pkmu_halo}, the errors presented there are computed using equation~\ref{error_equation}. The colored lines show the theoretical predictions made with our emulator methodology. 

Our estimated halo power spectra are depicted as solid color curves and are computed using equation~\ref{Galaxy_PS_equation_final}. It's important to note that we don't have precise values for our nuisance parameters. For this simple exercise, we fitted the equation to determine the best-fit values of our three nuisance parameters: $b_1$, $b_2$, and $\sigma_p$.

The best-fit parameters we found are as follows: $b_1=1.95$, $b_2=0.25$, and $\sigma_p=3.01$. These parameters result in a reduced $\chi^2$ less than 1, indicating a good fit to the data. It's worth mentioning that our models agree with all data points within their respective error bars.

 \section{Conclusions}
 \label{sec:Conclucion}

The expected precision of forthcoming LSS surveys is set to yield measurements of the redshift space power spectrum of tracers with unprecedented accuracy. To fully harness these measurements, we need theoretical templates capable of matching this high level of precision. This  becomes especially challenging on semi-nonlinear scales, where the application of models extending beyond second-order perturbation theory may be necessary.

We have previously introduced our hybrid model emulators, capable of accurately modeling the effects of {\it scale dependent} parameters such as $A_s$, $w_0$, and $w_a$ with sub-per~cent accuracy at semi-nonlinear scales. This is done by precisely measuring the power spectrum in a specified fiducial cosmology using an N-body simulation. Subsequently, we predict beyond second-order corrections through a set of scaling relations. This approach enables efficient exploration of variations in {\it scale dependent} parameters with just one N-body simulation built on a fiducial cosmology.

In this study, we introduce an extension to our methodology that enables us to model the impact of \textit{scale-dependent} parameters in the power spectrum. For this, we employ Gaussian process emulators to construct surrogate models of the necessary N-body simulations statistics. The crucial statistics for calculating the redshift space matter power spectrum in our model are $P_{\delta \delta}(k)$, $P_{\delta \Theta}(k)$, and $P_{\Theta \Theta}(k)$, along with three higher-order correction terms: $W(k,\mu)$, $A(k,\mu)$, and $M(k,\mu)$. Notably, among these higher-order corrections, $A(k,\mu)$ and $M(k,\mu)$ are independent and require specific emulators for their computation, whereas $W(k,\mu)$ does not require dedicated emulators, as it depends on $P_{\delta \delta}(k)$, $P_{\delta \Theta}(k)$, and $P_{\Theta \Theta}(k)$.

Throughout our work, we develop surrogate models for all these terms. Our models systematically explore the parameter space of the three \textit{scale dependent} parameters $w_{cdm}$, $w_b$, and $n_s$, within a 5 $\sigma$ range of their Planck 2018 predictions.

Gaussian Processes are machine learning algorithms that act as local interpolators. Therefore the precision of the model at a specific point relies on the proximity to the nearest point in the training set. Hence, it is crucial to thoroughly explore the parameter space. However, our N-body simulations are expensive to run and therefore we are interested in running as little of them as possible. To achieve this we use an Annealing algorithm designed to optimize the sparsity of symmetric Latin-hypercube grids of points in the parameter space.

We employ CAMB non-linear models to generate templates of the matter power spectrum, which are then used to assess the number of training points necessary for building accurate emulators. We find that approximately 11-13 points are adequate for achieving predictions with sub-per~cent accuracy in $P_{\delta\delta}$ and in all scales of interest. Based on this, we build a 13-point simulation design. Subsequently, N-body simulations are conducted on this design, which are used to train our N-body emulators for all the statistics presented above. Additionally, two extra points are chosen, distinct from those used for building emulators. Instead, these two points are used as test points, where we compare the predictions of our emulated statistics with the actual values obtained from the simulations. These two points are selected around the  1$\sigma$ and 5$\sigma$ regions of Planck 2018 measurements and we refer to them as $\theta_{1}$ and $\theta_{5}$ respectively. This selection is done to assess the accuracy of the methodology in both high-likelihood regions and at the very edge of our parameter space.

Once our N-body simulations are built we can check if in fact 13 simulations are needed to build accurate emulators. We note that by using only the most sparsely separated nine points our emulators we can achieve per~cent accuracy when modeling $P_{\delta\delta}$ in the regions of interest. However the predictions for $\theta_{5}$ are barley below 1~per~cent accuracy.  With 11 points, the accuracy significantly improves, dropping to around 0.01~per~cent. Nevertheless, we observe some minor systematic shifts at high and low values of $k$. Finally, when using all 13 models, our emulators exhibit very small errors, with no discernible deviations at the 0.01~per~cent accuracy level for $\theta_{5}$. As with our tests with CAMB, we note that 11 to 13 points seem to be enough to predict accurate emulators.

Throughout this work, we do not directly emulate the statistics that we require. Instead, we utilise principal value decomposition to represent them as a linear combination of principal components and their corresponding weights. Since the most significant principal components contain the majority of the power spectra information, we can select a few and employ our GP  to predict their associated weights. The remaining principal components are then discarded, simplifying the prediction process for our emulators. We have shown that by considering at least the first three principal components, we can accurately predict the power spectra of our test N-body simulation with sub-per~cent precision. In this study, all our statistics are computed including the first five principal components, as the computational cost of incorporating the final two is negligible, while providing a wider margin of error. 

We show exceptional accuracy in predicting the first order terms $P_{\delta\delta}$, $P_{\delta\Theta}$ and $P_{\Theta\Theta}$. At all scales of interest and for all three redshifts, our predictions maintain an accuracy level below $0.1\%$ for $\theta_1$, surpassing our target accuracy of by an order of magnitude, and staying below $1\%$ for $\theta_5$. We have also test the accuracy of our emulators on reproducing the higher order correction terms $A(k,\mu)$ and $M(k,\mu)$, we have noted that the accuracy of our emulators is such that the errors are negligible in the final measurement of the power spectrum at $k > 0.1$\,$h$\,Mpc$^{-1}$.

The methodology discussed so far allows us to explore the effect of \textit{scale dependent} variables on the redshift space matter power spectrum. In order to consider the effect of \textit{scale independent} parameters, the hybrid model introduces a set of scaling relations that predict the beyond second order corrections for our statistics of interest in a new cosmology as a function of their value in a given fixed cosmology and of the growth functions of the density and velocity fields $G_\delta$ and $G_\Theta$, which are treated as two free parameters of our methodology.

For $P_{\delta \delta}(k)$, $P_{\delta \Theta}(k)$, and $P_{\Theta \Theta}(k)$, the beyond-second-order corrections are combined with second-order PT templates to predict their values at the new target cosmology. Although the templates are relatively fast to run, taking around 15 seconds to build all three of them, the exploration of the parameter space may require millions of evaluations, therefore there is an incentive  to accelerate the evaluation time. To address this, we construct emulators for each of the necessary second-order terms. This is done using a grid of 50 training points and 100 test points, we note that these PT templates are much faster to run than our N-body simulations and therefore we can afford to use larger training and test sets.

Our PT emulators show high accuracy in predicting the required PT statistics of our test set, which are well below the per~cent threshold of accuracy. In fact, for most scales of interest, the accuracy consistently hovers around 0.1~per~cent and 0.01~per~cent for all redshifts. 

We observe that conducting 13 N-body simulations is a reasonably manageable task, and we can complete these simulations within a couple of weeks using our computer cluster. Therefore, our current methodology allows us to efficiently construct precise and efficient models of the galaxy power spectra in a manageable time frame. Once our emulators are trained, we can predict all of the inputs of our hybrid model in approximately 0.2\,seconds for a given redshift.

Comparatively, when testing the model using non-linear CAMB templates using a five parameter space, we find that achieving the same level of accuracy would require tripling the number of training points, which would require us to build three times more N-body simulations. We also note that the tight constraints placed by Planck on our three parameters make doubling the size of our priors of little effect to the accuracy of our emulators. 

To conclude our work, we introduce a mapping formulation connecting the power spectra of matter and galaxies. This mapping is a function of the bias parameters $b_{1}$ and $b_{2}$, serving as new independent parameters within our methodology. It is important to note that the emulator models presented in this work remain independent of this mapping, and in principle, our emulators could be integrated into a different bias formulation.

To assess our ability to replicate the power spectrum of a given galaxy survey with our mapping, we populate one of the N-body simulations with halos. Subsequently, we select halos within a narrow range of halo-masses to simulate the selection criteria of a specific tracer in a galaxy survey. We then optimize our bias parameters to ensure that our theoretical galaxy redshift space power spectrum accurately reproduces the power spectra of the selected halos. The models accurately replicate the simulated data, with the best-fit galaxy power spectra agreeing within errors with all the data points.

\section*{Acknowledgements}
This work was supported by the National Research Foundation of Korea (NRF) grant funded by the Korea government (MIST), Grant No. 2021R1A2C1013024.YZ acknowledges the supports from the National Natural Science Foundation of China (NFSC) through grant 12203107, the Guangdong Basic and Applied Basic Research Foundation with No.2019A1515111098, and the science research grants from the China Manned Space Project with NO.CMS-CSST-2021-A02. This work was supported by the high performance computing cluster Seondeok at the Korea Astronomy and Space Science Institute.

\section{Data Availability}

The data used in this work can be shared if requested from the authors.

\bibliographystyle{mnras}
\bibliography{Emulator}



\appendix

\section{Building a symmetric Latin hypercube}
\label{apendix:LatinHypercube}

We mentioned in Section~\ref{subsec:grid-selection}, that we are following the methodology of \cite{2009ApJ...705..156H} to build and optimize our Latin Hypercube (LH) designs. In this section we include a brief summary of this methodology for completeness, but refer to the original work for a more thorough discussion. 

We use a Latin Hypercube designs to generate a grid that attempts to explore all regions of the parameter space with a  similar density. This approach minimizes the likelihood of overlooking features or disproportionately emphasizing a specific subset of the space. We try to ensure that each point sampled provides unique information to the emulator, and therefore reduces the number of points required to build an emulator of sub-per~cent precision. 

\begin{figure*}
\includegraphics[width=55mm,height=55mm]{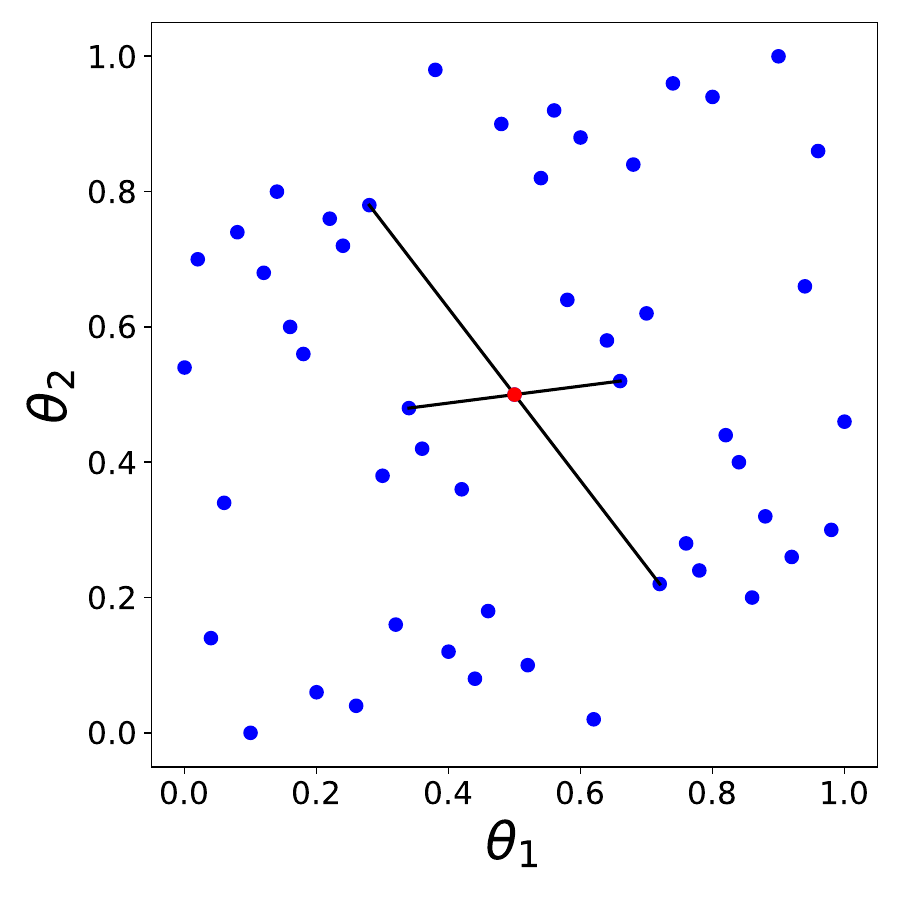}
\includegraphics[width=55mm,height=55mm]{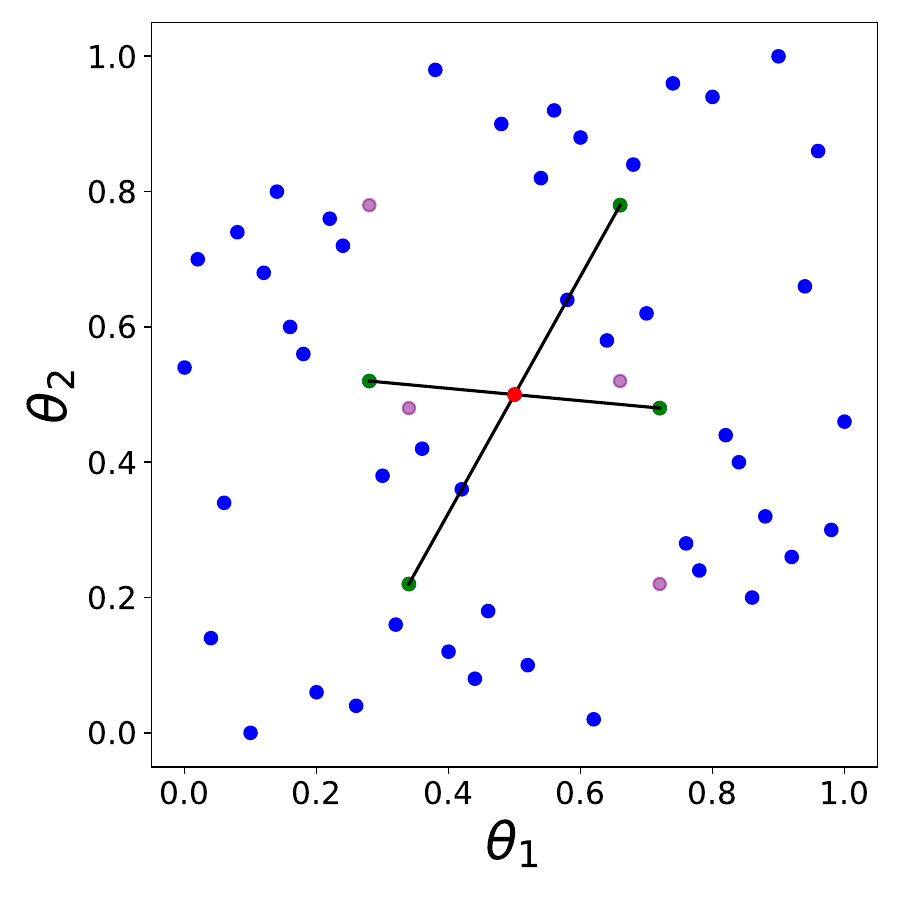}
\includegraphics[width=55mm,height=55mm]{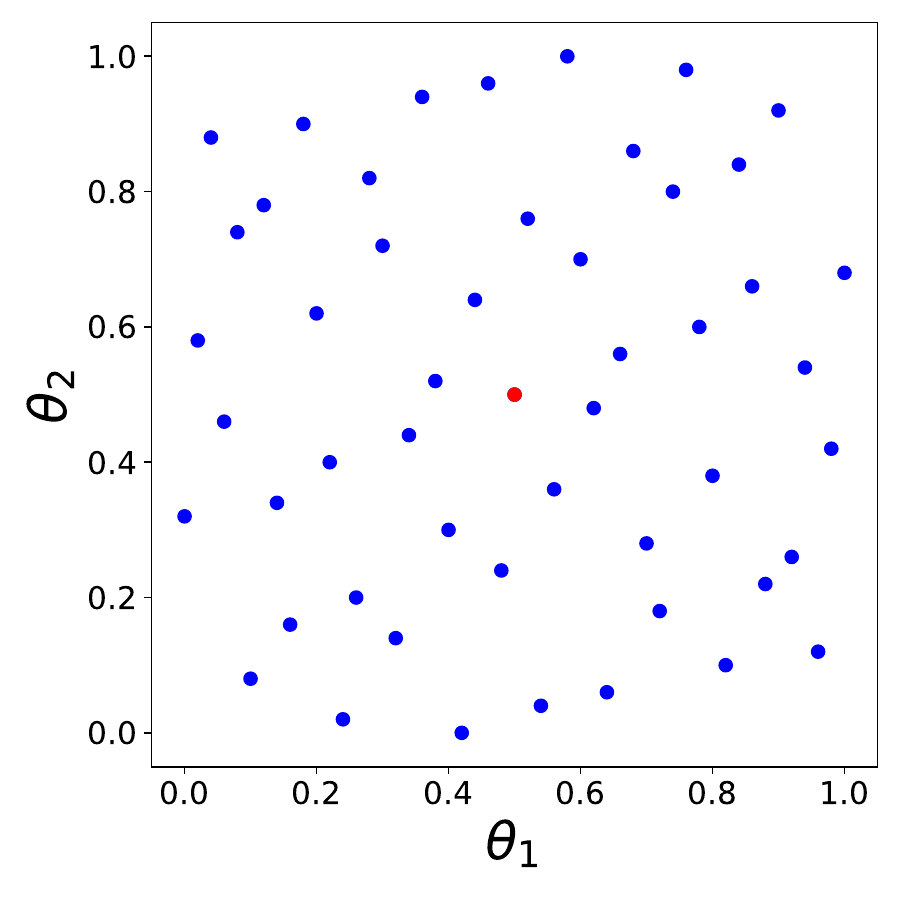}
\caption[ ]{{\bf Left}: Example of an SLH with 37 points in a two-dimensional normalized space with parameters $\theta_1$ and $\theta_2$. The red point represents the center of the grid and is one of the elements of the design. The black lines indicate that when two points in our design are reflected through the center, the reflections are also part of the design. \textbf{Center:} The plot shows the resulting SLH obtained by constructing a new design using the method explained in the main text. The purple lines represent the points that were removed from the design, while the green points indicate their replacements. \textbf{Right:} The optimized SLH design achieved by applying the Annealing methodology to the SLH shown in the left panel. The figure demonstrates that the design is significantly more sparse.}
\label{slh}
\end{figure*}

Consider a grid of points inside a parameter space priors. Such a grid is a Latin hypercube (LH) if each column and row of the grid has only one value. A LH is called symmetric if for any point in the grid the point that its reflection trough the center of the volume is also a part of the grid. The left panel of Fig.~\ref{slh} shows an example of a SLH of 37 points in 2 dimensions. The SLH constitutes the collection of all points (both red and blue). The center of the grid is represented in red, and is also a point of our design, the black lines exemplify  points in our SLH and their reflection trough the center, that are also part of the design. 

The left panel of the figure also shows that while the design has sampled different regions of the space the density is not very uniform and some regions of space are undersampled. In order to optimize our design we use an Annealing algorithm \cite{MORRIS1995381}, that explores the parameter space of possible grids that can be build by moving points in our original SLH design. 

To build a new SLH, we perform the following steps: first, we randomly select two points within a dimension of the design. Then we include the points that result out of switching the position of the points within that column. Next, we remove these points and their reflections. Then, we include the reflections of the newly selected points. An illustration of this process is presented in the middle panel of Fig.~\ref{slh}, where the purple dots represent the switched points and their reflections, and the green dots indicate their new positions. The black lines emphasize that the reflections of the new points, passing through the center, are included, while their previous reflections are discarded.

Once we have constructed a new SLH the algorithm tests if the point density is more evenly distributed for the new design, if so we move our algorithms walker into it. Otherwise, we make a decision whether to move or not based on a probability $P = e^{[D(\theta_i) - D(\theta_{i+1})]/t}$. Here, $D(\theta_i)$ represents a criterion used to assess the point density distribution at the $i^{th}$ step of our annealing algorithm, and $t$ is a hyperparameter referred to as the 'temperature' in the algorithm. Our criterion $D(\theta)$ is computed as the average inverse distance between all points.

After several steps the algorithm explores the parameter space of possible SLH design that we can construct. We then select the explored design with the smallest value of our criterion $D(\theta)$ as a candidate for a simulation design. The right panel of Fig.~\ref{slh}, shows the design selected by our Anneling algorithm, the figure shows that this design is much more evenly distributed that the original SLH in the left panel.

To create a new design, we begin by generating a random SLH. Subsequently, our algorithm is initiated with a value of $t=1000$, at each step, we keep track the best configuration found so far denoted as $\theta_{BF}$, which corresponds to the smallest value of $D(\theta_{BF})$. If no improvement has been observed in the last 10,000 steps, we decrease $t$ by one order of magnitude. We continue this process until no improvement has been detected in the last 100,000 total steps, regardless of the value of $t$. 

The final design obtained exhibits a slight dependence on the initial SLH. Keeping this in mind, we construct and optimize 40 individual random SLHs. Subsequently, we select the design with the smallest overall value of $D(\theta_{BF})$ as our final simulation design.

\section{High precision modeling of the density and velocity matter power spectrum}
\label{appendix:hybrid}

In this section, we outline the methodology we follow to compute the second-order theoretical templates of $\bar{P}^{PT}_{XY}(k, z)$ as defined in equation~\ref{N_body-regpt}. We will start by introducing our approach for calculating these templates within a fiducial cosmology. Additionally, we will discuss the scaling relations we use to adapt these calculations to a new cosmology. We maintain the same notation as used previously, where barred quantities correspond to the fiducial cosmology.

As previously mentioned, our theoretical templates are calculated using the RegPT paradigm of \cite{RegPT}. In this approach, the power spectrum is expanded in terms of multi-point propagators. At the two-loop order, the expression is as follows, 
\ba
\label{therietical_PXY}
&&\bar P^{PT}_{XY}(k,z) = \bar \Gamma_X^{(1)}(k,z)\bar \Gamma_Y^{(1)}(k,z)\bar P_{\rm lin}(k)
\nn
\\
&&\quad
+2\int\frac{d^3\vec q}{(2\pi)^3}\bar \Gamma_X^{(2)}(\vec q, \vec k-\vec q,z) \bar \Gamma_Y^{(2)}(\vec q, \vec k-\vec q,z)
\nn
\\
&&\quad\quad\quad
\times \bar  P_{\rm lin}(q) \bar  P_{\rm lin}(|\vec k-\vec q|) 
\nn
\\
&&\quad
+6\int\frac{d^3\vec pd^3\vec q}{(2\pi)^6}
\nn
\\
&&\quad\quad\quad
\times \bar \Gamma_X^{(3)}(\vec p,\vec q, \vec k-\vec p-\vec q,z)\bar \Gamma_Y^{(3)}(\vec p,\vec q, \vec k-\vec p-\vec q,z) 
\nn
\\
&&\quad\quad\quad
\times\bar  P_{\rm lin}(p) \bar  P_{\rm lin}(q) \bar  P_{\rm lin}(|\vec k-\vec p-\vec q|).
\label{eq:pk_RegPT}
\ea

Here  $\bar \Gamma_X^{(p)}$ is the regularised $(p+1)$-point propagator. Therefore in order to compute  we first need to estimate $\bar \Gamma_X^{(p)}$ for $p=1,2,3$ point propagators. The expression relevant at the two-loop order for the two point propagator $p=1$ is written as

\ba\label{eq:gammax1}
\bar \Gamma_X^{(1)}(k,z) = {\rm exp}\left(-\bar G_\delta^2\,\bar\gamma \right)\sum_n \bar G_X \bar G_\delta^{n-1} \bar{\cal C}^{(1)}_n(\bar \gamma).
\ea

Where  $\bar\gamma=k^2\bar \sigma^2_{\rm d}/2$ and $\bar \sigma_{\rm d}^2$ is the dispersion of displacement field,that we compute from the initial power spectrum as $\bar \sigma_{\rm d}^2=\int_0^{k/2} (dq/6\pi^2)\bar P_{\rm lin}(q)$.

The coefficients $\bar{\cal C}^{(1)}_n$ in equation~\ref{eq:gammax1} are expressed in terms of the standard perturbation theory results. These terms are such that $\bar{\cal C}^{(1)}_n=0$ when $n$ is an even number. For odd values of $n$ and when considering theoretical uncertainties, the remaining terms can be expressed as follows:

\ba
\bar{\cal C}^{(1)}_1(\bar \gamma) &=&1, \nn\\
\bar{\cal C}^{(1)}_3(\bar \gamma) &=& \bar \gamma + \bar \Gamma_{X,{\rm 1-loop}}^{(1)}(k),\nn\\
\bar{\cal C}^{(1)}_5(\bar \gamma) &=& \bar \gamma^2/2 + \bar \gamma\bar \Gamma_{X,{\rm 1-loop}}^{(1)}(k) + \bar \Gamma_{X,{\rm 2-loop}}^{(1)}(k)+\bar{\cal O}^{(1)}_{X,5},\nn\\
\bar{\cal C}^{(1)}_n(\bar \gamma) &=& \bar{\cal O}^{(1)}_{X,n}\,.
\ea

The function $\bar \Gamma_{X,{\rm n-loop}}^p$ represents the standard PT $(p+1)$-point propagator at the $n$-loop order, which is different from the regularised propagator, $\bar \Gamma_{X}^p$. Its explicit expression is given in \cite{RegPT,PhysRevD.89.023502}. Here, $\bar{\cal O}^{(1)}_{X,n}$ corresponds to the uncertainties or systematics in PT. In our methodology, these uncertainties are calibrated with N-body simulations. We assume that the uncertainties arise not only from the three-loop order but also from the two-loop order. This can be  partly attributed to the UV-sensitive behavior of the single-stream PT calculation.

The expressions for the three-point propagator are,

\ba
\bar \Gamma_X^{(2)}(k,z) &=& {\rm exp}\left(-\bar G_\delta^2\,\bar\gamma \right)\sum_n \bar G_X\,\bar G_\delta^{n-1} \bar{\cal C}^{(2)}_n\,\,;\nn \\
\bar{\cal C}^{(2)}_2(\bar \gamma) &=& \bar F_X^{(2)}(\vec q, \vec k-\vec q),\nn\\
\bar{\cal C}^{(2)}_4(\bar \gamma) &=& \frac{\bar\gamma}{2} \bar F_X^{(2)}(\vec q, \vec k-\vec q)+\bar\Gamma_{X,{\rm 1-loop}}^{(2)}(\vec q, \vec k-\vec q)+\bar{\cal O}^{(2)}_{X,4}\,, \nn\\ 
\bar{\cal C}^{(2)}_n(\bar \gamma) &=& \bar{\cal O}^{(2)}_{X,n}\,,
\ea

and $\bar{\cal C}^{(2)}_n=0$ for odd number of $n$. 

Finally, the expression of the four-point propagator, $\bar \Gamma_X^{(3)}(k,z)$, relevant to the two-loop order, are

\ba
\bar \Gamma_X^{(3)}(k,z) &=& {\rm exp}\left(-\bar G_\delta^2\,\gamma \right)\sum_n \bar G_X\,\bar G_\delta^{n-1} \bar{\cal C}^{(3)}_n\,\,; \\
\bar{\cal C}^{(3)}_3(\bar \gamma)&=&\bar F_X^{(3)}(\vec p,\vec q, \vec k-\vec p-\vec q) + \bar{\cal O}^{(3)}_{X,3}, \\
\bar{\cal C}^{(3)}_n(\bar \gamma) &=& \bar{\cal O}^{(3)}_{X,n}.
\ea

From the previous expressions, we notice that for a given fiducial cosmology, we need to compute $\bar{\cal C}^{(1)}_3(\bar \gamma)$, $\bar{\cal C}^{(1)}_5(\bar \gamma)$, $\bar{\cal C}^{(2)}_2(\bar \gamma)$, $\bar{\cal C}^{(2)}_4(\bar \gamma)$, and $\bar{\cal C}^{(3)}_3(\bar \gamma)$. These estimates can then be used to compute estimates of the $\bar \Gamma_X^{(p)}$ point propagators required in equation~\ref{eq:pk_RegPT}. In our methodology, these statistics are closely related to the RegPT terms that we emulate introduced in equation~\ref{RegPT_2loop}, as shown in \cite{RegPT}. Therefore, our RegPT emulator predictions can be transformed into predictions of the $\bar{\cal C}_n$ coefficients.

In Section~\ref{subsec:Model_description}, as we state that we  transform the multi-point propagators from our fiducial cosmology into a cosmology with different values of the \textit{scale independent} parameters, this is done with the following expression:
\begin{equation}
\label{point_prop_new_cosmo_1}
\Gamma_X^{(m)} = {\rm exp}\left(-G_X^2\,\bar \gamma \right)\sum_n G_X\, G_\delta^{n-1} \bar{\cal C}^{(m)}_n(\gamma)
\end{equation}

While $\gamma$ is computed trough the following scaling relation.
\begin{equation}
\label{point_prop_new_cosmo_2}
    \gamma=\left(\frac{G_\delta}{\bar G_\delta G_X} \right)^2 \bar \gamma
\end{equation}

Therefore, our estimates of the $\bar{\cal C}_n$ coefficients done with our RegPT emulators in a given fiducial cosmology can be computes into estimates of $P^{PT}_{XY}$ in a new cosmology.

\section{High-precision modeling of higher-order corrections}
\label{appendix:higher_order}

In this section we introduce the equations that are required to compute the higher order $A_g$ and $M_g$ terms from equation~\ref{Galaxy_PS_equation}  in our methodology. As we emphasize latter in this section the expressions computed here are also utilised to compute the power spectrum of the underlying matter distribution by setting the linear bias between tracers and matter to one $b_{1}=1$.

We begin by writing equation~\ref{terms} as a function of the density and velocity of tracers:
\bea
\label{eq:A_h}
A_g(k,\mu)&=& j_1\,\int d^3\bfx \,\,e^{i\bfk\cdot\bfx}\,\,\langle A_1A_2A_3\rangle_c\nonumber\\
&=&j_1\,\int d^3\bfx \,\,e^{i\bfk\cdot\bfx}\,\,\langle (u_{z,g}-u'_{z,g})\nonumber\\
&&(\delta_g+\nabla_zu_{z,g})(\delta'_g+\nabla_zu'_{z,g})\rangle_c\, 
\eea

And our $M$ term is written as
\bea
\label{eq:T_h}
M_g(k,\mu)&=& \frac{1}{2} j_1^2\,\int d^3\bfx \,\,e^{i\bfk\cdot\bfx}\,\,\langle A_1^2A_2A_3\rangle \nonumber \\
&=&\frac{1}{2} j_1^2\,\int d^3\bfx \,\,e^{i\bfk\cdot\bfx}\,\,\left\{\langle (u_{z,h}-u_{z,h}')^2  \right. \no \\ 
&& \left. (\delta_h+\nabla_zu_{z,h})(\delta_h'+\nabla_zu_{z,h}')\rangle\right\}. 
\eea

In the following step, we replace the  density of tracers of matter with the density of the underlying matter distribution using the linear expression $\delta_g(\vec r) = b_{1}\delta(\vec r)$. We note that we do not require a more complicated non-linear mapping (like equation~\ref{eq:bias_x}) as  $A$ and $M$ are already higher order corrections, and including extra terms would only complicate the equations without providing significant additional information. Additionally, we assume that $b_{v}=1$ thereby making $u_{z} = u_{z,g}$. Then equation~\ref{eq:A_h} for $A_{g}$ becomes:

\bea
\label{eq:A_h_terms}
&&A_g(k,\mu) \simeq\nonumber\\
&& j_1\,\left[b_1^2\int d^3\bfx \,\,e^{i\bfk\cdot\bfx}\,\,\langle u_z\delta\delta'\rangle_c+b_1\int d^3\bfx \,\,e^{i\bfk\cdot\bfx}\,\,\langle u_z\delta\nabla_zu'_z\rangle_c \right.\nonumber\\
&&+b_1\int d^3\bfx \,\,e^{i\bfk\cdot\bfx}\,\,\langle u_z\nabla_z u_z\delta'\rangle_c+\int d^3\bfx \,\,e^{i\bfk\cdot\bfx}\,\,\langle u_z\nabla_z u_z\nabla_z u'_z\rangle_c \nonumber\\
&&-b_1^2\int d^3\bfx \,\,e^{i\bfk\cdot\bfx}\,\,\langle \delta u_z'\delta'\rangle_c- b_1\int d^3\bfx \,\,e^{i\bfk\cdot\bfx}\,\,\langle \delta u_z'\nabla_zu_z'\rangle_c \nonumber\\
&&-b_1\int d^3\bfx \,\,e^{i\bfk\cdot\bfx}\,\,\langle \nabla_zu_zu_z'\delta'\rangle_c \nonumber\\
&&\left.-\int d^3\bfx \,\,e^{i\bfk\cdot\bfx}\,\,\langle \nabla_zu_z u'_z\nabla_z u'_z\rangle_c \right] \,.
\eea

And equation~\ref{eq:T_h} transforms into:

\bea
\label{eq:T_h_terms}
&&M_g(k,\mu) \simeq\nonumber\\
&&\frac{1}{2}j_1^2\left[b_1^2\int d^3\bfx \,\,e^{i\bfk\cdot\bfx}\langle u_z^2\delta\delta'\rangle + b_1 \int d^3\bfx \,\,e^{i\bfk\cdot\bfx}\langle u_z^2\delta\nabla_zu_z'\rangle\right. \nonumber\\
&&+b_1\int d^3\bfx \,\,e^{i\bfk\cdot\bfx}\langle u_z^2\nabla_zu_z\delta'\rangle + \int d^3\bfx \,\,e^{i\bfk\cdot\bfx}\langle u_z^2\nabla_zu_z\nabla_zu_z'\rangle \nonumber \\
&&-2b_1^2\int d^3\bfx \,\,e^{i\bfk\cdot\bfx}\langle u_z\delta u_z'\delta'\rangle - 2b_1 \int d^3\bfx \,\,e^{i\bfk\cdot\bfx}\langle u_z\delta u_z'\nabla_zu_z'\rangle\nonumber\\
&&-2b_1\int d^3\bfx \,\,e^{i\bfk\cdot\bfx}\langle u_z\nabla_zu_z u_z'\delta'\rangle \nonumber\\ 
&&- 2\int d^3\bfx \,\,e^{i\bfk\cdot\bfx}\langle u_z\nabla_zu_zu_z'\nabla_zu_z'\rangle 
+b_1^2\int d^3\bfx \,\,e^{i\bfk\cdot\bfx}\langle \delta u_z'^2\delta'\rangle\nonumber \\
&& + b_1 \int d^3\bfx \,\,e^{i\bfk\cdot\bfx}\langle \delta u_z'^2\nabla_zu_z'\rangle +b_1\int d^3\bfx \,\,e^{i\bfk\cdot\bfx}\langle \nabla_zu_zu_z'^2\delta'\rangle \nonumber\\ 
&& \left.+ \int d^3\bfx \,\,e^{i\bfk\cdot\bfx}\langle \nabla_zu_zu_z'^2\nabla_zu_z'\rangle \right].
\eea

By rewriting the expression in equation~\ref{eq:A_h_terms} and grouping together all terms related to the different powers of $b_{1}$ we can rewrite $A_{g}$ and $M_{g}$ as:

\bea
\label{A_g_app}
A_g(k,\mu)=b_1^2 A_{b^2}+b_1A_{b}+A_L. \nonumber\\
M_g(k,\mu)=b_1^2 M_{b^2}+b_1M_{b}+M_L.
\eea
Where:
\bea
&&A_b^2=j_1\left[\int d^3\bfx \,\,e^{i\bfk\cdot\bfx}\,\,\langle u_z\delta\delta'\rangle_c -\int d^3\bfx \,\,e^{i\bfk\cdot\bfx}\,\,\langle \delta u_z'\delta'\rangle_c \right]\nonumber\\
&&\nonumber\\ 
&&A_b=j_1\left[\int d^3\bfx \,\,e^{i\bfk\cdot\bfx}\,\,\langle u_z\delta\nabla_zu'_z\rangle_c +\int d^3\bfx \,\,e^{i\bfk\cdot\bfx}\,\,\langle u_z\nabla_z u_z\delta'\rangle_c\right. \nonumber\\
&&\left. -\int d^3\bfx \,\,e^{i\bfk\cdot\bfx}\,\,\langle \nabla_zu_zu_z'\delta'\rangle_c- \int d^3\bfx \,\,e^{i\bfk\cdot\bfx}\,\,\langle \delta u_z'\nabla_zu_z'\rangle_c\right] \nonumber\\
&&\nonumber\\ 
&&A_L=j_1\left[\int d^3\bfx \,\,e^{i\bfk\cdot\bfx}\,\,\langle u_z\nabla_z u_z\nabla_z u'_z\rangle_c \right.\nonumber\\ 
&&\left.-\int d^3\bfx \,\,e^{i\bfk\cdot\bfx}\,\,\langle \nabla_zu_z u'_z\nabla_z u'_z\rangle_c\right].\nonumber\\
\eea

Similarly we use equation~\ref{eq:T_h_terms} to define:

\bea
&&M_b^2=\frac{1}{2}j^{2}_1 \left[\int d^3\bfx \,\,e^{i\bfk\cdot\bfx}\langle u_z^2\delta\delta'\rangle-2\int d^3\bfx \,\,e^{i\bfk\cdot\bfx}\langle u_z\delta u_z'\delta'\rangle \right.\nonumber\\ 
&&\left. +\int d^3\bfx \,\,e^{i\bfk\cdot\bfx}\langle \delta u_z'^2\delta'\rangle\right]\nonumber\\
&&\nonumber\\ 
&&M_b=\frac{1}{2}j^{2}_1\left[\int d^3\bfx \,\,e^{i\bfk\cdot\bfx}\langle u_z^2\delta\nabla_zu_z'\rangle+\int d^3\bfx \,\,e^{i\bfk\cdot\bfx}\langle u_z^2\nabla_zu_z\delta'\rangle \right. \nonumber\\
&&- 2 \int d^3\bfx \,\,e^{i\bfk\cdot\bfx}\langle u_z\delta u_z'\nabla_zu_z'\rangle-2\int d^3\bfx \,\,e^{i\bfk\cdot\bfx}\langle u_z\nabla_zu_z u_z'\delta'\rangle\nonumber\\
 && \left. +\int d^3\bfx \,\,e^{i\bfk\cdot\bfx}\langle \delta u_z'^2\nabla_zu_z'\rangle +\int d^3\bfx \,\,e^{i\bfk\cdot\bfx}\langle \nabla_zu_zu_z'^2\delta'\rangle \right] \nonumber\\
&&\nonumber\\ 
&&M_L=\frac{1}{2}j^{2}_1\left[\int d^3\bfx \,\,e^{i\bfk\cdot\bfx}\langle u_z^2\nabla_zu_z\nabla_zu_z'\rangle \right. \nonumber \\ 
&&\left. -2\int d^3\bfx \,\,e^{i\bfk\cdot\bfx}\langle u_z\nabla_zu_zu_z'\nabla_zu_z'\rangle +\int d^3\bfx \,\,e^{i\bfk\cdot\bfx}\langle \nabla_zu_zu_z'^2\nabla_zu_z'\rangle \right].\nonumber\\
\eea

In this work, we measure these terms from individual N-body simulations. The measured statistics are then utilised to train individual emulators for $A_{b^2}$, $A_{b}$, $A_L$, $M_{b^2}$, $M_{b}$, and $M_L$ respectively. For each of these six terms, we build one emulator at all five of our redshifts of interest. It is important to highlight that if we set the linear bias $b_1$ to one, then $A=A_g$ and $M=M_g$.  Therefore, our emulators can be used to compute the higher-order corrections for the underlying matter distribution: $A$ and $M$. This methodology was followed to generate the results analysed in Section~\ref{subsec:highorder_performance}.

We now introduce the scaling relations we utilise to predict the values of higher-order terms in an alternative cosmological models with different values of the {\it scale independent} parameters. As have stated before, these relationships allow us to calculate the higher-order corrections in a new cosmology based on their values in a reference cosmology, as well as their dependence on the variables $G_\delta$ and $G_\Theta$. The underlying proposition is that the temporal evolution of each term is primarily determined by the growth factor dependence of $u_{z}$ and $\delta$ at the leading order. Using our standard notation where bared quantities correspond to a fiducial cosmology, we can write the equations to compute higher order corrections in a new cosmology as:
\bea \label{eq:estimatedAn}
A_g(k,\mu)
= j_1[b_1^2 \left(G_\delta/\bar G_\delta\right)^2\left(G_\Theta/\bar G_\Theta\right)\bar A_{b^2} \nonumber\\ 
+b_1\left(G_\delta/\bar G_\delta\right)\left(G_\Theta/\bar G_\Theta\right)^2\bar A_{b} \nonumber\\ 
+\left(G_\Theta/\bar G_\Theta\right)^3\bar A_L],
\nonumber\\ 
\nonumber\\ 
M_g(k,\mu)= \frac{1}{2}j_1^2[b_1^2 \left(G_\delta/\bar G_\delta\right)^2\left(G_\Theta/\bar G_\Theta\right)^2\bar M_{b^2} \nonumber\\ 
+b_1\left(G_\delta/\bar G_\delta\right)\left(G_\Theta/\bar G_\Theta\right)^3\bar M_{b}\nonumber\\ 
+\left(G_\Theta/\bar G_\Theta\right)^4\bar M_L].
\eea

\bsp	
\label{lastpage}
\end{document}